\documentclass[12pt,a4paper]{article}

\usepackage{amsmath}
\usepackage{amssymb}
\usepackage[nosort]{cite}
\usepackage[hyperref,bulletsep]{collect}
\def\gfxon{\usepackage[final]{graphicx}}

\gfxon

\newcommand{\Eqn}[1]{&\hspace{-0.5em}#1\hspace{-0.5em}&}

\font\numbers=cmss12
\font\upright=cmu10 scaled\magstep1
\def\stroke{\vrule height8pt width0.4pt depth-0.1pt}
\def\topfleck{\vrule height8pt width0.5pt depth-5.9pt}
\def\botfleck{\vrule height2pt width0.5pt depth0.1pt}
\def\Zmath{\vcenter{\hbox{\numbers\rlap{\rlap{Z}\kern 0.8pt\topfleck}\kern2.2pt
                   \rlap Z\kern 6pt\botfleck\kern 1pt}}}
\def\Qmath{\vcenter{\hbox{\upright\rlap{\rlap{Q}\kern
                   3.8pt\stroke}\phantom{Q}}}}
\def\Nmath{\vcenter{\hbox{\upright\rlap{I}\kern 1.7pt N}}}
\def\Cmath{\vcenter{\hbox{\upright\rlap{\rlap{C}\kern
                   3.8pt\stroke}\phantom{C}}}}
\def\Rmath{\vcenter{\hbox{\upright\rlap{I}\kern 1.7pt R}}}
\def\bbZ{\ifmmode\Zmath\else$\Zmath$\fi}
\def\bbQ{\ifmmode\Qmath\else$\Qmath$\fi}
\def\bbN{\ifmmode\Nmath\else$\Nmath$\fi}
\def\bbC{\ifmmode\Cmath\else$\Cmath$\fi}
\def\bbR{\ifmmode\Rmath\else$\Rmath$\fi}

\newcommand{\eps}{\epsilon}

\def\({\left(}
\def\){\right)}

\makeatletter
\let\old@makecaption=\@makecaption
\def\@makecaption{\small\old@makecaption}
\makeatother

\makeatletter \@addtoreset{equation}{section} \makeatother

\makeatletter
\let\old@startsection=\@startsection
\renewcommand{\@startsection}[6]{\old@startsection{#1}{#2}{#3}{#4}{#5}{#6\mathversion{bold}}}
\makeatother

\let\oldPhi=\Phi
\let\oldOmega=\Omega
\let\oldPsi=\Psi
\let\oldGamma=\Gamma
\let\oldDelta=\Delta
\let\oldSigma=\Sigma
\let\oldTheta=\Theta
\let\oldPi=\Pi
\renewcommand{\Phi}{\mathnormal{\oldPhi}}
\renewcommand{\Omega}{\mathnormal{\oldOmega}}
\renewcommand{\Psi}{\mathnormal{\oldPsi}}
\renewcommand{\Gamma}{\mathnormal{\oldGamma}}
\renewcommand{\Sigma}{\mathnormal{\oldSigma}}
\renewcommand{\Delta}{\mathnormal{\oldDelta}}
\renewcommand{\Theta}{\mathnormal{\oldTheta}}
\renewcommand{\Pi}{\mathnormal{\oldPi}}

\newcommand{\indup}[1]{_{\mathrm{#1}}}
\newcommand{\indups}[1]{_{\mathrm{\scriptscriptstyle #1}}}

\newcommand{\rep}[1]{{\mathbf{#1}}}
\newcommand{\matr}[2]{\left(\begin{array}{#1}#2\end{array}\right)}
\newcommand{\alg}[1]{\mathfrak{#1}}
\newcommand{\grp}[1]{\mathrm{#1}}

\newcommand{\sfrac}[2]{{\textstyle\frac{#1}{#2}}}
\newcommand{\half}{\sfrac{1}{2}}

\newcommand{\pexp}{\mathrm{P}\exp}

\newcommand{\superN}{\mathcal{N}}
\newcommand{\gym}{g\indups{YM}}

\newcommand{\order}[1]{\mathcal{O}(#1)}
\newcommand{\bigorder}[1]{\mathcal{O}\bigbrk{#1}}

\newcommand{\cdott}{\mathord{\cdot}}
\newcommand{\trans}{{\scriptscriptstyle\mathsf{T}}}

\newcommand{\Real}{\mathbb{R}}
\newcommand{\Comp}{\mathbb{C}}

\newcommand{\Integers}{\mathbb{Z}}

\newcommand{\lrbrk}[1]{\left(#1\right)}
\newcommand{\bigbrk}[1]{\bigl(#1\bigr)}
\newcommand{\Bigbrk}[1]{\Bigl(#1\Bigr)}
\newcommand{\brk}[1]{(#1)}

\newcommand{\bigcomm}[2]{\big[#1,#2\big]}

\newcommand{\state}[1]{\mathopen|#1\mathclose\rangle}
\newcommand{\bigstate}[1]{\bigl|#1\bigr\rangle}
\newcommand{\set}[1]{\{#1\}}

\newcommand{\opident}{\mathcal{I}}
\newcommand{\opperm}{\mathcal{S}}
\newcommand{\optrace}{\mathcal{K}}
\newcommand{\opproj}{\mathcal{P}}
\newcommand{\oprot}{\mathcal{J}}
\newcommand{\Rmatrix}{\mathcal{R}}
\newcommand{\hamdens}{\mathcal{H}}

\newcommand{\dil}{\mathbf{D}}
\newcommand{\genrot}{\mathbf{J}}
\newcommand{\genident}{\mathbf{I}}
\newcommand{\ham}{\mathbf{H}}
\newcommand{\optrans}{\mathbf{T}}
\newcommand{\bethegen}{\mathbf{B}}
\newcommand{\opcharge}{\mathbf{Q}}
\newcommand{\opshift}{\mathbf{U}}
\newcommand{\opmono}{\mathbf{\oldOmega}}
\newcommand{\oplax}[1][L]{\mathcal{#1}}
\newcommand{\indvec}{{\scriptscriptstyle\rep{V}}}
\newcommand{\indrep}{{\scriptscriptstyle\rep{R}}}
\newcommand{\indspin}{{\scriptscriptstyle\rep{S}}}

\newcommand{\dimn}{D}
\newcommand{\transfer}{T}
\newcommand{\charge}{Q}

\newcommand{\mono}{\Omega}
\newcommand{\gmat}{h}
\newcommand{\gvec}{\gmat_\indvec}
\newcommand{\gspin}{\gmat_\indspin}
\newcommand{\detshift}{\Psi}

\newcommand{\contour}[1][C]{\mathcal{#1}}

\newcommand{\bits}{R}
\newcommand{\bitp}{V}
\newcommand{\resolv}{G}
\newcommand{\resolvsl}{\makebox[0pt][l]{\hspace{0.06em}$/$}\resolv}
\newcommand{\rsing}{\tilde{\resolv}}
\newcommand{\rsingsl}{\makebox[0pt][l]{\hspace{0.06em}$/$}\rsing}
\newcommand{\Hresolv}{H}
\newcommand{\Hresolvsl}{\makebox[0pt][l]{\hspace{0.15em}$/$}\Hresolv}
\newcommand{\Hrsing}{\tilde{\Hresolv}}
\newcommand{\Hrsingsl}{\makebox[0pt][l]{\hspace{0.15em}$/$}\Hrsing}
\newcommand{\sheet}{p}
\newcommand{\sheetsl}{\makebox[0pt][l]{\hspace{0.06em}$/$}\sheet}
\newcommand{\qsheet}{q}
\newcommand{\qsheetsl}{\makebox[0pt][l]{\hspace{-0.00em}$/$}\qsheet}

\newcommand{\Xvec}{\vec{X}}
\newcommand{\gammavec}{\vec{\gamma}}
\newcommand{\intcurr}{a}
\newcommand{\grn}{m}

\newcommand{\nn}{\nonumber} 
\newcommand{\nln}{\nonumber\\}
\newcommand{\nl}{\nonumber\\&\hspace{-4\arraycolsep}&\mathord{}}
\newcommand{\nlnum}{\\&\hspace{-4\arraycolsep}&\mathord{}}
\newcommand{\earel}[1]{\mathrel{}&\hspace{-2\arraycolsep}#1\hspace{-2\arraycolsep}&\mathrel{}}
\newcommand{\eq}{\earel{=}}

\makeatletter
\newcommand{\newop}[2]{\def#1{\mathop{\operator@font #2}\nolimits}}
\newcommand{\newbin}[2]{\def#1{\mathbin{\operator@font #2}}}
\makeatother

\newop{\Re}{Re}
\newop{\Im}{Im}
\newop{\diag}{diag}
\newop{\rank}{rank}
\newop{\Tr}{Tr}
\newop{\sign}{sign}

\def\[{\begin{equation}}
\def\]{\end{equation}}
\def\<{\begin{eqnarray}}
\def\>{\end{eqnarray}}

\makeatletter
\def\mr@ignsp#1 {\ifx\:#1\@empty\else #1\expandafter\mr@ignsp\fi}%
\newcommand{\multiref}[1]{\begingroup
\xdef\mr@no@sparg{\expandafter\mr@ignsp#1 \: }%
\def\mr@comma{}%
\@for\mr@refs:=\mr@no@sparg\do{\mr@comma\def\mr@comma{,}\ref{\mr@refs}}%
\endgroup}
\makeatother

\newcommand{\hypref}[2]{\ifx\href\asklfhas #2\else\href{#1}{#2}\fi}

\newcommand{\secref}[1]{Sec.~\multiref{#1}}

\newcommand{\appref}[1]{App.~\multiref{#1}}

\newcommand{\figref}[1]{Fig.~\multiref{#1}}
\renewcommand{\eqref}[1]{(\multiref{#1})}

\newenvironment{bulletlist}{\begin{list}{$\bullet$}{\leftmargin1.5em\itemsep0pt}}{\end{list}}

\ifx\href\asklfhas\newcommand{\href}[2]{#2}\fi
\newcommand{\arxivno}[1]{\href{http://arxiv.org/abs/#1}{#1}}

\begin{document}

\thispagestyle{empty}
\begin{flushright}\footnotesize
\texttt{\arxivno{hep-th/0410253}}\\
\texttt{AEI 2004-087}\\
\texttt{LPTENS-04/44}\\
\texttt{NSF-KITP-04-115}\\
\texttt{PUTP-2138}\\
\end{flushright}
\vspace{1cm}

\renewcommand{\thefootnote}{\fnsymbol{footnote}}
\setcounter{footnote}{0}

\begin{center}
{\Large\textbf{\mathversion{bold}
Algebraic Curve for the SO(6) sector of AdS/CFT}\par}
\vspace{1cm}

\textsc{
N.~Beisert$^{a,b}$,
V.A.~Kazakov$^{c,}$\footnote{Membre de l'Institut Universitaire de France}
and K.~Sakai$^c$}
\vspace{5mm}

\textit{$^{a}$ Max-Planck-Institut f\"ur Gravitationsphysik,\\
Albert-Einstein-Institut,\\
Am M\"uhlenberg 1, 14476 Potsdam, Germany}
\vspace{2mm}

\textit{$^{b}$ Joseph Henry Laboratories, Princeton University,\\
Princeton, NJ 08544, USA}
\vspace{2mm}

\textit{$^{c}$
Laboratoire de Physique Th\'eorique\\
de l'Ecole Normale Sup\'erieure et l'Universit\'e Paris-VI,\\
Paris, 75231, France}\\[2mm]
\vspace{3mm}

\texttt{nbeisert@princeton.edu}\\
\texttt{kazakov,sakai@lpt.ens.fr}\par\vspace{1cm}

\vfill

\textbf{Abstract}\vspace{5mm}

\begin{minipage}{12.7cm}
We construct the general algebraic curve of degree four solving the
classical sigma model on $\mathbb{R}\times S^5$. 
Up to two loops it coincides with
the algebraic curve for the dual sector of scalar operators in
$\mathcal{N}=4$ SYM, also constructed here. 
We explicitly reproduce some particular solutions.

\end{minipage}

\vspace*{\fill}

\end{center}

\newpage
\setcounter{page}{1}
\renewcommand{\thefootnote}{\arabic{footnote}}
\setcounter{footnote}{0}


\section{Introduction}
\label{sec:Intro}

Since 't~Hooft's discovery of planar limit in field theories
\cite{'tHooft:1974jz}, the idea that the planar non-abelian gauge
theory could be exactly solvable, or integrable, always fascinated
string and field theorists. The analogy between planar graphs
of the 4D YM theory and the dynamics of string world sheets  of a
fixed low genus (described by some unidentified 2D CFT) already
pronounced in \cite{'tHooft:1974jz} lead to numerous attempts
aimed at the precise formulation of the YM string.

This circle of ideas lead to a dual, matrix model formulation
\cite{David:1985nj,Kazakov:1985ds} of  completely integrable toy
models of the string theory and 2D quantum gravity, having also
 their dual description in the usual world
sheet formalism
\cite{Knizhnik:1988ak,David:1988hj,Distler:1988jt}. Big planar
graphs of the matrix models find their description in terms of
the Liouville string theory proposed in \cite{Polyakov:1981rd}.
However the problem of such dual description is still open in the
original bosonic 4D YM theory.

Fortunately, these ideas are beginning to work in the 4D world 
thanks to supersymmetry. The work of Maldacena
\cite{Maldacena:1998re}, inspired by some earlier ideas and
observations \cite{Polyakov:1997tj,Gubser:1996de}, 
lead to a precise
formulation of the string/gauge duality 
at least in case of IIB superstrings on $AdS_5\times S^5$ and
the conformal $\superN=4$ SYM theory. 
A lot of work has been done since then to find the AdS/CFT dictionary
identifying the string analogs of physical operators and correlators
in the $\superN=4$ SYM theory, see \cite{Aharony:1999ti,D'Hoker:2002aw} for
reviews.

The two most recent important advances in this field
are the BMN correspondence \cite{Berenstein:2002jq} 
(see \cite{Pankiewicz:2003pg,Plefka:2003nb,Kristjansen:2003uy,Sadri:2003pr,Russo:2004kr} 
for reviews) and the semiclassical spinning strings duality \cite{Frolov:2003qc} 
(see \cite{Tseytlin:2003ii,Tseytlin:2004cj,Tseytlin:2004xa} for reviews).
These promise to enable quantitative comparisons between both
theories even though the duality is of a strong/weak type.
Let us only comment briefly on classical spinning strings
on $AdS_5\times S^5$ which were investigated 
in \cite{Gubser:2002tv,Frolov:2002av,Russo:2002sr,Minahan:2002rc,Frolov:2003xy,Arutyunov:2003uj}.
These were argued to be dual to long SYM operators
first investigated in \cite{Beisert:2003xu,Beisert:2003ea}
where a remarkable agreement was found up to two-loops.
Agreement for other particular examples 
\cite{Engquist:2003rn,Kristjansen:2004ei,Smedback:1998yn,Kristjansen:2004za}
as well as at the level of the Hamiltonian
\cite{Kruczenski:2003gt,Kruczenski:2004kw,Hernandez:2004uw,Stefanskijr.:2004cw,Hernandez:2004kr,Kruczenski:2004cn,Kazakov:2004qf,Kazakov:2004nh}
was obtained until discrepancies surfaced at three-loops,
first in the (near) BMN correspondence \cite{Callan:2003xr,Callan:2004uv,Callan:2004ev,Callan:2004dt}, 
later also for spinning strings \cite{Serban:2004jf}.
This problem, which might be the order-of-limits issue 
explained in \cite{Beisert:2004hm},
is not resolved at the moment. We will observe further evidence
of two-loop agreement/three-loop discrepancy
in this work.
\bigskip

Luckily, in the last few years the first
signs of integrability were observed on both sides of the duality.
The first, striking observation of integrability in $\superN=4$ SYM 
was made by Minahan and Zarembo \cite{Minahan:2002ve}. They
investigated the sector of single-trace local operators of
$\superN=4$ supersymmetric gauge theory composed from scalars
\[
\Tr \Phi_{m_1} \Phi_{m_2} \dots \Phi_{m_L}.
\]
It was found that the planar one-loop dilatation operator, which
measures their anomalous scaling dimensions, is the Hamiltonian of
an integrable spin chain. This chain has $\alg{so}(6)$ symmetry and
the spins transform in the vector representation. A basis of the
spin chain Hilbert space is thus given by the states
\[
\state{m_1,m_2,\ldots,m_L}
\]
which correspond, up to cyclic permutations, to single-trace local
operators. It was subsequently shown that integrability not only
extends to all local single-trace operators of $\superN=4$ SYM
\cite{Beisert:2003jj,Beisert:2003yb}
(extending earlier findings of integrability in gauge theory, 
c.f.~\cite{Belitsky:2004cz} for a review), 
but more surprisingly also
to higher loops (at least in some subsectors)
\cite{Beisert:2003tq,Beisert:2003ys}
(see also \cite{Klose:2003qc} for integrability in a related theory).
The hypothesis of all-loop integrability 
together with input from the BMN conjecture \cite{Berenstein:2002jq}
has allowed to make precise predictions of
higher-loop scaling dimensions 
\cite{Beisert:2003tq,Beisert:2003jb,Serban:2004jf,Beisert:2004hm}, 
which have just recently been verified 
by explicit computations \cite{Eden:2004ua}, 
see also \cite{Moch:2004pa,Kotikov:2004er}.
Integrability and the Bethe ansatz was also an essential tool 
in obtaining scaling dimensions for states dual to spinning strings.
For reviews of gauge theory results and integrability, 
see \cite{Beisert:2004ry,Beisert:2004yq}.
\smallskip

Integrability in string theory on $AdS_5\times S^5$,
whose sigma model was explicitly formulated in \cite{Metsaev:1998it,Metsaev:2000bj}, 
is based on a so-called \emph{Lax pair}, a family of flat
connections of the two-dimensional world sheet theory. Its existence
is a common feature of sigma models on coset spaces and it can be
used to construct Pohlmeyer charges
\cite{Pohlmeyer:1975nb,Luscher:1977rq}. These multilocal charges
have been discussed in the context of classical bosonic string
theory on $AdS_5\times S^5$ in \cite{Mandal:2002fs}, while a family
of flat connections for the corresponding superstring 
was identified in \cite{Bena:2003wd}. 
The Lax pair of the string sigma model 
was first put to use in \cite{Kazakov:2004qf}
in the case of the restricted target space
$\Real\times S^3$, where $\Real$ represents the time coordinate of
$AdS_5$. There the analytic properties of the monodromy 
of the flat connections around the closed string were investigated
and translated to integral equations similar to the
ones encountered in the algebraic Bethe ansatz of gauge theory.
This lead to the first rigorous proof of two-loop agreement
of scaling dimensions for an entire sector of states.
The possibility to quantize the sigma model by 
discretizing the continuous equations, 
by analogy with the finite chain Bethe ansatz for gauge theory, 
was suggested in \cite{Kazakov:2004qf}.
A concrete proposal for the quantization of these 
equations was given in \cite{Arutyunov:2004vx}.
It reproduces the near BMN results of
\cite{Callan:2003xr,Callan:2004uv,Callan:2004ev,Callan:2004dt}
and, even more remarkably, a generic $\sqrt[4]{\lambda}$ behavior 
at large coupling in agreement predicted in \cite{Gubser:1998bc}.
Curiously, this proposal appears to have a 
spin chain correspondence in the weak-coupling extrapolation
\cite{Beisert:2004jw}, which however does
not agree with gauge theory.
\bigskip

Integrability is usually closely related with the theory of
algebraic curves. 
In most of cases, integrable models of 2D field theory 
or integrable matrix models are completely characterized by
their algebraic curves. 
Often the algebraic curve unambiguously
defines also the quantum version of the model. We also know many
examples of 
algebraic curves characterizing the massive 4D $\superN=2$ SYM theories, 
starting from the famous Seiberg-Witten curve \cite{Seiberg:1994rs}, 
as well as for the $\superN=1$ SYM theories \cite{Dijkgraaf:2002fc}. 
However, the curves describe in that cases
only particular BPS sectors  of the gauge theories characterized by
massive moduli. The $\superN=4$ SYM CFT  gives us the first hope
for an  entirely integrable 4D gauge theory, including the non-BPS
states.

There is an increasing evidence that the full integrability on
both sides of AdS/CFT duality might be governed by similar algebraic curves. 
On the SYM side, such a curve for the quasi-momentum
can be built so far only for small 't~Hooft coupling. 
Its mere existence is due to the perturbative
integrability, confirmed up to three-loops \cite{Beisert:2003ys,Serban:2004jf} 
and hopefully existing for all-loops \cite{Beisert:2004hm} and even non-perturbatively. 
On the string sigma model side, we already know the entire classical curve,
also for the quasi-momentum, for the $\Real\times S^3$
\cite{Kazakov:2004qf} and $AdS_3\times S^1$ \cite{Kazakov:2004nh}
sectors of the theory, dual to $\alg{su}(2)$ and 
$\alg{sl}(2)$ sectors of the gauge theory, respectively.

In this paper will make a new step in the direction of construction
of the full algebraic curve of the classical $AdS_5\times S^5$
sigma model and of its perturbative counterpart on the SYM side
following from the one-loop integrability of the full SYM theory found
in \cite{Beisert:2003jj,Beisert:2003yb}. 
We will accomplish this program for the $\alg{so}(6)$ sector of SYM, 
which is closed at one-loop as well as in the thermodynamic limit \cite{Minahan:2004ds}, 
and its dual, the sigma model for the string on $\Real_t\times S^5$. 
We will show in both theories that the projection of the algebraic curve 
onto the complex plane is a Riemann surface with four sheets corresponding 
to the four-dimensional chiral spinor representation of $\grp{SU}(4)\sim \grp{SO}(6)$. 
In the SYM case such a curve solves the classical limit of the corresponding 
Bethe equations. 
We will identify and fix all the parameters of 
this curve on both sides of AdS/CFT.

On the gauge side, the main tool of our analysis will be 
transfer matrices in the algebraic Bethe ansatz framework
(see e.g.~\cite{Faddeev:1996iy} for an introduction).
These can be derived from a Lax-type formulation 
of the Bethe equations proposed in \cite{Krichever:1997qd} 
for the $\alg{su}(m)$ algebras. 
On the string side,
we will construct the so-called finite-gap solution 
\cite{Its:1975aa,Dubrovin:1976xx}
of the $\Real\times S^5$ sigma model, 
also based on the Lax method \cite{Novikov:1984id}.

This paper is organized as follows.
In \secref{sec:Spins} we review 
the vector $\alg{so}(6)$ spin chain
which is dual to the one-loop planar dilatation operator
of $\superN=4$ SYM in the sector of local operators
composed from scalar fields.
Special attention is paid to 
various transfer matrices and
their analytic properties.
In the thermodynamic limit
will then construct the generic solution 
in terms of an algebraic curve
and illustrate by means of two examples.
All this is meant to serve as an introduction 
to the treatment of the 
string sigma model in the sections to follow. 
We start in \secref{sec:Sigma} by investigating
the properties of the monodromy of the Lax pair
around the string. They are similar to
the ones encountered for the spin chain, but
there is an additional symmetry for the
spectral parameter.
These are used in \secref{sec:Alg.EqSM} to
reconstruct an algebraic curve associated
to each solution of the equations of motion. 
We will show that the algebraic curve
is uniquely defined by the analytic
properties of the monodromy matrix.
It thus turns out that solutions 
are completely characterized by their 
$\contour[B]$-periods (mode numbers)
and filling fractions (excitation numbers).
In \secref{sec:Bethe} we 
construct equations similar to the Bethe equations of
the spin chain. These are equivalent to the algebraic curve, 
but allow for a particle/scattering interpretation.%
\footnote{These equations were 
proposed independently by 
M.~Staudacher and confirmed by comparing to explicit
solutions of the string equations of motion
\cite{Staudacher:2004qq}.}
We conclude in \secref{sec:Concl}. 
In the appendix we present results which are
not immediately important for the AdS/CFT correspondence. 
Let us mention in particular \secref{sec:RS2} where 
we apply our formalism to the quite simple, 
yet interesting case of $\Real\times S^2$ and
compare to particular limits of known solutions.

\section{The $\alg{so}(6)$ Spin Chain}
\label{sec:Spins}

In this section we will review the integrable $\alg{so}(6)$ spin chain
with spins transforming in the vector representation
of the symmetry algebra.
Due to the isomorphism
of the algebras $\alg{so}(6)$ and $\alg{su}(4)$
we can rely on a vast collection of results on integrable
spin chain with unitary symmetry algebra.
We will use these firm facts to gain a better understanding of the
Bethe ansatz in the thermodynamic limit for higher-rank symmetry
groups. We identify some key properties of the resolvents which
describe the distribution of Bethe roots.
In the following chapters we will derive similar properties for
the sigma model which later on will be used
to (re)construct a similar Bethe ansatz for classical string theory.

\subsection{Spin Chains Operators}
\label{sec:Spins.Ops}

\paragraph{Dilatation Operator.}

Let us consider single-trace local operators of $\superN=4$ SYM
composed from $L$ scalars without derivatives.
These are isomorphic to states of a quantum $\alg{so}(6)$ spin chain
with spins transforming in the vector ($\rep{6}$) representation.
The planar one-loop dilatation generator of $\superN=4$ SYM
closes on these local operators,
it can thus be written in terms
of a spin chain Hamiltonian $\ham$ as follows
\[\label{eq:Dil}
\dil=L+g^2 \ham+\cdots,\qquad
g^2=\frac{\gym^2 N}{8\pi^2}\,.
\]
The Hamiltonian was derived in \cite{Minahan:2002ve},
it is given by the nearest-neighbor interaction $\hamdens$
\footnote{We shall distinguish between
global and local spin chain operators
by boldface and curly letters, respectively.
For their eigenvalues we shall use regular letters.}
\[\label{eq:Ham}
\ham=\sum_{p=1}^L\hamdens_{p,p+1},
\qquad
\hamdens
=2\opproj^{\rep{15}}+3\opproj^{\rep{1}}
=\opident-\opperm+\half\optrace^{\rep{6},\rep{6}}
.
\]
The spin chain operators
$\opproj^{\rep{20'}},\opproj^{\rep{15}},\opproj^{\rep{1}}$
project to the modules $\rep{1},\rep{15},\rep{20'}$ which
appear in the tensor product of two spins, $\rep{6}\times\rep{6}$.
These can be written using the operators $\opident,\opperm,\optrace^{\rep{6},\rep{6}}$
which are the identity, the permutation of two spins and
the trace
$(\optrace^{\rep{6},\rep{6}})^{ij}{}_{kl}=\delta^{ij}\delta_{kl}$
of two $\alg{so}(6)$ vectors, respectively
\[\label{eq:Proj}
\opproj^{\rep{20'}}=\half\opident+\half\opperm-\sfrac{1}{6}\optrace^{\rep{6},\rep{6}},\qquad
\opproj^{\rep{15}}=\half\opident-\half\opperm,\qquad
\opproj^{\rep{1}}=\sfrac{1}{6}\optrace^{\rep{6},\rep{6}}.
\]
%

\paragraph{R-matrices.}

Minahan and Zarembo have found out that
the Hamiltonian \eqref{eq:Ham} obtained from $\superN=4$ SYM
is integrable~\cite{Minahan:2002ve}:
It coincides with the Hamiltonian of a standard
$\alg{so}(\grn)$ spin chain investigated
by Reshetikhin \cite{Reshetikhin:1983vw,Reshetikhin:1985vd}.
In this section we focus on the case $\grn=6$,
but we will present generalizations of some expressions
in \appref{sec:SpinSOm}.
Integrability for a standard quantum spin chain means that the
(nearest-neighbor) Hamiltonian density
$\hamdens$
can be obtained via
\[\label{eq:HamFromR}
\Rmatrix(u)=
\opperm\bigbrk{1+iu\hamdens+\order{u^2}}
\]
from the expansion of an R-matrix $\Rmatrix(u)$
(see e.g.~\cite{Beisert:2004ry} 
for an introduction in the context of gauge theory)
which satisfies the Yang-Baxter relation
\[\label{eq:YBE}
\Rmatrix_{12}(u_1-u_2)\Rmatrix_{13}(u_1-u_3)\Rmatrix_{23}(u_2-u_3)
=
\Rmatrix_{23}(u_2-u_3)\Rmatrix_{13}(u_1-u_3)\Rmatrix_{12}(u_1-u_2).
\]
In addition to the YBE, we would like to demand the inversion formula
\[\label{eq:Rinverse}
\Rmatrix_{12}(u_1-u_2)\Rmatrix_{21}(u_2-u_1)=\opident.
\]
The R-matrix for two vectors of $\alg{so}(6)$ is given by
\cite{Reshetikhin:1983vw,Reshetikhin:1985vd}%
\footnote{This R-matrix coincides with the
one given in \cite{Minahan:2002ve,Reshetikhin:1983vw,Reshetikhin:1985vd}
up to an overall factor and a redefinition of $u$.
The redefinition is needed to comply with
\eqref{eq:Rinverse}.}
\<\label{eq:R66}
\Rmatrix^{\rep{6},\rep{6}}(u)\eq
\opproj^{\rep{20'}}
+\frac{u-i}{u+i}\,\opproj^{\rep{15}}
+\frac{(u-i)(u-2i)}{(u+i)(u+2i)}\,\opproj^{\rep{1}}
\nln\eq
\frac{i}{u+i}\,\opperm
+\frac{u}{u+i}\,\opident
-\frac{iu}{(u+i)(u+2i)}\,\optrace^{\rep{6},\rep{6}}.
\>
It yields, via \eqref{eq:HamFromR}, the spin chain Hamiltonian
\eqref{eq:Ham}.
For completeness, we shall also state the
R-matrices between a vector and a (anti)chiral spinor
\<\label{eq:R46}
\Rmatrix^{\rep{4},\rep{6}}(u)\eq
\opproj^{\rep{20}}
+\frac{u-\frac{3i}{2}}{u+\frac{3i}{2}}\,\opproj^{\rep{\bar 4}}
=\opident
-\frac{\sfrac{i}{2}}{u+\frac{3i}{2}}\,\optrace^{\rep{4},\rep{6}},
\nln
\Rmatrix^{\rep{\bar 4},\rep{6}}(u)\eq
\opproj^{\rep{\overline{20}}}
+\frac{u-\frac{3i}{2}}{u+\frac{3i}{2}}\,\opproj^{\rep{4}}
=\opident
-\frac{\sfrac{i}{2}}{u+\frac{3i}{2}}\,\optrace^{\rep{\bar 4},\rep{6}}.
\>
These also satisfy the Yang-Baxter equation
\eqref{eq:YBE} when we assign any of the three representations
$\rep{6},\rep{4},\rep{\bar 4}$ to the three spaces labeled by $1,2,3$
(the remaining R-matrices between $\rep{4}$ and $\rep{\bar 4}$ can be
found in \appref{sec:Rmatrices}).
Here we can again express the projectors in terms of
identity $\opident$ and
spin-trace $\optrace^{\rep{4},\rep{6}},\optrace^{\rep{\bar 4},\rep{6}}$
defined with Clifford gamma matrices by
$(\optrace^{\rep{4},\rep{6}})^{\beta j}{}_{\alpha i}=(\gamma^j\gamma_i)^\beta{}_\alpha$
and
$(\optrace^{\rep{\bar 4},\rep{6}})^{\dot \beta j}{}_{\dot \alpha i}=(\gamma^j\gamma_i)^{\dot\beta}{}_{\dot\alpha}$
\[\label{eq:ProjSpin}
\opproj^{\rep{20}}=\opident-\sfrac{1}{6}\optrace^{\rep{4},\rep{6}},\quad
\opproj^{\rep{\bar 4}}=\sfrac{1}{6}\optrace^{\rep{4},\rep{6}},\qquad
\opproj^{\rep{\overline{20}}}=\opident-\sfrac{1}{6}\optrace^{\rep{\bar 4},\rep{6}},\quad
\opproj^{\rep{4}}=\sfrac{1}{6}\optrace^{\rep{\bar 4},\rep{6}}.
\]
%

\paragraph{Transfer matrices.}

An R-matrix describes elastic scattering of two spins,
it gives the phase shift for both spins at the same time.
For a spin chain, it can also be viewed
as a (quantum) $\grp{SO}(6)$ lattice link variable.
If we chain up the link variables around the closed
chain, we obtain a Wilson loop.
The open Wilson loop is also known as the monodromy matrix
$\opmono_{a}(u)$, where $a$ labels the auxiliary space of the
Wilson line. The complex number $u$ of the Wilson loop
is the spectral parameter.
In the
$\rep{6}$ representation it is convenient to use the combination
\[\label{eq:Mono6}
\opmono^{\rep{6}}_{a}(u)=
\frac{(u+i)^L}{u^L}\,
\Rmatrix_{a,1}^{\rep{6},\rep{6}}(u-i) \Rmatrix_{a,2}^{\rep{6},\rep{6}}(u-i)
\cdots  \Rmatrix_{a,L}^{\rep{6},\rep{6}}(u-i).
\]
The closed Wilson loop is also known as the
transfer matrix
\[
\label{eq:Transfer6}
\optrans_{\rep{6}}(u)
=
\Tr\nolimits_{a} \opmono_{a}^{\rep{6}}(u).
\]
We can also write the monodromy and transfer matrices
in the spinor representations
\<
\label{eq:Mono4}
\opmono^{\rep{4}}_{a}(u)\eq
\frac{(u+\sfrac{i}{2})^L}{u^L}\,
\Rmatrix^{\rep{4},\rep{6}}_{a,1}(u-i)
\cdots
\Rmatrix^{\rep{4},\rep{6}}_{a,L}(u-i),
\qquad
\optrans_{\rep{4}}(u)=
\Tr_a \opmono^{\rep{4}}_{a}(u),
\nln
\opmono^{\rep{\bar 4}}_{a}(u)\eq
\frac{(u+\sfrac{i}{2})^L}{u^L}\,
\Rmatrix^{\rep{\bar 4},\rep{6}}_{a,1}(u-i)
\cdots
\Rmatrix^{\rep{\bar 4},\rep{6}}_{a,L}(u-i),
\qquad
\optrans_{\rep{\bar 4}}(u)=
\Tr_a \opmono^{\rep{\bar 4}}_{a}(u).
\qquad
\>
The transfer matrices commute due to the
Yang-Baxter relation \eqref{eq:YBE},
as one can easily convince oneself
by inserting the inversion
relation \eqref{eq:Rinverse}
between both Wilson loops.

\paragraph{Local Charges.}

The transfer matrices give rise to commuting charges
when expanded in powers of $u$.
Local charges $\opcharge_r$ are obtained from $\optrans_{\rep{6}}(u)$,
(the representation of the Wilson loop coincides with
the spin representation)
when expanded around $u=i$, i.e.
\[
\label{eq:LocalCharges}
\frac{(u+i)^L}{(u+2i)^L}\,\optrans_{\rep{6}}(u+i)
=
\opshift\,\exp
i\sum_{r=2}^\infty
u^{r-1}\opcharge_{r}
\]
The operator $\opshift$ is a global shift operator,
it shifts all spins by one site.
For gauge theory we are interested in the
subspace of states with zero momentum,
i.e.~with eigenvalue 
%
\[\label{eq:MomConstr}
U=1
\]
of $\opshift$.
This is the physical state condition.
The second charge $\opcharge_2$ is the Hamiltonian
\[\label{eq:Q2Ham}
\opcharge_2=\ham;
\]
this fact can be derived from \eqref{eq:HamFromR}.
Therefore all transfer matrices and charges are also conserved quantities.
The third charge $\opcharge_3$ leads to pairing of states,
a peculiar property of integrable spin chains \cite{Grabowski:1995rb,Beisert:2003tq}.

\paragraph{Global Charges.}

An interesting value of the spectral parameter is $u=\infty$
where one finds the generators
of the symmetry algebra, in our case of $\alg{so}(6)=\alg{su}(4)$.
Let us first note the
symmetry generators for different representations
\[\label{eq:SymGenRep}
\oprot^{\rep{6},\rep{6}}=\opperm-\optrace^{\rep{6},\rep{6}},
\qquad
\oprot^{\rep{4},\rep{6}}=\sfrac{1}{2}\opident-\sfrac{1}{2}\optrace^{\rep{4},\rep{6}},
\qquad
\oprot^{\rep{\bar 4},\rep{6}}=\sfrac{1}{2}\opident-\sfrac{1}{2}\optrace^{\rep{\bar 4},\rep{6}}.
\]
%
The two vector spaces on which these operators act
are interpreted as follows:
The first ($\rep{6},\rep{4},\rep{\bar 4}$)
specifies the parameters for the rotation which
acts on the second space ($\rep{6}$ in all cases).
To be more precise,
consider the operator $\oprot^{\alpha i}{}_{\beta j}$
where $\alpha,\beta$ belong to $\rep{6},\rep{4},\rep{\bar 4}$
while $i,j$ belong to the second $\rep{6}$.
In all three cases,
the indices $\alpha,\beta$ can be combined into an index of the
adjoint representation which determines the parameters
of the rotation. Note that the expressions
in \eqref{eq:SymGenRep} respect the symmetry properties of the
adjoint representation in the tensor products $\rep{6}\times\rep{6}$,
$\rep{4}\times\rep{\bar 4}$ and $\rep{\bar 4}\times\rep{4}$.

We now express the R-matrices in terms of these symmetry generators
and find
\<\label{eq:RRot}
\Rmatrix^{\rep{6},\rep{6}}(u)\eq
\frac{u}{u+i}\,\opident
+\frac{iu}{(u+i)(u+2i)}\,\oprot^{\rep{6},\rep{6}}
-\frac{2}{(u+i)(u+2i)}\,\opperm,
\nln
\Rmatrix^{\rep{4},\rep{6}}(u)\eq
\frac{u+i}{u+\frac{3i}{2}}\,\opident
+\frac{i}{u+\frac{3i}{2}}\,\oprot^{\rep{4},\rep{6}},
\nln
\Rmatrix^{\rep{\bar 4},\rep{6}}(u)\eq
\frac{u+i}{u+\frac{3i}{2}}\,\opident
+\frac{i}{u+\frac{3i}{2}}\,\oprot^{\rep{\bar 4},\rep{6}}.
\>
One then finds that the expansion at infinity
\<\label{eq:RInfty}
\frac{u+i}{u}\,\Rmatrix^{\rep{6},\rep{6}}(u-i)\eq
\opident+\frac{i}{u}\oprot^{\rep{6},\rep{6}}
+\order{1/u^2},
\nln
\frac{u+\sfrac{i}{2}}{u}\,\Rmatrix^{\rep{4},\rep{6}}(u-i)\eq
\opident+\frac{i}{u}\oprot^{\rep{4},\rep{6}}
+\order{1/u^2},
\nln
\frac{u+\sfrac{i}{2}}{u}\,\Rmatrix^{\rep{\bar 4},\rep{6}}(u-i)\eq
\opident+\frac{i}{u}\oprot^{\rep{\bar 4},\rep{6}}
+\order{1/u^2}.
\>
Here we have used the same shifts and prefactors as
in the construction of the monodromy matrices
\eqref{eq:Mono6,eq:Mono4}.
In all three cases, the monodromy matrix therefore has the
expansion
\[\label{eq:MonoInfty}
\opmono^{\indrep}_a(u)=
\genident_a+\frac{i}{u}\,\genrot^{\indrep}_a
+\order{1/u^2}
\]
at $u=\infty$ with the global rotation operators
\[\label{eq:RotGlob}
\genrot^{\indrep}_a=\sum_{p=1}^L\oprot^{\indrep,\rep{6}}_{a,p}.
\]
If we expand further around $u=\infty$ we will find
multi-local operators along the spin chain. These are the
generators of the Yangian, see e.g.~\cite{Bernard:1993ya,MacKay:2004tc}
and \cite{Dolan:2003uh,Dolan:2004ps,Agarwal:2004sz} in
the context of $\superN=4$ SYM.

\subsection{Bethe ansatz}
\label{sec:Spins.Bethe}

\paragraph{States.}

Consider a spin chain state
\[\label{eq:BetheState}
\bigstate{\set{u_{j,k}},L}\sim
\lrbrk{\prod_{j=1}^3\prod_{k=1}^{K_j} \bethegen_{j}(u_{j,k}) }\state{0,L}.
\]
The vacuum state $\state{0,L}$ is the tensor product of
$L$ spins in a highest weight configuration of the $\rep{6}$.
In other words, $\state{0,L}$ is the ferromagnetic vacuum with
all spins aligned to give a maximum total spin.
The operator $\bethegen_j(u)$, $u\in\Comp$, $j=1,2,3$, creates an excitation
with rapidity $u$ and
quantum numbers of the $j$-th simple root of $\alg{su}(4)$.
A state with a given weight $[r_1,r_2,r_3]$ (Dynkin labels) of $\alg{su}(4)$
has excitation numbers $K_j$ given by (c.f.~\cite{Beisert:2003yb}):
\<\label{eq:BetheExcite}
K_1\eq\half L-\sfrac{3}{4}r_1-\half r_2-\sfrac{1}{4}r_3,
\nln
K_2\eq \phantom{\half}L-\half r_1-\phantom{\half}r_2-\half r_3,
\nln
K_3\eq\half L-\sfrac{1}{4}r_1-\half r_2-\sfrac{3}{4}r_3.
\>

\paragraph{Transfer matrices.}

Now let us assume that the state $\state{\set{u_{j,k}},L}$
is an eigenstate of all transfer matrices $\optrans_{\indrep}(u)$
for all values of the spectral parameter $u$.
Then it can be shown that the eigenvalue of the transfer matrix
in the $\rep{6}$ representation is given by
(see \appref{sec:Antisym} for a derivation)
\<
\label{eq:BetheTransfer6}
\transfer_{\rep{6}}(u)\eq
\mathbin{\phantom{+}}
\phantom{\frac{\bits_1(u-i)}{\bits_1(u)}}\,
\frac{\bits_2(u-\sfrac{3i}{2})}{\bits_2(u-\sfrac{i}{2})}
\phantom{\frac{\bits_3(u-i)}{\bits_3(u)}}\,
\phantom{\frac{\bitp(u-i)}{\bitp(u)}}\,
\frac{\bitp(u+i)}{\bitp(u)}
\nl
+
\frac{\bits_1(u-i)}{\bits_1(u)}\,
\frac{\bits_2(u+\sfrac{i}{2})}{\bits_2(u-\sfrac{i}{2})}\,
\frac{\bits_3(u-i)}{\bits_3(u)}\,
\frac{\bitp(u-i)}{\bitp(u)}\,
\frac{\bitp(u+i)}{\bitp(u)}
\nl
+
\frac{\bits_1(u+i)}{\bits_1(u)}\,
\phantom{\frac{\bits_2(u+\sfrac{i}{2})}{\bits_2(u-\sfrac{i}{2})}}\,
\frac{\bits_3(u-i)}{\bits_3(u)}\,
\frac{\bitp(u-i)}{\bitp(u)}\,
\frac{\bitp(u+i)}{\bitp(u)}
\nl
+
\frac{\bits_1(u-i)}{\bits_1(u)}\,
\phantom{\frac{\bits_2(u+\sfrac{i}{2})}{\bits_2(u-\sfrac{i}{2})}}\,
\frac{\bits_3(u+i)}{\bits_3(u)}\,
\frac{\bitp(u-i)}{\bitp(u)}\,
\frac{\bitp(u+i)}{\bitp(u)}
\nl
+
\frac{\bits_1(u+i)}{\bits_1(u)}\,
\frac{\bits_2(u-\sfrac{i}{2})}{\bits_2(u+\sfrac{i}{2})}\,
\frac{\bits_3(u+i)}{\bits_3(u)}\,
\frac{\bitp(u-i)}{\bitp(u)}\,
\frac{\bitp(u+i)}{\bitp(u)}
\nl
+
\phantom{\frac{\bits_1(u-i)}{\bits_1(u)}}\,
\frac{\bits_2(u+\sfrac{3i}{2})}{\bits_2(u+\sfrac{i}{2})}
\phantom{\frac{\bits_3(u-i)}{\bits_3(u)}}\,
\frac{\bitp(u-i)}{\bitp(u)}\,.
\>
Note that the monodromy matrix
$\opmono^{\rep{6}}_a(u)$ is a $\rep{6}\times\rep{6}$ matrix in
the auxiliary space labelled by $a$.
This explains why the
transfer matrix as its trace consists of six terms.
For convenience, we have defined the functions
$\bits_j(u),\bitp(u)$
\[\label{eq:bitdef}
\bits_j(u)=\prod_{k=1}^{K_j}(u-u_{j,k}),
\qquad
\bitp(u)=u^L,
\]
which describe two and one-particle scattering, respectively.
Let us also state the eigenvalues of the
transfer matrices in the spinor representations
\<
\label{eq:BetheTransfer4}
\transfer_{\rep{4}}(u)\eq
\mathbin{\phantom{+}}
\frac{\bits_1(u-\frac{3i}{2})}{\bits_1(u-\frac{i}{2})}\,
\phantom{\frac{\bits_2(u-i)}{\bits_2(u)}}\,
\phantom{\frac{\bits_3(u-\frac{i}{2})}{\bits_3(u+\frac{i}{2})}}\,
\frac{\bitp(u+\frac{i}{2})}{\bitp(u)}
\nl
+
\frac{\bits_1(u+\frac{i}{2})}{\bits_1(u-\frac{i}{2})}\,
\frac{\bits_2(u-i)}{\bits_2(u)}\,
\phantom{\frac{\bits_3(u-\frac{i}{2})}{\bits_3(u+\frac{i}{2})}}\,
\frac{\bitp(u+\frac{i}{2})}{\bitp(u)}
\nl
+
\phantom{\frac{\bits_1(u+\frac{i}{2})}{\bits_1(u-\frac{i}{2})}}\,
\frac{\bits_2(u+i)}{\bits_2(u)}\,
\frac{\bits_3(u-\frac{i}{2})}{\bits_3(u+\frac{i}{2})}\,
\frac{\bitp(u-\frac{i}{2})}{\bitp(u)}
\nl
+
\phantom{\frac{\bits_1(u+\frac{i}{2})}{\bits_1(u-\frac{i}{2})}}\,
\phantom{\frac{\bits_2(u+i)}{\bits_2(u)}}\,
\frac{\bits_3(u+\frac{3i}{2})}{\bits_3(u+\frac{i}{2})}\,
\frac{\bitp(u-\frac{i}{2})}{\bitp(u)}
\>
and its conjugate
\<
\label{eq:BetheTransfer4bar}
\transfer_{\rep{\bar 4}}(u)\eq
\mathbin{\phantom{+}}
\phantom{\frac{\bits_1(u+\frac{i}{2})}{\bits_1(u-\frac{i}{2})}}\,
\phantom{\frac{\bits_2(u+i)}{\bits_2(u)}}\,
\frac{\bits_3(u-\frac{3i}{2})}{\bits_3(u-\frac{i}{2})}\,
\frac{\bitp(u+\frac{i}{2})}{\bitp(u)}
\nl
+
\phantom{\frac{\bits_1(u+\frac{i}{2})}{\bits_1(u-\frac{i}{2})}}\,
\frac{\bits_2(u-i)}{\bits_2(u)}\,
\frac{\bits_3(u+\frac{i}{2})}{\bits_3(u-\frac{i}{2})}\,
\frac{\bitp(u+\frac{i}{2})}{\bitp(u)}
\nl
+
\frac{\bits_1(u-\frac{i}{2})}{\bits_1(u+\frac{i}{2})}\,
\frac{\bits_2(u+i)}{\bits_2(u)}\,
\phantom{\frac{\bits_3(u-\frac{i}{2})}{\bits_3(u+\frac{i}{2})}}\,
\frac{\bitp(u-\frac{i}{2})}{\bitp(u)}
\nl
+
\frac{\bits_1(u+\frac{3i}{2})}{\bits_1(u+\frac{i}{2})}\,
\phantom{\frac{\bits_2(u-i)}{\bits_2(u)}}\,
\phantom{\frac{\bits_3(u-\frac{i}{2})}{\bits_3(u+\frac{i}{2})}}\,
\frac{\bitp(u-\frac{i}{2})}{\bitp(u)}\,.
\>
%

\paragraph{Bethe Equations.}

As they stand, the above expressions for $\transfer_{\indrep}(u)$
are rational functions of $u$.
{}From the definition
of $\optrans_{\indrep}(u)$ in \eqref{eq:Mono6,eq:Transfer6,eq:Mono4}
and $\Rmatrix^{\indrep,\rep{6}}(u)$ in \eqref{eq:R66}
it follows that $\transfer_{\indrep}(u)$ is a polynomial
in $1/u$
\footnote{The expansion in $1/u$ instead of the more common one in $u$
is due to our definitions.}
of degree at most $2L$
(for $\rep{R}=\rep{6}$; for $\rep{R}=\rep{4}$ or
$\rep{R}=\rep{\bar 4}$ the maximum degree is $L$).
This means that a state $\state{\set{u_{j,k}},L}$
cannot be an eigenstate of the transfer matrices if
$\transfer_{\indrep}(u)$ has poles anywhere in the complex plane
except the obvious singularity at $u=0$
from the definition of $\opmono^{\indrep}_a(u)$.
{}From the cancellation of poles for all $1/u\in\Comp$ one can derive
a set of equations which in effect allowed rapidities
$\set{u_{j,k}}$ to make up an eigenstate. These are precisely the Bethe
equations \cite{Reshetikhin:1983vw,Reshetikhin:1985vd,Ogievetsky:1986hu}
\<
\label{eq:BetheEquations}
\frac{\bits_1(u_{1,k}+i)}{\bits_1(u_{1,k}-i)}\,
\frac{\bits_2(u_{1,k}-\sfrac{i}{2})}{\bits_2(u_{1,k}+\sfrac{i}{2})}\,
\phantom{\frac{\bits_3(u_{1,k}-\sfrac{i}{2})}{\bits_3(u_{1,k}+\sfrac{i}{2})}}
\eq -1,
\nln
\frac{\bits_1(u_{2,k}-\sfrac{i}{2})}{\bits_1(u_{2,k}+\sfrac{i}{2})}\,
\frac{\bits_2(u_{2,k}+i)}{\bits_2(u_{2,k}-i)}\,
\frac{\bits_3(u_{2,k}-\sfrac{i}{2})}{\bits_3(u_{2,k}+\sfrac{i}{2})}
\eq -\frac{\bitp(u_{2,k}+\sfrac{i}{2})}{\bitp(u_{2,k}-\sfrac{i}{2})}\,,
\nln
\phantom{\frac{\bits_1(u_{3,k}-\sfrac{i}{2})}{\bits_1(u_{3,k}+\sfrac{i}{2})}}
\frac{\bits_2(u_{3,k}-\sfrac{i}{2})}{\bits_2(u_{3,k}+\sfrac{i}{2})}\,
\frac{\bits_3(u_{3,k}+i)}{\bits_3(u_{3,k}-i)}\,
\eq -1.
\>
Effectively, they ensure
that $T_{\rep{6}}(u),T_{\rep{4}}(u)$ and $T_{\rep{\bar 4}}(u)$
are all analytic for $1/u\in\Comp$.
Using the identity
\[\label{eq:DoubleProduct}
\prod_{k'=1}^{K_{j'}}\bits_j(u_{j',k'}+a)=
\prod_{k=1}^{K_{j}}\prod_{k'=1}^{K_{j'}}(u_{j',k'}-u_{j,k}+a)
=
(-1)^{K_jK_{j'}}\prod_{k=1}^{K_{j}}\bits_{j'}(u_{j,k}-a)
\]
it is easy to see that the product of all Bethe equations
yields the constraint
\[\label{eq:MomConstr2}
1=\prod_{k=1}^{K_2}\frac{\bitp(u_{2,k}+\sfrac{i}{2})}{\bitp(u_{2,k}-\sfrac{i}{2})}
=\frac{\bits_2(-\sfrac{i}{2})^L}{\bits_2(+\sfrac{i}{2})^L}\,.
\]
%

\paragraph{Local Charges.}

In $\transfer_{\rep{6}}(u)$ all terms but one are
proportional to $\bitp(u-i)=(u-i)^L$.
Thus the first $L$ terms in the expansion
in $u$ around $i$ are determined by this one
term alone
(unless there are singular roots
at $u=+\frac{i}{2}$ which would lower the bound)
\[\label{eq:TransferExpand}
\frac{\bitp(u+i)}{\bitp(u+2i)}\,
\transfer_{\rep{6}}(u+i)
=\frac{\bits_2(u-\sfrac{i}{2})}{\bits_2(u+\sfrac{i}{2})}
+\order{u^L}.
\]
According to \eqref{eq:LocalCharges} this is
precisely the combination for the expansion in
terms of local charges.
Comparing \eqref{eq:TransferExpand} to
\eqref{eq:LocalCharges} we obtain for the global
shift and local charge eigenvalues $U,Q_r$
\[\label{eq:ChargeEigenvalues}
U=\prod_{k=1}^{K_2}\frac{u_{2,k}-\sfrac{i}{2}}{u_{2,k}+\sfrac{i}{2}}\,,
\qquad
Q_r=\frac{i}{r-1}\sum_{k=1}^{K_2}
\lrbrk{\frac{1}{(u_{2,k}+\sfrac{i}{2})^{r-1}}-\frac{1}{(u_{2,k}-\sfrac{i}{2})^{r-1}}}.
\]
Note that the eigenvalue of the second charge $\charge_2$ is
the energy $E=\charge_2$, eigenvalue of the Hamiltonian,
see \eqref{eq:Q2Ham}.
The momentum $U$ must satisfy $U^L=1$ due to 
\eqref{eq:MomConstr2} in agreement with the
fact that the shift operator obeys $\opshift^L=1$.

The expansion in terms of local charges is a distinctive feature
of $\transfer_{\rep{6}}(u)$, which is in the
same representation as the spins.
For $\transfer_{\rep{6}}(u)$ we can expand around $u=\pm i$
and only one of the six terms does contribute in the
leading few powers as in \eqref{eq:TransferExpand}.
In contradistinction, at least two terms contribute
to the expansion of
$\transfer_{\rep{4}}(u)$ and $\transfer_{\rep{\bar 4}}(u)$
at every point $u$.
Therefore, neither $\transfer_{\rep{4}}(u)$ nor $\transfer_{\rep{\bar 4}}(u)$
can be used to yield \emph{local} charges,
which are the sums of the magnon charges as in \eqref{eq:ChargeEigenvalues}.

\subsection{Thermodynamic Limit}
\label{sec:Spins.Thermo}

In the thermodynamic limit the length $L$ of the spin chain as
well as the number of excitations $K_j$ approach infinity while
focusing on the low-energy spectrum
\cite{Sutherland:1995aa,Beisert:2003xu}.
Let us now rescale the parameters
\[\label{eq:ThermoLimit}
\set{K_j,u,r_j,D,g}\mapsto L\set{K_j,u,r_j,D,g},
\]
while $E\mapsto E/L$.
The Bethe roots $u_{j,k}$ condense on (not necessarily connected)
curves $\contour_{j}$ in the complex plane with a density function $\rho_j(u)$,
i.e.
\[\label{eq:ThermoCuts}
\sum_{k=1}^{K_j}\ldots \to L \int_{\contour_j} du\,\rho_j(u)\,\ldots\,.
\]
This fixes the normalization of the densities to
\[
\label{eq:ThermoNorm}
\int_{\contour_j}du\,\rho_j(u)=K_j.
\]

It is useful to note the following limits of
fractions involving $\bits$ and $\bitp$
\[
\label{eq:ThermoBits}
\frac{\bits_j(u+a)}{\bits_j(u+b)}
\to
\exp\bigbrk{(b-a)\resolv_j(u)},
\qquad
\frac{\bitp(u+a)}{\bitp(u+b)}
\to
\exp\lrbrk{\frac{a-b}{u}}
\]
with the resolvent
\[
\label{eq:ThermoResolv}
\resolv_j(u)=\int_{\contour_j}\frac{dv\,\rho_j(v)}{v-u}\,.
\]
We can now determine the limit of the transfer matrices. Let us
start with the fundamental representation
\eqref{eq:BetheTransfer4}, we obtain
\[\label{eq:ThermoTransfer4a}
\transfer_{\rep{4}}(u)
\to
\exp\bigbrk{i\sheet_1(u)}
+\exp\bigbrk{i\sheet_2(u)}
+\exp\bigbrk{i\sheet_3(u)}
+\exp\bigbrk{i\sheet_4(u)}.
\]
The four exponents $p_{1,2,3,4}(u)$ read
\<
\label{eq:ThermoSheets}
\sheet_1(u)\eq \rsing_1(u),
\nln
\sheet_2(u)\eq \rsing_2(u)-\rsing_1(u),
\nln
\sheet_3(u)\eq \rsing_3(u)-\rsing_2(u),
\nln
\sheet_4(u)\eq \phantom{\rsing_4(u)}-\rsing_3(u),
\>
where we have defined the singular resolvents $\rsing_j(u)$ as
\<
\label{eq:ThermoResolvSing}
\rsing_1(u)\eq \resolv_1(u)+1/2u,
\nln
\rsing_2(u)\eq \resolv_2(u)+1/u,
\nln
\rsing_3(u)\eq \resolv_3(u)+1/2u.
\>
Note that the exponents add up to zero
\[\label{eq:SheetSum}
p_1(u)+p_2(u)+p_3(u)+p_4(u)=0.
\]
The limit of a transfer matrix
in an arbitrary representation $\rep{R}$
now reads simply
\[
\label{eq:ThermoTransfers}
\transfer_{\indrep}(u)\to \sum_{k=1}^{R} \exp\bigbrk{i\sheet^{\indrep}_k(u)}.
\]
The functions $\sheet^{\rep{4}}(u)=\sheet(u)$ are related to the
transfer matrix in the fundamental representation.
{}From \eqref{eq:BetheTransfer6,eq:BetheTransfer4bar}
we can derive up similar functions $\sheet^{\rep{6}}(u),\sheet^{\rep{\bar 4}}(u)$
for the vector and conjugate fundamental representation.
For each component of the multiplet
there is an exponent $\sheet^{\indrep}_k(u)$
\<
\label{eq:Sheets644}
\sheet^{\rep{4}}\eq(\sheet_1,\sheet_2,\sheet_3,\sheet_4),
\nln
\sheet^{\rep{6}}\eq(\sheet_1+\sheet_2,\,\,\sheet_1+\sheet_3,\,\,\sheet_1+\sheet_4,\,\,
                    \sheet_2+\sheet_3,\,\,\sheet_2+\sheet_4,\,\,\sheet_3+\sheet_4),
\nln\eq(\sheet_1+\sheet_2,\,\,\sheet_1+\sheet_3,\,\,\sheet_1+\sheet_4,\,\,
                    -\sheet_1-\sheet_4,\,\,-\sheet_1-\sheet_3,\,\,-\sheet_1-\sheet_2),
\nln
\sheet^{\rep{\bar 4}}\eq(\sheet_1+\sheet_2+\sheet_3,\,\,
                         \sheet_1+\sheet_2+\sheet_4,\,\,
                         \sheet_1+\sheet_3+\sheet_4,\,\,
                         \sheet_2+\sheet_3+\sheet_4)
\nln\eq(-\sheet_4,-\sheet_3,-\sheet_2,-\sheet_1).
\>
%




\subsection{Properties of the Resolvents}
\label{sec:Spins.Props}

\paragraph{Bethe Equations and Sheets.}

\begin{figure}
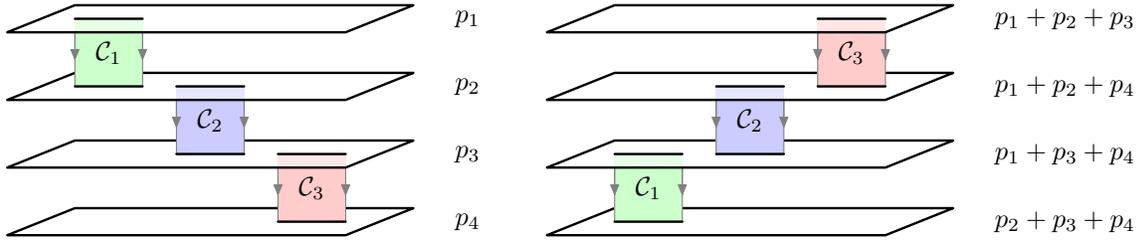
\centering
\includegraphics{bks.sheets.4.eps}\qquad
\includegraphics{bks.sheets.4bar.eps}
\caption{Transfer matrix in $\rep{4}$ and $\rep{\bar 4}$
representation.}
\label{fig:Sheets44bar}
\end{figure}

We know that $\transfer_{\rep{4}}(u)$ is a polynomial in $1/u$.
It therefore has no singularities except at $u=0$
and it should remain analytic in the thermodynamic limit.
This is ensured by the Bethe equations \eqref{eq:BetheEquations}
whose limit reads
\<\label{eq:ThermoEquations}
2\rsingsl_1(u)-\rsing_2(u)=
\sheetsl_1(u)-\sheetsl_2(u)\eq 2\pi n_{1,a},
\qquad u\in \contour_{1,a},
\nln
2\rsingsl_2(u)-\rsing_1(u)-\rsing_3(u)
=\sheetsl_2(u)-\sheetsl_3(u)\eq 2\pi n_{2,a},
\qquad u\in \contour_{2,a},
\nln
2\rsingsl_3(u)-\rsing_2(u)
=\sheetsl_3(u)-\sheetsl_4(u)\eq 2\pi n_{3,a},
\qquad u\in \contour_{3,a}.
\>
Here we have split up the curves $\contour_{j}$ into their
connected components $\contour_{j,a}$ with
\[\label{eq:ThermoContours}
\contour_j=\contour_{j,1}\cup\cdots\cup \contour_{j,A_j}
\]
and introduced a mode number $n_{j,a}$ for each curve to
select the branch of the logarithm that was used to bring
the equations \eqref{eq:BetheEquations}
into the form \eqref{eq:ThermoEquations}.
Furthermore $\rsingsl$ and $\sheetsl$ are the principal
values of $\rsing$ and $\sheet$, respectively,
at a cut, e.g.
\[\label{eq:Principal}
\rsingsl_j(u)=\half \rsing_j(u-\eps)+\half \rsing_j(u+\eps).
\]
Let us explain the meaning of the Bethe equations in words.
The first one implies that
a cut in $p_1(u)$ or $p_2(u)$ at $\contour_{1,a}$
can be analytically continued
by the function $p_2(u)$ and $p_1(u)$, respectively
(up to a shift by $\pm 2\pi n_{1,a}$).
For the transfer matrices
in \eqref{eq:ThermoTransfers} neither the interchange
between $p_1$ and $p_2$ nor a shift by an integer multiple of
$2\pi$ has any effect.
Similarly, $p_2$ and $p_3$ or $p_3$ and $p_4$ are connected
by cuts along $\contour_2$ or $\contour_3$
as depicted in \figref{fig:Sheets44bar}.
Therefore the transfer matrices are analytic
except at $u=0$.
In total, the functions $\sheet_{1,2,3,4}(u)$ (modulo $2\pi$)
make up four sheets of a Riemann surface,
an algebraic curve of degree four.
The function $\sheet=(\sheet_1,\sheet_2,\sheet_3,\sheet_4)$ is not
single valued due to the ambiguities by multiples of $2\pi$.
In $d\sheet$ the (constant) ambiguities drop out.
The differential $d\sheet$ therefore is a holomorphic function
on the algebraic curve except at the singular points $u=0$ on each sheet,
see \eqref{eq:ThermoResolvSing,eq:Sheets644}.

The configurations of sheets and their connections are displayed in
\figref{fig:Sheets44bar,fig:Sheets6}.%
\footnote{The cuts are not necessarily along the real axis
as might be suggested by the figures. 
In fact, for compact spin representations, they usually cross the
real axis at right angles.}
\begin{figure}\centering
\includegraphics{bks.sheets.6.eps}
\caption{Transfer matrix in $\rep{6}$ representation.}
\label{fig:Sheets6}
\end{figure}%
The sheet function
$\sheet^{\indrep}_k(u)$ is obtained by summing up
the outgoing singular resolvents $\rsing_j(u)$
and subtracting the incoming ones, c.f.~\eqref{eq:ThermoSheets,eq:Sheets644}.

Let us note here that the equations \eqref{eq:ThermoEquations}
are reminiscent of the saddle point equations of \cite{Kostov:1991cg} for the
RSOS type multi-matrix models. However the potential part is
different and, most importantly, the right hand sides of
\eqref{eq:ThermoEquations} would be zero for RSOS models. 

\paragraph{Local Charges.}

The expansion of $\transfer_{\rep{6}}(u)$ at $u=i$ gives the local charges.
In the thermodynamic limit, this point is scaled to $u=0$ and from
\eqref{eq:TransferExpand,eq:ThermoBits} we find
\cite{Arutyunov:2003rg}
\[\label{eq:LocalThermo}
\rsing_2(u)=p_1(u)+p_2(u)=\frac{1}{u}+\sum_{r=1}^\infty u^{r-1}\charge_r,
\]
where $\charge_r$ has been rescaled by $L^{r-1}$.
The first charge $\charge_1$ is the total momentum
around the spin chain which should equal
\[\label{eq:MomThermo}
\charge_1=2\pi n_0
\]
for gauge theory states.
The second charge 
\[\label{eq:EngThermo}
\charge_2=E=(D-1)/g^2.
\]
is the energy eigenvalue of the Hamiltonian.
The other two resolvents are non-singular
\[\label{eq:RsingNoSing}
\rsing_1(u),\rsing_3(u)=\order{u^0}
\]
and their expansion (thus) does not correspond to local quantities.
For convenience, we also display the expansion of the sheet functions
at zero
\<
\label{eq:ThermoZeroSheet}
+\sheet_1(u),+\sheet_2(u),-\sheet_3(u),-\sheet_4(u)\eq
\frac{1}{2u}+\order{u^{0}},
\>
%

\paragraph{Global Charges.}

The charges of the symmetry algebra are obtained from the
mono\-dro\-my matrix at $u=\infty$, see \secref{sec:Spins.Ops}.
When we expand the resolvents $\resolv_j(u)$ at infinity
\[\label{eq:ThermoInfty}
\resolv_j(u)=-\frac{1}{u}\int_{\contour_j}dv\,\rho_j(v)+\order{1/u^2}
=-\frac{K_j}{u}+\order{1/u^2}.
\]
we find the fillings of the cuts \eqref{eq:ThermoNorm}, which are
related to the representation
$[r_1,r_2,r_3]$ of the state via \eqref{eq:BetheExcite}.
The singular resolvents directly relate to the Dynkin labels
as follows ($L=1$ after rescaling)
\<\label{eq:ThermoInftySing}
\rsing_1(u)\eq\frac{1}{u}\lrbrk{\half-K_1}+\order{1/u^{2}}=
\frac{1}{u}\lrbrk{\sfrac{3}{4}r_1+\half r_2+\sfrac{1}{4}r_3}+\order{1/u^{2}},
\nln
\rsing_2(u)\eq\frac{1}{u}\lrbrk{\,1-K_2}+\order{1/u^{2}}=
\frac{1}{u}\lrbrk{\half r_1+r_2+\half r_3}+\order{1/u^{2}},
\nln
\rsing_3(u)\eq\frac{1}{u}\lrbrk{\half-K_3}+\order{1/u^{2}}=
\frac{1}{u}\lrbrk{\sfrac{1}{4}r_1+\half r_2+\sfrac{3}{4}r_3}+\order{1/u^{2}}.
\>
For convenience, we also display the expansion of the sheet functions
at infinity
\<
\label{eq:ThermoInftySheet}
p_1(u)\eq
\frac{1}{u}\lrbrk{+\sfrac{3}{4}r_1+\half r_2+\sfrac{1}{4}r_3}+\order{1/u^{2}},
\nln
p_2(u)\eq
\frac{1}{u}\lrbrk{-\sfrac{1}{4}r_1+\half r_2+\sfrac{1}{4}r_3}+\order{1/u^{2}},
\nln
p_3(u)\eq
\frac{1}{u}\lrbrk{-\sfrac{1}{4}r_1-\half r_2+\sfrac{1}{4}r_3}+\order{1/u^{2}},
\nln
p_4(u)\eq
\frac{1}{u}\lrbrk{-\sfrac{1}{4}r_1-\half r_2-\sfrac{3}{4}r_3}+\order{1/u^{2}}.
\>
%

\subsection{Algebraic curve}
\label{sec:Spins.Curve}

Let us now try to restore the function $p(x)$ from the information
derived in the previous subsection,%
\footnote{In the $\alg{su}(2)$ case the
corresponding hyperelliptic curve was constructed 
in \cite{Kazakov:2004qf} using the method proposed 
in \cite{Reshetikhin:1983ab}.}
namely, from the Riemann-Hilbert
equations \eqref{eq:ThermoEquations}
\[\label{eq:CurveCuts}
\sheetsl_{k}(x)-
\sheetsl_{k+1}(x)=2\pi n_a
\qquad
\mbox{for }x\in\contour_a,
\]
where $\contour_a$ connects sheets $k$ and $k+1$
and from the behavior at the various sheets
\eqref{eq:ThermoSheets}
at $x\to\infty$ \eqref{eq:ThermoInftySheet}
\[\label{eq:CurveInfty}
p_k\sim \frac{1}{u}+\order{1/u^2}
\]
as well as $x\to 0$
\eqref{eq:ThermoZeroSheet}
\[\label{eq:CurveZero}
+p_1,+p_2,-p_3,-p_4=\frac{1}{2u}+0\log u+\order{u^0}
\]

{}From the discussion in \secref{sec:Spins.Props}
we know that $\exp(ip)$ is a single valued holomorphic function on the Riemann
surface with four sheets
except at the points $0$ and $\infty$.
It is however not an algebraic curve because it has an essential singularity
of the type $\exp(i/u)$ at $u=0$.
While $p$ only has pole-singularities, it is defined only modulo $2\pi$.
This problem is overcome in the derivative $p'$ which is
has a double pole $1/u^2$ at $u=0$,
but no single pole $1/u$, neither at $u=0$ nor at $u=\infty$.

All this suggests that there exists a
function $p(u)$, the quasi-momentum, such that its derivative
\[\label{eq:ydef}
y(u)=u^2\,\frac{dp}{du}(u).
\]
satisfies a quartic algebraic equation
\[\label{eq:algeq}
F(y,u)=P_4(u)\,y^4+P_2(u)\,y^2+P_1(u)\,y+P_0(u)= P_4(u)\prod_{k=1}^4
\bigbrk{y-y_k(u)}=0.
\]
For a solution with finitely many cuts we
may assume the coefficients $P_k(u)$
to be polynomials in $u$.
The term $y^3$ is absent
because $p_1+p_2+p_3+p_4=0$.
We have adjusted $y$ to approach a constant limiting value
at $x=0$ as well as at $x=\infty$.
It follows that all the polynomials $P_k(u)$
have the same order $2A$ and a non-vanishing constant coefficient.
Altogether the function $F(y,u)$ which determines the curve
is parameterized by $8A+4$ coefficients minus
one overall normalization.

Let us now investigate the analytic structure of the
solution of $F(y,u)=0$ and compare it to the structure
of $p$. In general we can expect that $p$ behaves like
$\sqrt{u-u^\ast}$ at a branch point $u^\ast$,
consequently $y\sim 1/\sqrt{u-u^\ast}$.
To satisfy the equation $F(y,u)=0$ at $y=\infty$
we should look for zeros of $P_4(u)/P_2(u)$.
Incidentally, we find precisely the correct
behavior for $y$ due to the missing of the $y^3$ term.%
\footnote{A pole on a single sheet could never
be cancelled in $p_1+p_2+p_3+p_4=0$.
In contrast, a branch singularity $+\alpha/\sqrt{u-u^\ast}$
will be cancelled by an accompanying singularity
$-\alpha/\sqrt{u-u^\ast}$ on the sheet which is
connected along the branch cut.}
For a generic $P_2(u)$, the branch points
are thus the roots of $P_4(u)$
\[\label{eq:branchpoints}
P_4(u)=\prod_{a=1}^{A} (u-a_a)(u-b_a).
\]
Therefore, $A$ is the number of cuts and $a_a, b_a$ are the branch points.
The algebraic equation \eqref{eq:algeq} potentially
has further cuts.
The associated singularities are of the undesired form
$y\sim (u-u^\ast)^{r+1/2}$ or $p\sim (u-u^\ast)^{r+3/2}$ and
we have to ensure their absence.
Their positions can be obtained as
roots of the discriminant of the quartic equation
\<\label{eq:Discr}
R\eq
-4P_1^2 P_2^3
+ 16P_0 P_2^4
- 27P_1^4 P_4
+ 144P_0 P_1^2 P_2 P_4
- 128P_0^2 P_2^2 P_4
+ 256P_0^3 P_4^2
\nln
\eq
P_4^5\,
(y_1-y_2)^2
(y_1-y_3)^2
(y_1-y_4)^2
(y_2-y_3)^2
(y_2-y_4)^2
(y_3-y_4)^2
.
\>
All solutions of $R(u)=0$ with odd multiplicity
give rise to undesired branch cuts, in other words we have to demand that
the discriminant is a perfect square
\[\label{eq:DisSquare}
R(u)=Q(u)^2
\]
with a polynomial $Q(u)$.%
\footnote{An equivalent condition is: All solutions to the 
equations $dF(y,u)=0$ and $P_4(u)\neq 0$
lie on the curve $F(y,u)=0$.
The condition $dF(y,u)=0$ $\Rightarrow$ $F(y,u)=0$ eliminates
branch points and $P_4(u)\neq 0$ preserves the desired ones.} 
This fixes $5A$ coefficients and we
remain with only $3A+3$ free coefficients.

First of all we can fix the coefficients of the
double pole in $p'$ at $u=0$
according to \eqref{eq:CurveZero}.
This fixes three coefficients, $P_4(0)=-8P_2(0)=16P_0(0)$ and $P_1(0)=0$.

The function $p(u)$ has to be single-valued (modulo $2\pi$) on the curve.
We can put the $\contour[A]$-cycles to zero%
\footnote{Even though we should assume multiples of $2\pi$ as the periods,
the cuts can be chosen in such a way as to yield
single-valued functions $p_k$ \cite{Kazakov:2004qf}.}
\[\label{eq:Acycles}
\oint_{\contour[A]_a} dp=0.
\]
The cycle $\contour[A]_a$ surrounds the cut $\contour_a$.
Note that there are only $A-3$ independent $\contour[A]$-cycles
in agreement with the genus of the algebraic curve, $A-3$.
The sum of all $\contour[A]$-cycles on each of the three
independent sheets can be joined to a cycle around
the punctures at $u=0$. Here we expect a double pole,
but not a single pole, \eqref{eq:CurveZero}%
\[\label{eq:Acycles0}
\oint_{0} dp_k=0.
\]
The $\contour[A]$-cycles together with the
absence of single poles at $u=0$ yield
$A$ constraints.

Next we consider the $\contour[B]$-periods.
The cycle $\contour[B]_a$ connects
the points $u=\infty$ of two sheets $k,k+1$
going through a cut $\contour_a$ which connects these
sheets.%
\footnote{Here the property $p'\sim 1/x^2$, \eqref{eq:CurveInfty},
is useful to mark the points $u=\infty$.}
We now rewrite the Bethe equations \eqref{eq:CurveCuts} as
$A$ integer $\contour[B]$-periods
\[\label{eq:Bcycles}
\int_{\contour[B]_a} dp=2\pi n_a
\]
where $n_a$ is the mode number associated to the cut $\contour_a$.

We can now integrate $p'(u)$ and obtain $p(u)$. The integration constants
are determined by the value at $u=\infty$, \eqref{eq:CurveInfty}.
At this point we are left with precisely $A$ undefined coefficients.
These can be identified with the filling
fractions
\[\label{eq:FillingInt}
K_a=
-\frac{1}{2\pi i}
\oint_{\contour[A]_a} p(u)\, du.
\]
In the integral representation these correspond to the
quantities
\[\label{eq:FillingCont}
K_a=\int_{\contour_a} \rho(u)\, du.
\]

When all filling fractions $K_a$ and integer mode numbers $n_a$
are fixed, we can calculate in principle any function of physical interest.
In particular, $\rsing_2=p_1+p_2=-p_3-p_4$ gives an infinite set of
local charges \eqref{eq:LocalThermo}, including the anomalous
dimension.

Note that so far we have not considered the momentum constraint
\eqref{eq:MomThermo} which serves as a physicality
condition for gauge theory states.%
\footnote{Spin chain states which do not obey \eqref{eq:MomThermo}
are perfectly well-defined, they merely have no 
correspondence in gauge theory.}
This reduces the number of independent continuous parameters by one since
$n_0$ is discrete.
We can even express the constraint fully in terms
of $K_a$ and $n_a$ 
\[
n_0=\sum_{a=1}^A n_a K_a \in \Integers
\]
by integrating the Bethe equations 
\eqref{eq:ThermoEquations} over all cuts.

\subsection{Examples}
\label{sec:Spins.Exam}

Here we will discuss the algebraic solutions of
the $\alg{so}(6)$ integrable spin chain
found in \cite{Engquist:2003rn}.
The filling fractions $(K_1,K_2,K_3)$ of the three
solutions are given by
$(\half\alpha,\alpha,\half\alpha)$,
$(\half\alpha,\alpha,0)$ and
$(1-\alpha,1-\half\alpha,0)$.
For the first two cases $(i,ii)$, there are
two symmetric cuts $\contour_\pm$ for $\resolv_2$ stretching
from points $\pm a$ to $\pm b$, while the
cut for $\resolv_1$ (and $\resolv_3$)
stretches all along the imaginary axis.
The latter cut is however
not a genuine branch cut, because
both branch points coincide at $u=\infty$.
It merely permutes the sheets $p_1$ and $p_2$
(or $p_3$ and $p_4$) when crossing the imaginary axis.
Therefore this cut effectively screens
all the charges behind the imaginary axis
(looking from either side of it).
In our treatment we shall lift this unessential branch cut,
the structure of cuts and sheets will
thus be different as we will describe below.
The solution for case $(iii)$ was found to be equivalent
to the case $(ii)$ upon analytic continuation.
For us this means that the algebraic curve underlying the
solution is actually the same, only the
labelling of its sheets is modified.

\paragraph{Frolov-Tseytlin Circular String.}

Let us start with case ($ii/iii$) for which there
are no excitations
of type $3$, i.e.~$\resolv_3=0$.
The string analog of this solution was
originally studied in \cite{Frolov:2003qc}.
Therefore, the sheet $p_4=-1/2u$ is detached
from the other sheets and the curve factorizes as follows
\[\label{eq:FactorCurveSU3}
\bigbrk{y(u)-1/2u^2}
\lrbrk{
\tilde P_3(u)z(u)^3
+\tilde P_1(u)z(u)
+\tilde P_0(u)}=0.
\]
Here we have introduced the shifted
variable $z_k=y_k+1/6$
which ensures that $z_1+z_2+z_3=0$
due to $y_1+y_2+y_3+y_4=0$ and $y_4=1/2$.
Now the cubic equation for $z$ corresponds
to a spin chain with $\alg{su}(3)$ symmetry and
spins in the fundamental representation.
It can be solved analogously to the
$\alg{su}(4)$ case.

\begin{figure}
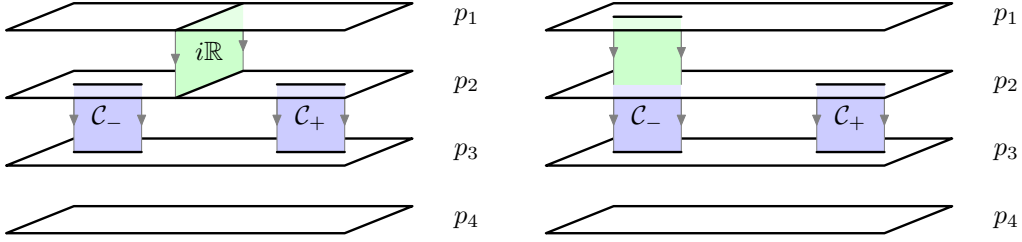
\centering
\includegraphics{bks.sol.asym.old.eps}\qquad
\includegraphics{bks.sol.asym.new.eps}
\caption{Frolov-Tseytlin string.
The diagram on the left depicts
the cuts as described in
\protect\cite{Engquist:2003rn}.
The cut along the imaginary line
($i\Real$) interchanges the
two involved sheets $(p_1,p_2)$.
When we remove this inessential branch cut, we obtain
the diagram on the right.
Here the cut $\contour_-$ goes directly from
$p_1$ to $p_3$ right through $p_2$ effectively
screening half of it from $\contour_+$.}
\label{fig:Sol.Asym}
\end{figure}

The simplest solution
which does not reduce further to
$\alg{su}(2)$ requires two cuts
which do not connect the same two sheets.
The corresponding curve obviously has degree three and
genus zero, i.e.~it is algebraic.
This agrees precisely with the solution
of case $(ii/iii)$ in \cite{Engquist:2003rn},
which however appears to have three cuts,
see the diagram in \figref{fig:Sol.Asym} on the left.
As emphasized above, the cut between
sheets $p_1$ and $p_2$ can be lifted
by interchanging the two when crossing the imaginary axis.
Here we have to make the choice on which side
we should flip $p_1$ and $p_2$, we choose the left one.
Now the cut $\contour_-$ extends directly from
$p_3$ to $p_1$,
see the diagram in \figref{fig:Sol.Asym} on the right.
Alternatively, one could choose
$\contour_+$ to extend between $p_3$ and $p_1$
while $\contour_-$ remains between $p_3$ and $p_2$.
The solution was originally found
in \cite{Engquist:2003rn}, here we will
demonstrate the properties discussed in
\secref{sec:Spins.Curve}.
The coefficients describing the algebraic curve
in \eqref{eq:FactorCurveSU3} are given by
\<
\tilde P_3(u)\eq
27
\left(
(1-\alpha)
+(-8+36\alpha-27\alpha^2) (\pi n u)^2
+16 (\pi n u)^4 \right),
\\
\tilde P_1(u)\eq
-9
\left(
(1-\alpha  )
+(-8+24\alpha-15\alpha^2)(\pi n u)^2
+(16-48\alpha+36\alpha^2)(\pi n u)^4
\right),
\nln
\tilde P_0(u)\eq
-2\left(
(1 -\alpha)
+(-8+18\alpha-9\alpha^2)(\pi n u)^2
+(16-72\alpha+108\alpha^2-54\alpha^3)(\pi n u)^4
\right).
\nonumber
\>
The discriminant (modulo factors of $\tilde P_0$)
is indeed a perfect square
\<
\tilde R(u)\eq 4\tilde P_1(u)^3+27\tilde P_0(u)^2\tilde P_3(u)
\\\nonumber\eq
-8503056\alpha^2 (\pi n u)^6
\left(
(4 - 9\alpha + 5\alpha^2 )
+(- 16 + 60\alpha - 72\alpha^2 + 27\alpha^3) (\pi n u)^2
\right)^2.
\>
For the solution we find the expansion around $u=0$
\[
z_{1,2}=-\sfrac{1}{3}+2\alpha(\pi nu)^2+\order{u^3},
\qquad
z_{3}=\sfrac{2}{3}-4\alpha(\pi nu)^2+\order{u^4}.
\]
and around $u=\infty$
\[
z_{1,2}=-\sfrac{1}{3}+\half\alpha+\order{1/u},
\qquad
z_{3}=\sfrac{2}{3}-\alpha+\order{1/u^2}.
\]
After integrating to the function $p$
we get the expansions
\[
p_{1,2}=\frac{1}{2u}\pm f(\alpha,n)+2\alpha(\pi n)^2 u+\order{u^2},
\qquad
p_{3}=-\frac{1}{2u}-4\alpha(\pi n)^2 u+\order{u^3}.
\]
and
\[
p_{1,2}=\frac{1-\alpha}{2u}+\order{1/u^2},
\qquad
p_{3}=\frac{-1+2\alpha}{2u}+\order{1/u^3}.
\]
Now $p_3+p_4=-\resolv_2$ is a physical sheets
and $p_4$ is trivially $p_4=-1/2u$.
Therefore we can read off the energy as
the negative coefficient of $u$ in the expansion of $p_3$
around $u=0$.

{}From here it is not obvious to see that $n$ must
be integer. This fact can be derived from
demanding that the $\contour[B]$-periods are integers.

\paragraph{Pulsating String.}

\begin{figure}
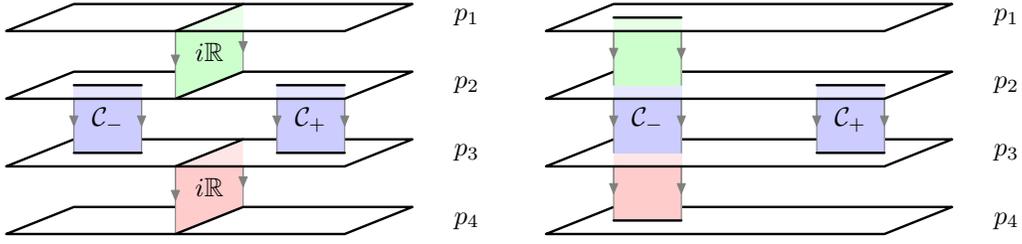
\centering
\includegraphics{bks.sol.sym.old.eps}\qquad
\includegraphics{bks.sol.sym.new.eps}
\caption{Pulsating string solution.
The diagram on the left depicts
the cuts as described in
\protect\cite{Engquist:2003rn}.
The cuts along the imaginary line
($i\Real$) interchange the
two involved sheets $(p_1,p_2)$
and $(p_3,p_4)$. When we remove these
inessential branch cuts, we obtain
the diagram on the right. Here
the cut $\contour_-$
goes directly from
$p_1$ to $p_4$ right through
$p_2$ and $p_3$ effectively
screening it from
$\contour_+$.
}
\label{fig:Sol.Sym}
\end{figure}

\begin{figure}\centering
\includegraphics{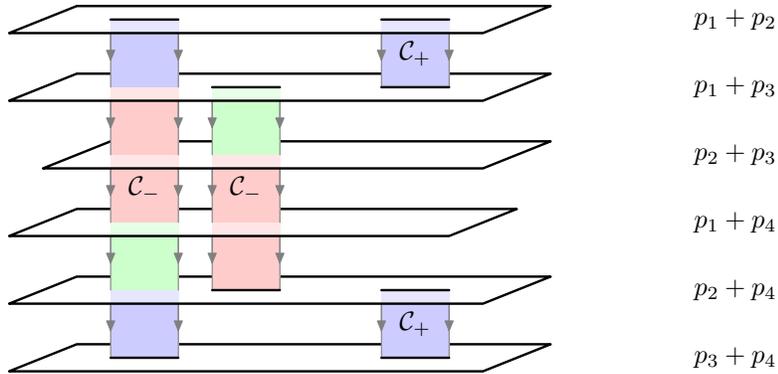}
\caption{Six-sheeted version of the pulsating string.
Note that on the outer two sheets
$p_1+p_2$ and $p_3+p_4$,
the physical sheets,
both cuts $\contour_+$ and $\contour_-$ can be seen.
In particular, these sheets do not change if we
``expand{}'' $\contour_+$ instead of $\contour_-$
in \protect\figref{fig:Sol.Sym}.
Here the screening works because on the two sheets
which $\contour_+$ connects,
a cut $\contour_-$ starts in the same direction,
this effectively cancels the forces on $\contour_+$.
The middle two sheets are free.}
\label{fig:Sol.Sym6}
\end{figure}

The case $(i)$ is analogous to the
case $(ii/iii)$.
The string analog of this solution was
originally found in \cite{Minahan:2002rc}.
In \cite{Engquist:2003rn} the
solution consisted of four cuts, two
of which can however be removed,
see \figref{fig:Sol.Sym}.
We see that the cut $\contour_+$
connects $p_2$ with $p_3$, while
$\contour_-$ connects $p_1$ with
$p_4$ directly.
For the above reasons the two cuts are completely independent.
The only connection is due to the
momentum constraint, energy and
higher local charges which are measured on 
$p_1+p_2$.
It is also interesting to look at the structure
of cuts in the vector representation,
see \figref{fig:Sol.Sym6}.
Here, two sheets are decoupled.
For the above reasons, the curve \eqref{eq:algeq}
factorizes in two
\[
\bigbrk{\tilde P^+_2(u)\,y(u)^2-\tilde P^+_0(u)}
\bigbrk{\tilde P^-_2(u)\,y(u)^2-\tilde P^-_0(u)}=0.
\]
{}From the solution in
\cite{Engquist:2003rn}
we can deduce the coefficients
\<
\tilde P^\pm_{2}(u)\eq
4\bigbrk{1 \pm 4(1-\alpha) \pi nu + 4(\pi nu)^2},
\nln
\tilde P^\pm_{0}(u)\eq
-\bigbrk{1 \pm 2(1-\alpha) \pi nu}^2 .
\>
This defines the algebraic curve corresponding
to the pulsating string at one-loop.

\subsection{Higher Loops}
\label{sec:Spins.Higher}

Before we turn to string theory, 
we make a digression towards higher loops.
In \cite{Minahan:2004ds} Minahan considered 
the $\alg{so}(6)$ sector at higher loops.
While this sector is actually not closed
at higher loops, he argued that it closes in 
the thermodynamic limit. He then combined the
one-loop Bethe equations for
the $\alg{su}(6)$ sector 
\eqref{eq:ThermoEquations}
with the higher-loop Bethe equations
for the $\alg{su}(2)$ subsector
\cite{Serban:2004jf,Beisert:2004hm}
\<\label{eq:HigherEquationsA}
2\bar\resolvsl_1(u)-\bar\resolv_2(u)\eq 2\pi n_{1,a},
\qquad u\in \bar\contour_{1,a},
\nln
2\bar\resolvsl_2(u)-\bar\resolv_1(u)-\bar\resolv_3(x)
+F\indup{gauge}(u)
\eq 2\pi n_{2,a},
\qquad u\in \bar\contour_{2,a},
\nln
2\bar\resolvsl_3(u)-\bar\resolv_2(u)
\eq 2\pi n_{3,a},
\qquad u\in \bar\contour_{3,a}.
\>
with 
\[
F\indup{gauge}(u)=\frac{1}{\sqrt{u^2-2g^2}}\,.
\]
At this point, we have added a bar to the resolvent, because
there will be another (unbarred) resolvent $\resolv_j(u)$
which is more closely related to physical quantities.
The proposed generalization 
amounts to a modification of
the singular term in $\rsing_j(u)$, 
c.f.~\eqref{eq:ThermoResolvSing}%
\footnote{We have used the inverse of the
Cartan metric for $\alg{su}(4)$,
see \eqref{eq:CartanSU4}, to specify the coefficient of the 
singular term, i.e.~$M_{2j}^{-1}=(\half,1,\half)$ }
\[
\bar\rsing_j(u)= \bar\resolv_j(u)+M_{2j}^{-1}\,\frac{1}{\sqrt{u^2-2g^2}}\,,
\]
The Bethe equations for $\bar\rsing_1(u)$ are just the
same as the ones at one-loop \eqref{eq:ThermoEquations}.
Note that the pole at $u=0$ in \eqref{eq:ThermoResolvSing}
has turned into a cut, but at $g=0$ we recover 
the one-loop expressions.
To understand the structure of the equations better, 
we should unfold the cut by a suitable coordinate 
transformation. This is achieved by the map $u\to x$ given 
in \cite{Beisert:2004hm}
\[
x(u)=\half u+\half\sqrt{u^2-2g^2}\,,
\qquad
u(x)=x+\frac{g^2}{2x}\,.
\]

To read off the physical information from the solution of 
the Bethe equations we will introduce a new resolvent 
in the $x$-plane
\[
\resolv_j(x)=\int_{\contour_j}\frac{dy\,\rho_j(y)}{1-g^2/2y^2}
\,\frac{1}{y-x}\,.
\]
where the densities are related as
$du\,\rho_j(u)=dx\,\rho_j(x)$.
The total momentum $U=1$ and the anomalous dimension $E=(D-L)/g^2$ are 
read off from $\resolv_2(x)$ as 
\[
U=\exp\bigbrk{iG_2(0)},
\qquad
E=G'_2(0).
\]
The relation to the $u$-resolvent $\bar\resolv_j$ is given by
\[
\bar\resolv_j(x+g^2/2x)=\resolv_j(x)+\resolv_j(g^2/2x)-\resolv_j(0).
\]
Let us now transform the singular resolvent 
to the $x$-plane as $\rsing_j(x)=\bar\rsing_j(u(x))$,
we obtain
\[\label{eq:HigherSing}
\rsing_j(x)=\resolv_j(x)+\resolv_j(g^2/2x)-\resolv_j(0)+M_{2j}^{-1}\,\frac{x}{x^2-\half g^2}\,.
\]
Note that now the singular term changes sign under the transformation
$x\to g^2/2x$ while rest remains invariant, i.e.
we have the transformation law
\[\label{eq:HigherSym}
\rsing_j(g^2/2x)=\rsing_j(x)-2M_{2j}^{-1}\frac{x}{x^2-\half g^2}
\,.
\]
The Bethe equations 
\<\label{eq:HigherEquations}
2\rsingsl_1(x)-\rsing_2(x)
=\sheetsl_1(x)-\sheetsl_2(x)
\eq 2\pi n_{1,a},
\qquad x\in \contour_{1,a},
\nln
2\rsingsl_2(x)-\rsing_1(x)-\rsing_3(x)
=\sheetsl_2(x)-\sheetsl_3(x)
\eq 2\pi n_{2,a},
\qquad x\in \contour_{2,a},
\nln
2\rsingsl_3(x)-\rsing_2(x)
=\sheetsl_3(x)-\sheetsl_4(x)
\eq 2\pi n_{3,a},
\qquad x\in \contour_{3,a}.
\>
now imply that $\sheet(x)$ corresponds
to an algebraic curve of degree four
as explained in \secref{sec:Spins.Curve}.
This curve has slightly different properties 
\eqref{eq:HigherSing,eq:HigherSym} 
as compared to the one-loop case.

\section{Classical Sigma-Model on $\Real\times S^{\grn-1}$}
\label{sec:Sigma}

In this section we will investigate the analytic 
properties of the monodromy of the Lax pair around the closed string.
The string is the two-dimensional non-linear sigma model on 
$\Real\times S^{\grn-1}$ supplemented by the Virasoro constraints. 
This is an interesting model, because (classically)
it is a consistent truncation of the superstring on 
$AdS_5\times S^5$.

\subsection{The Sigma-Model}
\label{sec:Sigma.Basics}

Consider the two-dimensional sigma model on $\Real\times S^{\grn-1}$.
Let $\{X_i\}_{i=0}^\grn$ denote the target space coordinates.
While $X_0$ can take any value on $\Real$, the other
coordinates $\{X_i\}_{i=1}^\grn$ satisfy a constraint
\[\label{eq:Xconstraint}
{X_1}^2+\cdots+{X_\grn}^2=1.
\]
Let us also introduce the following vector notation
\[\label{eq:Xvector}
\Xvec=\left(
\begin{array}{c}
X_1\\ \vdots\\ X_\grn
\end{array}
\right),
\quad
\Xvec^2=1
\]
and a matrix $\gvec$ associated with $\Xvec$ by
\[\label{eq:gofX}
\gvec=1-2\Xvec\Xvec^\trans\qquad(\mbox{i.e.}\quad
\gvec{}_{ij}=\delta_{ij}-2X_iX_j).
\]
The matrix describes a reflection along the
vector $\Xvec$, 
$\det \gvec=-1$,
it satisfies $\gvec^{}\gvec^\trans=1$, i.e.~it is
orthogonal, $\gvec\in \grp{O}(\grn)$.
Furthermore it is symmetric, $\gvec^{}=\gvec^{\trans}$,
and therefore it equals its own inverse
\[\label{eq:ginv}
\gvec^{-1}=\gvec^{}.
\]

The action of a bosonic string rotating on the $S^5$ sphere
and restricted to a time-like geodesic $\Real$ of $AdS_5$
is given by
\[\label{eq:SigmaAction}
S_{\sigma}=-\frac{\sqrt{\lambda}}{ 4\pi}\int d\sigma\,d\tau\,
\Bigbrk{\partial_a \Xvec\cdot \partial^a \Xvec
 - \partial_a X_0\,\partial^a X_0
+\Lambda\bigbrk{\Xvec^2-1}},
\]
where $\Lambda$ is a Lagrange multiplier
that constrains $\Xvec$ to the unit sphere
\eqref{eq:Xconstraint}.
Here it is useful to introduce light-cone coordinates
\[\label{eq:LightCone}
\sigma_\pm=\half(\tau\pm\sigma), \qquad
\partial_\pm=\partial_\tau\pm\partial_\sigma.
\]
Then the equations of motion read
\[\label{eq:SigmaEOM}
\partial_+\partial_- \Xvec
+(\partial_+ \Xvec\cdot\partial_- \Xvec)\Xvec=0,
\qquad
\partial_+\partial_- X_0=0.
\]
A solution for the time coordinate which we use
to fix the residual gauge of the string is
\[\label{eq:TimeGauge}
X_0(\tau,\sigma)=\frac{\dimn\,\tau}{\sqrt{\lambda}}=
\frac{\dimn(\sigma_++\sigma_-)}{\sqrt{\lambda}}\,,
\]
where $\dimn$ is the dimension in the AdS/CFT
interpretation.
In addition to the action,
the string must satisfy the Virasoro constraints
\[\label{eq:Virasoro}
(\partial_\pm \Xvec)^2=(\partial_\pm X_0)^2,
\]
which read
\[\label{eq:Virasoro2}
(\partial_\pm \Xvec)^2=\frac{D^2}{\lambda}
\]
in the gauge \eqref{eq:TimeGauge}.

\subsection{Flat and Conserved Currents}
\label{sec:Sigma.Currents}

Let us next define the right current $j_\indvec$ and the
left current $\ell_\indvec$ by
\[ \label{eq:CurrentMatrix}
j_\indvec := \gvec^{-1}d\gvec^{},\qquad
\ell_\indvec := -d\gvec^{}\gvec^{-1}.
\]
Due to the special property \eqref{eq:ginv} these currents coincide.
In this case they are simply expressed in terms of $\Xvec$ as
\[\label{eq:SigmaCurrent}
j_\indvec=\ell_\indvec=2(\Xvec d\Xvec^\trans-d\Xvec\Xvec^\trans),
\]
or in terms of their components,
\[\label{eq:CurrentComponent}
(j_\indvec)_{a,ij}=(\ell_\indvec)_{a,ij}=2\bigbrk{X_i\partial_aX_j-X_j\partial_aX_i}.
\]
{}From \eqref{eq:CurrentMatrix} it is clear that $j_\indvec$ satisfies the flatness condition
\[\label{eq:CurrentFlat}
dj_\indvec+j_\indvec\wedge j_\indvec=0.
\]

The kinetic term of the Lagrangian
of the sigma model on $S^{\grn-1}$
is given by $\Tr j_{\indvec}\wedge\ast j_\indvec$.
It is invariant under global right (and left) multiplication to $\gvec$
and $j_\indvec$ is the associated conserved current
\[\label{eq:CurrentConserved}
d(\ast j_\indvec)=0\qquad (\mbox{i.e. }\partial_a j_\indvec^a=0).
\]
This relation can be verified using the equations of motion
\eqref{eq:SigmaEOM},
in fact it is equivalent to them.

The current $j_\indvec$ is an element of $\alg{so}(\grn)$ 
in the vector representation. 
Similarly, we can write a current in the spinor representation.
Let $\{\gamma_i\}$ form the basis of the Clifford algebra
of $\grp{SO}(\grn)$,
i.e.~they satisfy
\[\label{eq:Clifford}
\{\gamma_i,\gamma_j\}=2\delta_{ij}.
\]
Now we introduce the matrix
\[\label{eq:gspin}
\gspin=\gammavec\cdot\Xvec
\]
in the spinor representation.
It satisfies
\[\label{eq:gspininv}
\gspin^{}=\gspin^{-1},
\]
therefore it can be regarded as
a spinor equivalent of $\gvec$.
As above this gives rise
to equal right and left currents
$j_\indspin=\ell_\indspin$. These
are in fact equivalent to
$j_\indvec$ through
$j_\indspin\sim \gamma_i\gamma_j(j_\indvec)_{ij}$.

\subsection{Lax Pair and Monodromy Matrix}
\label{sec:Sigma.Lax}

Having a flat and conserved current $j$, one can construct
a family of flat currents
\[\label{eq:LaxCurrent}
\intcurr(x)=\frac{1}{1-x^2}\, j+\frac{x}{1-x^2}\,{\ast j}
\]
parameterized by the spectral parameter $x$.
These give rise to a pair of Lax operators
$(\oplax[M],\oplax)=d+\intcurr$
\<\label{eq:LaxPair}
\oplax(x)\eq
\partial_\sigma+\intcurr_\sigma(x)
=
\partial_\sigma
+\frac{1}{2}\left(\frac{j_+}{1-x}-\frac{j_-}{1+x}\right),
\nln
\oplax[M](x)\eq
\partial_\tau+\intcurr_\tau(x)
=
\partial_\tau
+\frac{1}{2}\left(\frac{j_+}{1-x}+\frac{j_-}{1+x}\right),
\>
where we make use of $\ast (j_\tau,j_\sigma)=(j_\sigma,j_\tau)$
and define $j_\pm=j_\tau\pm j_\sigma$.
The conservation and flatness conditions
for $j$ are interpreted as
the flatness condition for $\intcurr(x)$
for all values of $x$
\[\label{eq:LaxFlat}
d\intcurr(x)+\intcurr(x)\wedge \intcurr(x)=0
\qquad
\mbox{or}\qquad
\bigcomm{\oplax(x)}{\oplax[M](x)}=0.
\]
Here we have made use of the relations
${\ast b}\wedge c=-b\wedge \ast c$ and
$\ast b \wedge\ast c=-b\wedge c$
for one-forms $b,c$ in two dimensions.

We can now compute the monodromy of the operator $d+\intcurr(x)$ around the
closed string. This is the Wilson line along
the curve $\gamma(\sigma,\tau)$ which winds once around
the string and which is starting and ending at the point $(\tau,\sigma)$
\[\label{eq:MonoLoop}
\mono(x,\tau,\sigma)=\pexp \int_{\gamma(\tau,\sigma)} \bigbrk{-\intcurr(x)}.
\]
As the current $\intcurr(x)$ is flat, the actual shape of the curve $\gamma$
is irrelevant.
The monodromy $\mono(x)$ depends on the starting point $(\tau,\sigma)$
through the defining equations of the Wilson line,
\[\label{eq:shiftbase}
d\mono(x)+\bigcomm{\intcurr(x)}{\mono(x)}=0.
\]
which generates a similarity transformation.
Physical information should be invariant
under the choice of specific points on the world sheet.
Therefore, the monodromy $\mono(x)$ is not physical, but
only its conjugacy class, i.e.~the set of its eigenvalues.
For our purposes this means that neither the curve nor its starting
point is relevant. We can thus choose the curve
$\gamma$ to be given by $\tau=0$, $\sigma\in [0,2\pi]$.
The monodromy matrix becomes
\[\label{eq:SigmaMonodromy}
\mono(x)=\pexp\int_0^{2\pi}d\sigma\,
\frac{1}{2}\left(\frac{j_+}{x-1}+\frac{j_-}{x+1}\right).
\]
where the path ordering symbol $\mathrm{P}$ puts the
values of $\sigma$ in decreasing order from left to right.

\subsection{Eigenvalues of the Monodromy Matrix}
\label{sec:Sigma.Mono}

Let us now choose the current
in the vector representation $j=j_\indvec$.
Since $j_\indvec^\trans=-j_\indvec$ and $x\in\Comp$,
$\mono^\indvec$ is a complex orthogonal matrix%
\footnote{
In fact it satisfies the reality condition
$(\mono^\indvec(x^\ast))^\ast=\mono^\indvec(x)$,
i.e.~the complex values in $\mono^\indvec(x)$ are
introduced only through a complex $x$.
}
\[\label{eq:MonoOrtho}
\mono^\indvec\mono^\indvec{}^\trans=1.
\]
%
Only the conjugacy class of $\mono^\indvec(x)$,
characterized by its eigenvalues, corresponds to
physical observables.
$\Omega^\indvec\in \grp{SO}(\grn,\Comp)$ is diagonalized
into the following general form
\[\label{eq:OmegaEigen}
\mono^\indvec(x)\simeq
\left\{
\begin{array}{ll}
\diag\left(e^{iq_1(x)},e^{-iq_1(x)},e^{iq_2(x)},e^{-iq_2(x)},
\ldots,e^{iq_{[\grn/2]}(x)},e^{-iq_{[\grn/2]}(x)}\right)
&\mbox{for $\grn$ even},\\
\diag\left(e^{iq_1(x)},e^{-iq_1(x)},e^{iq_2(x)},e^{-iq_2(x)},
\ldots,e^{iq_{[\grn/2]}(x)},e^{-iq_{[\grn/2]}(x)},1\right)
&\mbox{for $\grn$ odd},
\end{array}
\right.
\]
where we express the eigenvalues in terms of
quasi-momenta $\{q_k(x)\}_{k=1}^{[\grn/2]}$.
However, there still remains the freedom of permutation
of the eigenvalues, switching the sign and
adding integer multiples of $2\pi i$.
A more convenient quantity is the characteristic polynomial
\<\label{eq:MonoDet}
\detshift^\indvec(\alpha)\eq
\det(\alpha-\mono^\indvec)
\nln\eq
\left\{
\begin{array}{ll}
\prod_{k=1}^{[\grn/2]}(\alpha-e^{iq_k})(\alpha-e^{-iq_k}),
  &\mbox{for $\grn$ even}\\
(\alpha-1)\prod_{k=1}^{[\grn/2]}(\alpha-e^{iq_k})(\alpha-e^{-iq_k}),
  &\mbox{for $\grn$ odd}
\end{array}
\right.
\nln\earel{\equiv}
\sum_{k=0}^\grn(-1)^{\grn-k}\,\alpha^k\,\transfer_{\indvec[k]}.
\>
It is an $\grn$-th order polynomial and each coefficient
$\transfer_{\indvec[k]}$ is a symmetric polynomial of the eigenvalues.
$\transfer_{\indvec[k]}$ is the trace of the monodromy matrix in
the $k$-th antisymmetric tensor product of the vector representation.
Note that
\[\label{eq:TransferAnti}
\transfer_{\indvec[\grn]}=\transfer_{\indvec[0]}=1,\quad
\transfer_{\indvec[\grn-k]}=\transfer_{\indvec[k]}
\]
in agreement with representation theory.



For the monodromy matrix in the spinor representation we find
\[\label{eq:monospineigen}
\mono^\indspin\simeq
\diag\bigbrk{\exp(\pm\sfrac{i}{2}q_1\pm\sfrac{i}{2}q_2\cdots \pm\sfrac{i}{2}q_{[\grn/2]})}.
\]
with all $2^{[\grn/2]}$ choices of signs.
When $n$ is even, we can reduce $\mono^\indspin$ further into its
chiral and antichiral parts $\mono^{\indspin\pm}$
\[\label{eq:MonoChiral}
\mono^\indspin=\matr{cc}{\mono^{\indspin+}&0\\0&\mono^{\indspin-}}.
\]
For $\mono^{\indspin+}$ and $\mono^{\indspin-}$ we should
only take those eigenvalues with an even or odd number
of plus signs in \eqref{eq:monospineigen}, respectively.

\subsection{Analyticity}
\label{sec:Sigma.Anal}

The monodromy matrix $\mono(x)$ depends analytically on $x$ 
except at the two singular points $x=\pm 1$, see \secref{sec:Sigma.Sing}. 
This, however, does not directly imply the same
for its eigenvalues or the $q_k$'s. 
Most importantly, at those points $\set{x^\ast_b}$ 
where two eigenvalues $e^{iq_k},e^{iq_l}$ degenerate
we should expect the generic behavior 
\[\label{eq:StartBranch}
q_{k,l}(x)=q_{k,l}(x^\ast_b)\pm\alpha_b\sqrt{x-x^\ast_b}+\order{x-x^\ast_b}
\]
with some coefficients $\alpha_b$.
This square-root singularity not only violates analyticity locally at
the point $x^\ast_b$, but also requires a square-root branch cut
originating from it. At the branch cut, the eigenvalues are permuted.
Furthermore, $q_k$ is defined only modulo $2\pi$. 
Finally, the labelling of $q_k$'s is defined at our will.
Although the $q_k$'s could fluctuate randomly from 
one point to the next according to these two ambiguities, 
we shall assume $q_k$ to be analytic except at 
two singular points and at a (finite) number of branch cuts $\contour_{a}$.
At the cuts, the $q_k$'s can be permuted and
shifted by multiples of $2\pi$.
Such a transformation is captured by the equations
\[
\qsheetsl_k(x)\mp\qsheetsl_l(x)=2\pi n_a\qquad
\mbox{for }x\in\contour_a.
\]
Here $\qsheetsl(x)$ means the principal part
of $q_k(x)$ across the cut
\[
\qsheetsl_k(x)=\half q_k(x+i\epsilon)
+\half q_k(x-i\epsilon).
\]
The integer $n_a$ is called the mode number of $\contour_a$
and it is assumed to be constant along the cut.
Without loss of generality we can restrict ourselves to a subset of allowed permutations, 
$q_k$ with $q_{k+1}$, i.e.
\[\label{eq:SigmaBethe0}
\qsheetsl_k(x)-\qsheetsl_{k+1}(x)=2\pi n_{k,a}\qquad
\mbox{for }x\in\contour_{k,a}.
\]
These must be supplemented by (for $\grn$ even)
\[\label{eq:SigmaBetheEven}
\qsheetsl_{[\grn/2]-1}(x)+\qsheetsl_{[\grn/2]}(x)=2\pi n_{[\grn/2],a}\qquad
\mbox{for }x\in\contour_{[\grn/2],a}
\]
or (for $\grn$ odd)
\[\label{eq:SigmaBetheOdd}
2\qsheetsl_{[\grn/2]}(x)=2\pi n_{[\grn/2],a}\qquad
\mbox{for }x\in\contour_{[\grn/2],a}.
\]
%

\subsection{Asymptotics}
\label{sec:Sigma.Asymp}

Let us investigate the expansion at $x=\infty$.
In leading order, the family of flat connections
\[\label{eq:LaxInfty}
\intcurr(x)=-\frac{1}{x}\,{\ast j}+\order{1/x^2}\]
yields the dual of the conserved current $j$.
The expansion of the monodromy matrix $\mono(x)$ at $x=\infty$ thus yields
\[\label{eq:SigmaMonoInfty}
\mono(x)=I+\frac{1}{x}\int_{0}^{2\pi} d\sigma \,j_{\tau}+\order{1/x^2}
=I+\frac{1}{x}\,\frac{4\pi J}{\sqrt{\lambda}}+\order{1/x^2}.
\]
Here, the conserved charges $J$ of the sigma-model
\[\label{eq:SigmaCharge}
J=\frac{\sqrt{\lambda}}{4\pi}
\int_{0}^{2\pi} d\sigma \,j_\tau,
\]
appear as the first order in the expansion in $1/x$.
The eigenvalues of $J^\indvec$ in the vector representation
are given by
\[\label{eq:ChargeEigen}
J^\indvec\simeq
\left\{
\begin{array}{ll}
\diag\left(iJ_1,-iJ_1,iJ_2,-iJ_2,
\ldots,iJ_{[\grn/2]},-iJ_{[\grn/2]}\right)
&\mbox{for $\grn$ even},\\
\diag\left(iJ_1,-iJ_1,iJ_2,-iJ_2,
\ldots,iJ_{[\grn/2]},-iJ_{[\grn/2]},0\right)
&\mbox{for $\grn$ odd}.
\end{array}
\right.
\]
The charge eigenvalues $J_k$ are related to the Dynkin labels
$[s_1,s_2,\ldots]$ for even $\grn$ by
\[\label{eq:ChargeDynkinEven}
J_{k}=\sum_{j=k}^{[\grn/2]}s_j-\half s_{[\grn/2]-1}-\half s_{[\grn/2]},
\]
and for odd $\grn$ by
\[\label{eq:ChargeDynkinOdd}
J_{k}=\sum_{j=k}^{[\grn/2]}s_j-\half s_{[\grn/2]}.
\]
When we compare
\eqref{eq:SigmaMonoInfty,eq:OmegaEigen,eq:ChargeEigen}
we find the expansion of the $q_k$'s
at $x=\infty$.
Since $q_k$'s are defined as in \eqref{eq:OmegaEigen},
there is a freedom of
choosing signs and the ordering of $q_k$'s
as well as the branch of the logarithm.
We fix them so that $q_k$ has asymptotic behavior
\[\label{eq:qasymp}
q_k(x)=\frac{1}{x}\,\frac{4\pi J_k}{\sqrt{\lambda}}
+\order{1/x^2}.
\]
In particular we fix the branch of all the logarithms
such that all $q_k(x)$ vanish at $x=\infty$.

\subsection{Singularities}
\label{sec:Sigma.Sing}

To study the asymptotics of the monodromy matrix
\eqref{eq:SigmaMonodromy} at $x\to \pm 1$ we
assume that
at all values of $\sigma$
the unitary matrix $u_\pm(\sigma)$
diagonalizes $j_\pm(\sigma)$
\footnote{We thank Gleb Arutyunov for
explaining to us the expansion.}
\[\label{eq:udiagj}
u_\pm(\sigma)\,j_\pm(\sigma)\,u_\pm^{-1}(\sigma)=j_\pm^{\,\diag}(\sigma).
\]
We furthermore assume that the function $u_\pm(x,\sigma)$ is some
analytic continuation of $u_\pm(\sigma)$ at
$x=\pm 1$, i.e.
\[\label{eq:uregular}
u_\pm(x,\sigma)=u_\pm(\sigma)+\order{x\mp 1}.
\]
When we now do a gauge transformation using $u_\pm(x,\sigma)$
we find that
\[\label{eq:LaxDiag}
u_\pm(x)\,\oplax(x)\,u_\pm(x)^{-1}=
\partial_\sigma
-\frac{1}{2}\,\frac{j_\pm^{\,\diag}}{x\mp 1}+\bigorder{(x\mp 1)^0}
\]
where the gauge term $\partial_\sigma u_\pm \,u_\pm^{-1}$
is of order $\order{(x\mp 1)^0}$. The higher-order
off-diagonal terms in $\oplax(x)$
can be removed order by order by adding
the appropriate terms to $u_\pm(x)$.
Therefore $u_\pm(x)$ can be used to completely
diagonalize $\oplax(x)$ for all $\sigma$.
Thus we can drop the path ordering and write
\[\label{eq:MonoDiag}
u_\pm(x,2\pi)\,\mono^\indvec(x)\, u_\pm^{-1}(x,0)
=\exp\(\frac{1}{2}\int_0^{2\pi} d\sigma\,
\frac{j^{\,\diag}_\pm}{x\mp 1}+\bigorder{(x\mp 1)^0}\)
\]
In other words, to compute the leading singular behavior of the eigenvalues of
$\mono^\indvec(x)$, it suffices to integrate the eigenvalues of $j_\pm$.
In \appref{sec:Charges} we will compute the next few terms in
this series and thus find some local commuting charges of the
sigma model.

{}From \eqref{eq:SigmaCurrent} we infer that the current $j_\indvec$
in the vector representation
has only two non-zero eigenvalues.
They are imaginary, have equal absolute value
but opposite signs.
The absolute value is determined through the Virasoro constraint
\eqref{eq:Virasoro},
$\brk{\partial_\pm \Xvec}^2
=\brk{\partial_\pm X_0}^2$.
We then find using
\eqref{eq:SigmaCurrent}
\[\label{eq:SigmaPoles}
-\sfrac{1}{8}\Tr\(j_{\indvec,\pm}\)^2
=\bigbrk{\partial_\pm \Xvec}^2
=\bigbrk{\partial_\pm X_0}^2
=\frac{\dimn^2}{\lambda}\,\brk{\partial_\pm \tau}^2
=\frac{\dimn^2}{\lambda}\,.
\]
%
%
The diagonalized matrix $j_{\indvec\,\pm}^{\,\diag}$ thus takes
the form
\[\label{eq:jdiag}
 j_{\indvec,\pm}^{\,\diag}=\frac{2i\dimn}{\sqrt{\lambda}}\  \diag(+1,-1,0,0,0,0,\ldots)\]
and hence does not depend on $\sigma$. This leads
to the following asymptotic formula for the eigenvalues of the
monodromy matrix:
\[\label{eq:qsing}
 q_k =
\delta_{k1}\,\frac{2\pi\dimn}{\sqrt{\lambda}\,(x\mp 1)}+
\bigorder{(x\mp 1)^0}
\qquad \mbox{for }x\to\pm 1 .
\]
This formula allows to relate the asymptotic data at $x\to\pm 1$
with the energy of a classical state of the string and hence identify
the anomalous dimension of an operator due to the AdS/CFT
correspondence.

\subsection{Inversion Symmetry}
\label{sec:Sigma.Sym}

For a sigma model with right and left currents
$j=\gmat^{-1}d\gmat$ and $\ell=-d\gmat\, \gmat^{-1}$,
the families of associated flat currents
$\intcurr(x)$ and $\intcurr_\ell(x)$ are related by an inversion of the
spectral parameter $x$:
\<\label{eq:LaxInverse}
\gmat\bigbrk{d+\intcurr(x)}\gmat^{-1}\eq
\gmat \,d\gmat^{-1}+\frac{1}{1-x^2}\,\gmat j\gmat^{-1}+\frac{x}{1-x^2}\,\gmat\,{\ast j\gmat^{-1}}
\nln\eq
d+\ell-\frac{1}{1-x^2}\,\ell-\frac{x}{1-x^2}\,{\ast \ell}
\nln\eq
d-\frac{x^2}{1-x^2}\,\ell-\frac{x}{1-x^2}\,{\ast \ell}
\nln\eq
d+\frac{1}{1-1/x^2}\,\ell+\frac{1/x}{1-1/x^2}\,{\ast \ell}
\nln\eq
d+\intcurr_\ell(1/x).
\>
In our $S^{\grn-1}$ model where $j=\ell$ this is particularly interesting
since it implies the symmetry
\[\label{eq:LaxSym}
\gvec^{}\,\oplax^{}_\indvec(x)\,\gvec^{-1}=\oplax^{}_\indvec(1/x)
\qquad\mbox{and}\qquad
\gspin^{}\,\oplax^{}_\indspin(x)\,\gspin^{-1}=\oplax^{}_\indspin(1/x)
\]
and consequently
\[\label{eq:MonoSym}
\gvec(2\pi)\,\mono^\indvec(x)\,\gvec(0)^{-1}=\mono^\indvec(1/x)
\qquad\mbox{and}\qquad
\gspin(2\pi)\,\mono^\indspin(x)\,\gspin(0)^{-1}=\mono^\indspin(1/x).
\]
%
For a closed string we have
$\gmat(0)=\gmat(2\pi)$, therefore $\mono(x)$ and $\mono(1/x)$ are related
by a similarity transformation and thus have the same set of eigenvalues.
For the vector representation
it means that each of the quasi-momenta $\{q_1,-q_1,q_2,-q_2,\ldots\}$
transforms into one of $\{q_1,-q_1,q_2,-q_2,\ldots\}$
under the $x\leftrightarrow1/x$ symmetry.

When $\grn$ is even, the spinor representation is can be reduced into
its chiral and antichiral components.
We know that a gamma matrix $\gammavec$ interchanges both
chiralities.
Therefore the matrix $\gspin$ inverts chirality
while $j_\indspin=\gspin^{-1}d\gspin^{}$ preserves it.
It follows that the inversion symmetry
relates monodromy matrices of opposite chiralities
\[\label{eq:MonoSymSpin}
\gspin(2\pi)\,\mono^{\indspin\pm}(x)\,\gspin(0)^{-1}=\mono^{\indspin\mp}(1/x).
\]
%


To be consistent with the singular behavior \eqref{eq:qsing},
$q_1$ has to transform as
\[\label{eq:q1Sym}
q_1(1/x)=4\pi n_0-q_1(x).
\]
The integer constant $n_0$ reflects
the difference of branches of the logarithm
at $x=0$ and $x=\infty$.
We need a factor of $4\pi$, because for
spinor representations \eqref{eq:monospineigen} we find the exponentials
$\exp(\pm \sfrac{i}{2}q_1)$, which must not change sign.
Furthermore, for even $\grn$ we require that chiral
and antichiral representations are interchanged.
This is achieved by an odd number of sign flips
for all the $q_k$.
A possible transformation rule for the
quasi-momenta that works for all
values of $\grn$ is%
\footnote{We cannot exclude different 
transformation rules at this point. 
It would be interesting to see whether 
there exist such solutions 
(which are not merely obtained by a relabelling of the eigenvalues).}
\[\label{eq:qkSym}
q_k(1/x)=(1-2\delta_{k1})\, q_k(x)+4\pi n_0\delta_{k1},
\]
i.e.~$q_1$ is flips sign while all other
$q_k$ are invariant.
We shall assume that it is the correct
rule although there might be other
consistent choices for $\grn>4$.
Note that there are no additional constant shifts
for $q_k$, $k\neq 1$, because these would be in conflict with
the even transformation rule.

\subsection{The Sigma-Model on $\Real\times S^3$}
\label{sec:RS3}

To test out results, we will consider the case $\Real\times S^3$ which
was extensively studied in \cite{Kazakov:2004qf}.
The isometry group $\grp{SO}(4)$ of $S^3$ is locally isomorphic
to $\grp{SU}(2)\indup{L}\times \grp{SU}(2)\indup{R}$,
which played a key role in the preceding
discussion.
Now we need not to make use of the isomorphism,
nevertheless it is useful to interpret
our formulation also in terms 
of $\grp{SU}(2)\indup{L}\times \grp{SU}(2)\indup{R}$.

The spinor representations $\rep{2}\indup{L},\rep{2}\indup{R}$ of $\grp{SO(4)}$
can be viewed as the fundamental representations
of $\grp{SU(2)}\indup{L}$, $\grp{SU(2)}\indup{R}$, respectively.
Monodromy matrices in these representations
are then viewed as the $SU(2)$ monodromy matrices.
They are diagonalized as
\[
\mono^{\indspin+}\sim\diag(e^{ip_{\rm L}},e^{-ip_{\rm L}}),
\quad
\mono^{\indspin-}\sim\diag(e^{ip_{\rm R}},e^{-ip_{\rm R}}).
\]
Now we have two independent quasi-momenta $p_{\rm L},p_{\rm R}$,
due to the reducibility of $SO(4)$.
They can be related to the quasi-momenta $q_1,q_2$ by
\[\label{eq:pLRdef}
p_{\rm L}=\half q_1+\half q_2,\qquad
p_{\rm R}=\half q_1-\half q_2
\]
where we put constant terms for convenience.

The inversion symmetry now gives rise to
\[\label{eq:pLRinv}
p(x):=p_{\rm R}(x)=-p_{\rm L}(1/x)+2\pi n_0.
\]
This means one can assemble the two quasi-momenta
to a single quasi-momentum $p(x)$
without inversion symmetry.
This special fact played a crucial role in
the previous analysis in \cite{Kazakov:2004qf}.
In fact this $p(x)$ is the quasi-momentum
discussed there.

Let us deduce properties of $p(x)$ from our general results.
The pole structure reads
\[\label{eq:pLsing}
p(x) =
\frac{\pi\dimn}{\sqrt{\lambda}\,(x\mp 1)}+
\bigorder{(x\mp 1)^0}.
\]
It exhibits the asymptotic behavior
\[\label{eq:pLasymp}
p(x)=\frac{2\pi r\indup{R}}{\sqrt{\lambda}}\,\frac{1}{x}
+\order{1/x^2}
\qquad \mbox{for }x\to\infty
\]
while the asymptotic behavior of $p_{\rm L}(x)$
for $x\to\infty$ is interpreted as
\[\label{eq:pLasymp2}
p(x)=2\pi n_0-\frac{2\pi r\indup{L}}{\sqrt{\lambda}}\,x
+\order{x^2}
\qquad \mbox{for }x\to 0.
\]
The Dynkin labels $r\indup{L}=J_1+J_2$ and
$r\indup{R}=J_1-J_2$ 
specify 
the quantum numbers of 
$\grp{SU(2)}\indup{L}$ and 
$\grp{SU(2)}\indup{R}$;
they equal twice the invariant $\grp{SU}(2)$ spin.
Both analyticity conditions 
\eqref{eq:SigmaBethe0,eq:SigmaBetheEven}
are now simply given by
\[\label{eq:BeqS3}
2\sheetsl(x)
=2\pi n_{a},
\qquad x\in \contour_{a}.
\]
All of these agree exactly with
the previous results \cite{Kazakov:2004qf}.

\section{Algebraic Curve for the Sigma-Model on $\Real\times S^{5}$}
\label{sec:Alg.EqSM}

In this section we will show that the generic solution to
the string sigma model on $\Real\times S^5$ is 
uniquely characterized by a set of 
mode numbers and fillings. 
These are related to certain cycles of
the derivative of the quasi-momentum, $q'_k(x)$, 
which is an algebraic curve of degree four. 

\subsection{SO(6) vs. SU(4)}

The isometry group $\grp{SO}(6)$ of $S^5$ is
locally isomorphic to $\grp{SU}(4)$.
This enables us to formulate the model
in terms of the $\alg{su}(4)$ algebra
and the spinor representation
which turns out to simplify the
structure of the algebraic curve.
Here we will translate the properties obtained
in the previous section in terms of the
quasi-momentum $p$ 
corresponding to the spinor representation
instead of $q$ which corresponds to the vector representation.

The chiral spinor representation $\rep{4}$ of $\grp{SO}(6)$
is equivalent to the fundamental representation of $\grp{SU}(4)$.
Therefore $\mono^{\indspin+}$ can be regarded as
the $\grp{SU}(4)$ monodromy matrix,
which is diagonalized as
\[
\mono^{\indspin+}\sim\diag(e^{ip_1},e^{ip_2},e^{ip_3},e^{ip_4})
\]
with $p_1+p_2+p_3+p_4=0$.
The quasi-momenta $p_k$ are identified as
\<\label{pqRel}
p_1\Eqn{=}\tfrac{1}{2}(\phantom{+}q_1+q_2-q_3),
\nln
p_2\Eqn{=}\tfrac{1}{2}(\phantom{+}q_1-q_2+q_3),
\nln
p_3\Eqn{=}\tfrac{1}{2}(-q_1+q_2+q_3),
\nln
p_4\Eqn{=}\tfrac{1}{2}(-q_1-q_2-q_3)
\>
in our general notation.

\begin{figure}\centering
\includegraphics{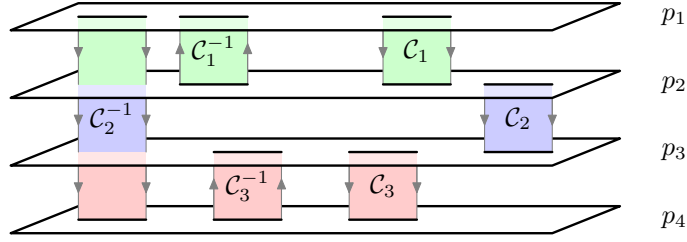}
\caption{Structure of sheets and branch cuts 
in the $\rep{4}$ representation
for the sigma model on $S^5$.
There are three types of cuts,
$\contour_{1,2,3}$, corresponding
to the simple roots of $\grp{SU}(4)$.
For each cut $\contour$, there is a mirror cut
$\contour^{-1}$. Whether or not it connects
the same two sheets depends on the type of cut.
The total number of cuts including the mirror images
is denoted by $A_1,A_2,A_3$.}
\label{fig:Sigma4}
\end{figure}

The inversion symmetry \eqref{eq:qkSym}
in terms of $p_k$ is now written as%
\footnote{This is based on the assumption \eqref{eq:qkSym}.
Other possibilities for $\grn=6$ are
$p_{1,2}(1/x)\in 2\pi \Integers-p_{1,2}(x)$
and/or $p_{3,4}(1/x)\in 2\pi \Integers-p_{3,4}(x)$.
This will slightly change the counting of individual constraints below,
but the overall number of moduli will remain the same.}
\[\label{eq:CurveSym}
p_{1,2}(1/x)=2\pi n_0-p_{2,1}(x),\qquad
p_{3,4}(1/x)=-2\pi n_0-p_{4,3}(x).
\]
This leads to the structure of branch cuts as depicted in 
\figref{fig:Sigma4}. 
The cuts of type $\contour_1,\contour_2,\contour_3$
correspond to the three simple roots 
of $\grp{SU}(4)$. While cuts $\contour_1$ and $\contour_3$
connect sheets $1,2$ and $3,4$, respectively,
cuts $\contour_2$ may connect either the sheets $2,3$ 
or the sheets $1,4$ due to the symmetry \eqref{eq:CurveSym}.
The total number of cuts of either type will be denoted by $A_1,A_2,A_3$,
mirror cuts are assumed to be explicitly included.

\begin{figure}
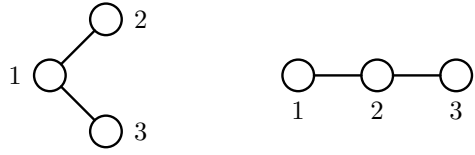
\centering
\parbox{3cm}{\centering\includegraphics{bks.dynkin.so6.eps}}
\qquad
\parbox{3cm}{\centering\includegraphics{bks.dynkin.su4.eps}}
\caption{Dynkin diagrams of $\grp{SO}(6)$ and $\grp{SU}(4)$.}
\label{fig:so6su4}
\end{figure}

The pole structure \eqref{eq:qsing} reads
%
\[\label{eq:pkSing}
 p_{1,2}(x)=-p_{3,4}(x) =
\frac{\pi\dimn}{\sqrt{\lambda}\,(x\mp 1)}+
\bigorder{(x\mp 1)^0}
\qquad \mbox{for }x\to\pm 1 .
\]
while the asymptotic behavior 
at $x=\infty$ \eqref{eq:qasymp}
now reads
\<
\label{eq:pkAsymp}
p_1(x)\eq
\frac{1}{x}\frac{4\pi}{\sqrt{\lambda}}
\lrbrk{\phantom{+}\sfrac{3}{4}r_1+\sfrac{1}{2}r_2+\sfrac{1}{4}r_3}+\cdots
=\frac{1}{x}\frac{2\pi}{\sqrt{\lambda}}
\lrbrk{\phantom{+}J_1+J_2-J_3}+\cdots
,
\nln
p_2(x)\eq
\frac{1}{x}\frac{4\pi}{\sqrt{\lambda}}
\lrbrk{-\sfrac{1}{4}r_1+\sfrac{1}{2}r_2+\sfrac{1}{4}r_3}+\cdots
=\frac{1}{x}\frac{2\pi}{\sqrt{\lambda}}
\lrbrk{\phantom{+}J_1-J_2+J_3}+\cdots
,
\nln
p_3(x)\eq
\frac{1}{x}\frac{4\pi}{\sqrt{\lambda}}
\lrbrk{-\sfrac{1}{4}r_1-\sfrac{1}{2}r_2+\sfrac{1}{4}r_3}+\cdots
=\frac{1}{x}\frac{2\pi}{\sqrt{\lambda}}
\lrbrk{-J_1+J_2+J_3}+\cdots
,
\nln
p_4(x)\eq
\frac{1}{x}\frac{4\pi}{\sqrt{\lambda}}
\lrbrk{-\sfrac{1}{4}r_1-\sfrac{1}{2}r_2-\sfrac{3}{4}r_3}+\cdots
=\frac{1}{x}\frac{2\pi}{\sqrt{\lambda}}
\lrbrk{-J_1-J_2-J_3}+\cdots
.
\>
Here the Dynkin labels $[r_1,r_2,r_3]$ of $\grp{SU}(4)$ are
related to the Dynkin labels $[s_1;s_2,s_3]$ and 
charges $(J_1,J_2,J_3)$ of $\grp{SO}(6)$
by
\<
r_1\eq s_2=J_2-J_3,\nln
r_2\eq s_1=J_1-J_2,\nln
r_3\eq s_3=J_2+J_3.
\>
This is due to the difference in the labelling of simple roots
between the Lie algebras of $\grp{SU}(4)$ and $\grp{SO}(6)$:
The labels $1$ and $2$ are interchanged
(see \figref{fig:so6su4}).

\subsection{Branch Cuts}

The monodromy matrix $\mono^{\indspin+}(x)$
has similar analytic properties as the one for the
spin chain of \secref{sec:Spins}.
Therefore, as discussed in \secref{sec:Spins.Curve}, 
the derivative of the quasi-momentum, $p'=(p'_1,p'_2,p'_3,p'_4)$,
is again an algebraic curve of degree four.

First of all, let us define 
\[\label{eq:CurveDefY}
y_k(x)=\lrbrk{x-1/x}^2 x\, p_k'(x).\]
This removes the poles at $x=\pm 1$ \eqref{eq:pkSing}
and leads to a simple transformation rule under the symmetry \eqref{eq:CurveSym}.
We can now write $y$ as the solution to an algebraic equation of the same
type as \eqref{eq:algeq}
\[\label{eq:algeqSM}
F(y,x)=P_{4}(x)\,y^4+P_{2}(x)\,y^2+P_{1}(x)\,y+P_{0}(x)=0.
\]
As explained in \secref{sec:Spins.Curve} we know that
the branch points are given by the roots of $P_{4}(x)$,
let us assume there are $A$ cuts
\[\label{eq:SMbranch}
P_{4}(x)\sim\prod_{a=1}^A \bigbrk{x-a_a}\bigbrk{x-b_a}. \]
%
Together with $P_3(x)=0$ this can easily be seen to yield
a $1/\sqrt{x-a_a}$ and $1/\sqrt{x-b_a}$ behavior at
$a_a,b_a$ as expected from \eqref{eq:StartBranch}.
We need to remove all further branch points,
which are generically of the type $\sqrt{x-x^\ast}$.
These would lead to unexpected $(x-x^\ast)^{3/2}$ behavior
in $p(x)$, cf. \secref{sec:Sigma.Anal}.
Their positions $x^\ast$ can be obtained as
roots of the discriminant $R$ of the quartic equation
\[\label{eq:Discr2}
R=
-4P_1^2 P_2^3
+ 16P_0 P_2^4
- 27P_1^4 P_4
+ 144P_0 P_1^2 P_2 P_4
- 128P_0^2 P_2^2 P_4
+ 256P_0^3 P_4^2.
\]
This means that the discriminant must be a perfect square
\[\label{eq:DisSquare2}
R(x)=Q(x)^2.
\]
%

\subsection{Asymptotics}

The asymptotics $p(x)\sim 1/x$ at $x=\infty$,
\eqref{eq:pkAsymp} translate to 
\[
y(x)\sim x \qquad\mbox{at }x=\infty.
\]
for $y$ as defined in \eqref{eq:CurveDefY}.
This requires that $P_k(x)\sim x^{-k}P_0(x)$ 
for the highest-order terms. 
Similarly, the asymptotics $p(x)\sim \mbox{const}.+x$ at $x=0$ obtained
through the symmetry \eqref{eq:CurveSym} translate to
\[
y(x)\sim 1/x \qquad\mbox{at }x=0.
\]
This requires that $P_k(x)\sim x^{k}P_0(x)$ for the lowest-order terms.
In total this leads to polynomials of the form
\[\label{eq:Pkcoeffs}
P_k(x)=\ast x^k+\cdots+\ast x^{2A+8-k}.
\]
Consequently, the discriminant \eqref{eq:Discr2} takes the form
\[
R(x)=\ast x^8+\cdots+\ast x^{10A+32}.
\]
%

\subsection{Symmetry}

The symmetry of the quasi-momentum $p$ in \eqref{eq:CurveSym} 
translates to 
\[\label{eq:CurveSymY}
y_{1,2}(1/x)= y_{2,1}(x),
\qquad
y_{3,4}(1/x)= y_{4,3}(x).
\]
In order the solution to the algebraic equation \eqref{eq:algeqSM}
have this symmetry, the polynomials must transform according to%
\footnote{We could also assume $P_k(1/x)=-x^{-2A-8} P_k(x)$,
but it turns out to be too restrictive.}
\[\label{eq:PkSym}
P_k(1/x)=x^{-2A-8} P_k(x).
\]
Similarly, the resolvent satisfies
\[\label{eq:RSym}
R(1/x)=x^{-10A-40}R(x).
\]
In other words, the coefficients 
of the polynomials are the same when read backwards and forwards.
Note that in \eqref{eq:SMbranch} we have not made 
the symmetry for $P_4(x)$ manifest. 
It requires that $a_a=1/a_b$ and $b_a=1/b_b$ for a pair of
cuts $\contour_{a,b}$ which interchange under the symmetry.%
\footnote{In principle we should
also allow symmetric cuts with $b_a=1/a_a$.
Apparently these do not occur for solutions
which correspond to gauge theory states
at weak coupling. At weak coupling
one cut should grow to infinity while
the other shrinks to zero. This is
not compatible with symmetric cuts.}

The symmetry of $F(y,x)$, however, merely guarantees that
$y_k(1/x)=y_{\pi(k)}(x)$ with some permutation $\pi(k)$.
In the most general case,
there can only be the trivial permutation $\pi(k)=k$.
This can be seen by looking at the fixed points $x=\pm 1$ of the
map $x\to 1/x$. If $y_1(\pm 1)\neq y_2(\pm 1)$
there is no chance that the permutation in \eqref{eq:CurveSymY}
is realized. To permit \eqref{eq:CurveSymY} we need to make sure that 
\[\label{eq:CurveAllowSym}
y_{1}(x)=y_2(x)\quad\mbox{and}\quad y_{3}(x)=y_4(x)
\qquad\mbox{for}\qquad x=+1\quad\mbox{and}\quad x=-1.
\]
This yields four constraints on the coefficients of $F(y,x)$.
At this point the trivial permutation $\pi(k)=k$ is
still an option. However, now 
the choice between $\pi(1)=1$ and $\pi(1)=2$
is merely a discrete one, there are no further
constraints which remove a continuous degree of freedom.
In fact, as the the solution to $F(y,x)=0$ degenerates
into two pairs at $x=\pm 1$, the discriminant 
must have a quadruple pole at these points,
i.e.~we can write
\[\label{eq:Rcoeffs}
R(x)=x^8(x^2-1)^4\bigbrk{\ast x^0+\cdots+\ast x^{10A+16}}.
\]
%

\subsection{Singularities}

Let us now consider the poles at $x=\pm 1$.
The expansion of a generic solution for $y$ 
yields
\[
p'_k(x)=\frac{\alpha^\pm_k}{(x\mp 1)^2}+\frac{\beta^\pm_k}{x\mp 1}+\bigorder{(x\mp 1)^0}.
\]
We have already demanded that 
$\alpha^\pm_{1}=\alpha^\pm_{2}$ and
$\alpha^\pm_{3}=\alpha^\pm_{4}$. 
The symmetry furthermore requires 
$\beta^\pm_{1}=-\beta^\pm_{2}$ and
$\beta^\pm_{3}=-\beta^\pm_{4}$.
As the sum of all sheets must be zero, $p_1+p_2+p_3+p_4=0$,
it moreover follows that
$\alpha^\pm_{1,2}=-\alpha^\pm_{3,4}$ whereas
$\beta^\pm_{1,2}$ and $\beta^\pm_{3,4}$ are independent.
This means there are three independent coefficients each for the
singular behavior at $x=\pm 1$.
Now the residue of $p$ at $x=\pm 1$ is proportional
to the energy or dimension $D$. This we cannot
fix as it will be the (hopefully) unique result of the calculation.
However, we know that the residues at both $x=\pm 1$ are equal,
\eqref{eq:pkSing},
which gives one constraint on the $\alpha$'s. Furthermore,
there is no logarithmic behavior in $p$, \eqref{eq:pkSing}, therefore all $\beta$'s must
be zero which gives four constraints.
In total there are five constraints from the poles at $x=\pm 1$.

\subsection{A-Cycles}
\label{sec:Acycle}

The eigenvalues $\exp(ip_k(x))$ of
$\mono^{\indspin+}(x)$ are holomorphic functions of $x$.
This however does not exclude the possibility 
of cuts where the argument $p_k(x)$ jumps by multiples
of $2\pi$ but is otherwise smooth.
Such cuts originate from logarithmic or
branch-cut singularities;
they are required when the closed integral 
around the singularity does not vanish.
We know that there are no logarithmic singularities,
therefore we merely need to ensure that
\[
\oint_{\contour[A]_a} dp\in 2\pi \Integers
\]
where the cycles $\contour[A]_a$ surrounds 
a cut $\contour_a$, see \figref{fig:cycles}.
\begin{figure}\centering
\includegraphics{bks.cycles.eps}
\caption{Branch cut $\contour_{a}$ between
sheets $k$ and $k+1$ with associated
$\contour[A]$-cycles and $\contour[B]$-period.}
\label{fig:cycles}
\end{figure}%
As was shown in \cite{Kazakov:2004qf}
we can even demand that all
$\contour[A]$-cycles are zero, which conveniently 
reduces the number of cuts.

Assume first $\contour_a$ connects
sheets $1,2$ or sheets $3,4$.%
\footnote{In \figref{fig:Sigma4} we have illustrated how
the sheets are connected by the cuts.} 
Then there is another
cut $\contour_b$ as the image of $\contour_a$ 
under \eqref{eq:CurveSym} between the same two sheets.
The values of the $\contour[A]$-cycles are
related
\[
\oint_{\contour[A]_a} dp_1
=
-\oint_{\contour[A]_b=1/\contour[A]_a} dp_2
=
\oint_{\contour[A]_b} dp_1.
\]
%
The two signs flips are explained as follows:
The first is related to the symmetry
\eqref{eq:CurveSym} and
the second to changing the sheet back to $1$.
For a cut which connects sheets $2,3$, the mirror
image will connect sheets $1,4$. In this case
we have
\[\label{eq:Asym2}
\oint_{\contour[A]_a} dp_2
=
-\oint_{1/\contour[A]_a} dp_1
=
-\oint_{\contour[A]_b} dp_1
\]
In particular this means that there is only
one constraint
\[
\oint_{\contour[A]_a} dp=0
\]
for each pair of cuts.
Moreover, the cycle around all cuts
on sheet $1$ is just the negative
of the corresponding one on sheet $2$;
equivalently for sheets $3$ and $4$.
As there are no further single poles on any
sheet, the cycle around all cuts can be contracted
and must be zero. The total number of constraints
from $\contour[A]$-cycles is thus $\half A-2$.

\subsection{B-Periods}
\label{sec:Bcycle}

We know that the set of eigenvalues $\exp(ip_k(x))$ of 
$\mono^{\indspin+}(x)$ depends analytically on $x$.
Their labeling $k=1,2,3,4$, however, is artificial.
This allows for the presence of cuts $\contour_a$ 
where the $p_k$ permute,
see \secref{sec:Sigma.Anal}.
In addition they can also shift by multiples
of $2\pi$ without effect on $\exp(ip_k(x))$.
This shift can be expressed through the
integral of $dp$ along the curve $\contour[B]_a$
which connects the points $x=\infty$ on
the involved sheets through the cut $\contour_a$.
We know that $p(x)$ is analytic along $\contour[B]_a$
except at the intersection of $\contour[B]_a$ with $\contour_a$.
Moreover we assume that $p(\infty)=0$ on both sheets,
therefore the period
\[
\int_{\contour[B]_a} dp\in 2\pi \Integers
\]
describes the shift in $p(x)$ at $\contour_a$ and must be 
a multiple of $2\pi$.

Note that the symmetry $x\to 1/x$ does not map
$\contour[B]$-periods directly to $\contour[B]$-periods
due to the explicit reference to the point $x=\infty$.
First, we should therefore consider the integral
\[
\int_0^\infty dp_k=p_k(\infty)-p_k(0)=-p_k(0)
\]
where we have made use of our choice $p_k(\infty)=0$.
{}From \eqref{eq:CurveSym} it follows that
\[
p_{1,2}(0)=-p_{3,4}(0)=2\pi n_0
\]
which is the momentum constraint. It
reduces the number of degrees of freedom by one, because
$n_0$ must be integer.
Now consider a $\contour[B]$-period between sheets
$1,2$ or sheets $3,4$. Due to the symmetry
\<\label{eq:Bsym1}
\int_{\contour[B]_a} dp
\eq
\int_{\infty}^{x_a} dp_1
+\int_{x_a}^{\infty} dp_2
=
-\int_{0}^{1/x_a} dp_2
-\int_{1/x_a}^{0} dp_1
\nln
\eq
-\int_{0}^{\infty} dp_2
-\int_{\infty}^{x_b} dp_2
-\int_{x_b}^{\infty} dp_1
-\int_{\infty}^{0} dp_1
\nln
\eq
p_2(0)-p_1(0)
+\int^{x_b}_{\infty} dp_1
+\int^{\infty}_{x_b} dp_2
=\int_{\contour[B]_b} dp.
\>
we see that the cycles $\contour[B]_a$ and $\contour[B]_b$
have the same value.
Equivalently, for $\contour[B]_a$ between sheets
$2,3$ which is related to
$\contour[B]_a$ between sheets $1,4$
\<\label{eq:Bsym2}
\int_{\contour[B]_a} dp
\eq
\int_{\infty}^{x_a} dp_2
+\int_{x_a}^{\infty} dp_3
=
-\int_{0}^{1/x_a} dp_1
-\int_{1/x_a}^{0} dp_4
\nln
\eq
-\int_{0}^{\infty} dp_1
-\int_{\infty}^{x_b} dp_1
-\int_{x_b}^{\infty} dp_4
-\int_{\infty}^{0} dp_4
\nln
\eq
p_1(0)
-p_4(0)
-\int_{\infty}^{x_b} dp_1
-\int_{x_b}^{\infty} dp_4
=
4\pi n_0-\int_{\contour[B]_b} dp.
\>
Note that by demanding that both values of cycles $\contour[B]_a$
and $\contour[B]_b$ are multiples of $2\pi$,
it follows that $n_0$ is integer.%
\footnote{This is different from spin chain for which
there are states with non-integer total momentum are
perfectly well-defined.}
So by fixing
\[
\int_{\contour[B]_a} dp=2\pi n_a
\]
we automatically determine the value of mirror period $\contour[B]_b$.
Consequently, the $\contour[B]$-periods together with the momentum
constraint fix $\half A+1$ coefficients.

\subsection{Fillings}
\label{sec:Fillings}

The polynomial $F(y,x)$ has $8A+22$ coefficients in total, see
\eqref{eq:Pkcoeffs}.
Of them $4A+9$ are incompatible with the symmetry \eqref{eq:PkSym} 
and another $4$ are constrained by
enabling non-trivial permutations of the $y_k$,
see \eqref{eq:CurveAllowSym}.
The overall normalization of $F(y,x)$ is irrelevant for the 
$F(y,x)=0$, this removes one degree of freedom.
The discriminant $R$,
\eqref{eq:Rcoeffs}, has $5A+8$ non-trivial pairs of roots
related by the symmetry \eqref{eq:RSym}.
These should all have even multiplicity,
\eqref{eq:DisSquare2},
which fixes $\frac{5}{2}A+4$ coefficients.
The residues of the poles and absence
of logarithmic singularities at $x=\pm 1$ leads
to $5$ constraints. The $\contour[A]$-cycles 
and $\contour[B]$-periods yield
$\sfrac{1}{2}A-2$ and 
$\sfrac{1}{2}A+1$ constraints, respectively.
In total there are $\half A$ continuous degrees of 
freedom remaining. 
These can be used to assign one filling to each pair of cuts.
We define the filling of a cut $\contour_a$ as
%
\[\label{eqDefFil}
K_a=
-\frac{\sqrt{\lambda}}{8\pi^2 i}\oint_{\contour[A]_a}
dx \lrbrk{1-\frac{1}{x^2}}p(x)
=\frac{\sqrt{\lambda}}{8\pi^2 i}\oint_{\contour[A]_a}
\lrbrk{x+\frac{1}{x}} dp.
\]
The second form which directly relates to $dp$ 
is obtained by partial integration.

In addition to the fillings we define one further similar
quantity which we call the ``length{}''
\[\label{eq:Ldef}
L 
= 
D+ \frac{\sqrt{\lambda}}{4\pi^2 i} 
\sum_{a=1}^{A/2} 
\oint_{\contour[A]_{a}} \frac{dx}{x^2} \,\bigbrk{p_1(x)+p_2(x)}
= 
D - \frac{\sqrt{\lambda}}{4\pi^2 i} 
\sum_{a=1}^{A/2} 
\oint_{\contour[A]_{a}} \frac{1}{x} \,\bigbrk{dp_1(x)+dp_2(x)}.
\]
Note that the sum $\sum_{a=1}^{A/2}$ extends only over
one cut from each pair of cuts related by the inversion symmetry.%
\footnote{This definition of length is ambiguous but in the comparison
to gauge theory it becomes clear which cut to select from each pair.
A potential self-symmetric cut should be counted with weight $1/2$.}
The length is related to the fillings by the constraint
\[
n_0 L=
\sum_{a=1}^{A_1/2} n_{1,a} K_{1,a}
+\sum_{a=1}^{A_2/2} n_{2,a} K_{2,a}
+\sum_{a=1}^{A_3/2} n_{3,a} K_{3,a}
=
\sum_{a=1}^{A/2} n_{a} K_{a}
\]
which means that 
among $\set{L,K_a}$ there are only
$A$ independent continuous parameters:
$A-1$ independent fillings $K_a$ 
and one expansion parameter $\lambda/L^2$.
This matches the counting for one-loop gauge theory since the
loop counting parameter $\lambda/L^2$ is absent.
Note that the case $n_0=0$ forces us to view the length as 
fundamental rather than depending on $A$ independent 
fillings. 
To derive the constraint, consider the integral
\[
\frac{\sqrt{\lambda}}{64\pi^3 i} 
\oint_{\infty}dx \lrbrk{1-\frac{1}{x^2}} 
\sum_{k=1}^4 p_k^2(x).
\]
On the one hand it is immediately zero due to $p_k(x)\sim 1/x$ at $x\to \infty$.
On the other hand we can split up the contour of integration around
the singularities and cuts and obtain the constraint
\<
0\eq
-\frac{\sqrt{\lambda}}{4\pi}\bigbrk{p'_1(0)+p'_2(0)}n_0
+\sum_{a=1}^{A}
\frac{\sqrt{\lambda}}{64\pi^3 i} 
\oint_{\contour[A]_a}dx \lrbrk{1-\frac{1}{x^2}}
\sum_{k=1}^4 p_k^2(x)
\nln\eq
n_0 L-n_0\sum_{a=1}^{A_2/2}K_{2,a}
-\half \sum_{a=1}^{A} n_a K_a
=
n_0 L
-\sum_{a=1}^{A/2} n_{a} K_{a}.
\>
We have made use of the identity 
\[
\frac{\sqrt{\lambda}}{64\pi^3 i}
\oint_{\contour[A]_{a}} 
dx \lrbrk{1-\frac{1}{x^2}}
\sum_{k=1}^4 p_k^2(x)
=-\half n_{a} K_{a}
\]
which one gets after pulling the contour $\contour[A]_a$ tightly
around the cut $\contour_a$.
Furthermore, the value of $p'_1(0)+p'_2(0)$ follows from the
residue at $x=\infty$, see the following subsection. 
Finally, the filling and mode number of the inverse cut 
$\contour_{2,a'}=1/\contour_{2,a}$ are given by
$K_{2,a'}=-K_{2,a}$, $n_{2,a'}=2n_0-n_{2,a}$, 
c.f.~\secref{sec:Acycle,sec:Bcycle}.
Similarly, the fillings and mode numbers for
the other types of cuts are invariant under inversion.

\subsection{Global Charges}
\label{sec:Globals}

Now let us compute the global charges at $x=\infty$,
see \eqref{eq:pkAsymp}.
These are obtained as the cycles of $p_k(x)\, dx$ around $x=\infty$
which we can also write as the sum of cycles 
around all singularities on the same sheet
(c.f.~\figref{fig:Sigma4} for the structure of cuts)
\<\label{eq:SigmaDynkin}
r_1
\eq
\frac{\sqrt{\lambda}}{8\pi^2 i}\oint_\infty dx\, \bigbrk{p_1-p_2}
=
\frac{\sqrt{\lambda}}{4\pi}\bigbrk{p'_1-p'_2}(0)
=
\sum_{a=1}^{A_2/2} K_{2,a}
-2\sum_{a=1}^{A_1/2} K_{1,a},
\nln
r_2
\eq
\frac{\sqrt{\lambda}}{8\pi^2 i}\oint_\infty dx\, \bigbrk{p_2-p_3}
=
\frac{\sqrt{\lambda}}{4\pi}\bigbrk{p'_4-p'_1}(0)
=
L
+\sum_{a=1}^{A_1/2} K_{1,a}
-2\sum_{a=1}^{A_2/2} K_{2,a}
+\sum_{a=1}^{A_3/2} K_{3,a}
,
\nln
r_3
\eq
\frac{\sqrt{\lambda}}{8\pi^2 i}\oint_\infty dx\, \bigbrk{p_3-p_4}
=
\frac{\sqrt{\lambda}}{4\pi}\bigbrk{p'_3-p'_4}(0)
=
\sum_{a=1}^{A_2/2} K_{2,a}
-2\sum_{a=1}^{A_3/2} K_{3,a}.
\>
Here we have made use of the symmetry
to write our findings in terms of the
fillings $K_{k,a}$. 
As in gauge theory, the fillings
are directly related to the
Dynkin labels $r_k$ and the length $L$.
Let us also note the particularly useful combination
\[
\half r_1
+r_2
+\half r_3
=
\frac{\sqrt{\lambda}}{8\pi^2 i}\oint_\infty dx\, \bigbrk{p_1+p_2}
=
-\frac{\sqrt{\lambda}}{4\pi}\bigbrk{p'_1(0)+p'_2(0)}
=
L
-\sum_{a=1}^{A_2/2} K_{2,a}.
\]
%

\subsection{Comparison to Gauge Theory}

Here we will show that the algebraic curve of the SYM theory 
in the $\alg{so}(6)$ sector coincides with the algebraic curve of the string 
sigma model on $\Real\times S^5$ at one loop, in accordance with the 
proposal of \cite{Minahan:2004ds}.

Let us compare the analytical data defining the curves 
in the Frolov-Tseytlin limit $\lambda/L^2\to 0$. We will define for 
convenience a rescaled variable $u=(\sqrt{\lambda}/4\pi L)\,x$, 
which makes it similar to the spectral parameter $u$ of  \secref{sec:Spins} 
(in \secref{sec:Bethe.Weak} we refine the relationship for higher loops). 
In this limit, for each pair of mutually symmetric branch points, 
$(u_a,\lambda/16\pi^2 L^2 u_a)$, 
one goes to zero and one remains finite.
This means half of the cuts approach $x=\infty$ and half of them
approach $x=0$. We will use this distinction to select half of the
cuts: The sums $\sum_{a=1}^{A/2}$ introduced in \secref{sec:Fillings}
refer to the long cuts with $x\to\infty$ which remain finite
in the $u$-plane. The other half of the cuts becomes infinitely
short in the limit and needs to be handled separately.
We are thus left with half of the cuts having no symmetry 
with respect to inversion, as in the case of the SYM curve. 

Both curves enjoy the following common properties:
\begin{bulletlist}  
\item
It is easy to see that the equation \eqref{eq:algeqSM} 
becomes \eqref{eq:algeq} in this limit, 
for similar definitions of $y(u)$.

\item
Four sheets for the quasi-momentum (which we call $p(u)$ even 
after rescaling) in the $u$-projection, connected by finite cuts, 
as discussed in \secref{sec:Spins,sec:Alg.EqSM}.

\item The same condition of zero $\contour[A]$-cycles.

\item  The same set of equations \eqref{eq:ThermoEquations} 
and \eqref{eq:pkBE} defining the symmetric part of $p(u)$ 
on the cuts and hence the same $\contour[B]$-periods \eqref{eq:Bcycles}  
of the cuts.

\item
The same asymptotics for $p(u)$ at $u\to\infty$ for all the sheets, 
given through the $\grp{SU}(4)$ charges by the 
formulas \eqref{eq:ThermoInftySheet} and \eqref{eq:pkAsymp},
when we rescale $r_k\to L r_k$ in \eqref{eq:pkAsymp}.  

\end{bulletlist}

To understand the expansion of $p(u)$ at $u=0$ we
need to take the inverse cuts into account which 
approach $u=0$. For this purpose, 
we shall define a contour $\contour$ in the $x$-plane which 
encircles the poles at $x=\pm 1$ and all the short cuts $1/\contour_a$. 
Equivalently, this may be considered a contour which excludes
$x=\infty$ and all the long/finite cuts $\contour_a$.
After rescaling $\contour$ merely encircles the point $u=0$ 
in the $u$-plane which can be used to obtain the expansion 
of $p(u)$ as follows
\[
\frac{\partial^{r-1} p_k}{\partial u^{r-1}}(0) 
=\lrbrk{\frac{4\pi L}{\sqrt{\lambda}}}^{r-1}\frac{1}{2\pi i}
\oint_{\contour} p_k(x)\frac{dx}{x^{r}}\,.
\]
Using the identities and definitions in \secref{sec:Fillings,sec:Globals} 
we find the useful relations
\<
L\eq \frac{\sqrt{\lambda}}{8\pi^2 i}\oint_{\contour}
dx\lrbrk{1-\frac{1}{x^2}}\bigbrk{p_1(x)+p_2(x)},
\nln
D\eq
\frac{\sqrt{\lambda}}{8\pi^2 i}\oint_{\contour}
dx\lrbrk{1+\frac{1}{x^2}}\bigbrk{p_1(x)+p_2(x)},
\nln
2\pi n_0
\eq
\frac{1}{2\pi i}\oint_{\contour}
\frac{dx}{x}
\bigbrk{p_1(x)+p_2(x)}.
\>
This determines the expansion of $p_1(u)+p_2(u)$ as follows
\[
p_1(u)+p_2(u)=
\frac{D+L}{2L}\,\frac{1}{u}
+2\pi n_0
+\frac{8\pi^2 L^2}{\lambda}\,\frac{D-L}{L}\,u
+\order{u^2}.
\]
The residues of the poles at $u=0$ are obtained in a similar manner
\[
p_{1,2}(u)=-p_{3,4}(u)=
\lrbrk{
\sfrac{1}{2}
+\order{\lambda/L^2}
}
\frac{1}{u}
+\ldots
\]
where $\order{\lambda/L^2}$ represents various integrals which are
suppressed by in the one-loop approximation.

The two curves therefore have
\begin{bulletlist}

\item 
the same poles at $u=0$, $p_{1,2}(u)=-p_{3,4}=1/2u+\order{u^0}$, 
c.f.~\eqref{eq:ThermoZeroSheet} in gauge theory.
The extra poles at zero for the sigma model come from 
the poles at $u=\pm \sqrt{\lambda}/4\pi L$ 
when $\lambda/L^2\to 0$. 
The small cuts contribute to the residue only at higher loop orders.

\item
the same expansion $p_1(u)+p_2(u)=1/u+2\pi n_0+uE+\ldots$ at $u=0$.
We make use of
$D=L+\order{\lambda/L}$ to match the residue of $1/u$.
Integrality of $n_0$ corresponds to cyclicity of the trace $U=1$, or
\eqref{eq:LocalThermo,eq:MomThermo} in SYM. 
The anomalous dimension $E=(D-L)/Lg^2$ with $g^2=\lambda/8\pi^2 L^2$
extracted from both curves also coincides.

\end{bulletlist}

These properties define the one-loop algebraic curves 
and their relation to the physical data unambiguously
and consequently they coincide.
\medskip

At two loops the full proof of the equivalence of two curves 
was only for the $\alg{su}(2)$ sector \cite{Kazakov:2004qf}, 
the only one where the two-loop dilatation operator is 
actually calculated \cite{Beisert:2003tq}. 
But we can borrow the idea of \cite{Minahan:2004ds} 
where the closure of the $\alg{so}(6)$ sector was 
demonstrated in higher loops in the classical limit and  
the Bethe equation in the second loop was guessed. 
We can do the comparison of the curves at two loops along 
the same guidelines. 
In the next section it will be done using the Bethe equations.

As we also know \cite{Serban:2004jf}, 
at three loops the curves do not match already 
in the $\alg{su}(2)$ sector,
most probably, due to yet unidentified 
non-perturbative corrections arising on the way 
from the weak to strong coupling.

%
%

\section{Bethe Ansatz for the Sigma-Model on $\Real\times S^{\grn-1}$}
\label{sec:Bethe}

Having constructed the algebraic curve for the classical string
on $\Real\times S^5$ and having convinced ourselves that
we have identified all relevant parameters,
we proceed by constructing an integral
representation of the curve (for all $\Real\times S^{\grn-1}$). 
The obtained equations are similar to the Bethe equations 
for integrable spin chains in the thermodynamic limit 
which in fact form a Riemann-Hilbert problem.
We finally compare the obtained equations to the one derived for
gauge theory and find agreement up to two loops.

\subsection{Simple Roots}
\label{sec:Bethe.Dynkin}

\begin{figure}\centering
\parbox{7cm}{\centering\includegraphics{bks.dynkin.soeven.eps}}
\qquad
\parbox{7cm}{\centering\includegraphics{bks.dynkin.soodd.eps}}
\caption{Dynkin diagram of $\grp{SO}(\grn)$ for even and odd $\grn$.}
\label{fig:dynkin}
\end{figure}

To reveal the group theory structure of the equations, we
will now introduce singular resolvents $\Hrsing_k(x)$
which can be associated to the simple roots of
$\alg{so}(\grn)$.
See \figref{fig:dynkin} for the
Dynkin diagram of the algebra and the labelling of the
simple roots.
They are related to the quasi-momenta $q_k(x)$ by
\[\label{eq:ToDynkin}
\Hrsing_k=\sum_{j=1}^k q_{j}.
\]
with the exceptions for the simple roots associated to spinors
for even $\grn$
\[\label{eq:ToDynkinEven}
\Hrsing_{[\grn/2]-1}=\sum_{j=1}^{[\grn/2]-1}\half  q_{j}-\half q_{[\grn/2]},
\qquad
\Hrsing_{[\grn/2]}=\sum_{j=1}^{[\grn/2]-1}\half  q_{j}+\half q_{[\grn/2]}
\]
and for odd $\grn$
\[\label{eq:ToDynkinOdd}
\Hrsing_{[\grn/2]}=\sum_{j=1}^{[\grn/2]}\half  q_{j}.
\]

Let us now collect the facts about
the analytic properties of $\Hrsing_k(x)$.
First of all we know that the expansion at $x=\infty$
is related to the representation $[s_1,s_2,\ldots]$ of the state.
Using the Cartan matrix $M_{kj}$ of $\grp{SO}(\grn)$
(see \appref{sec:Cartan}) it can be summarized as
\[\label{eq:Hatinfty}
\Hrsing_k(x)=\frac{1}{x}
\sum_{j=1}^{[\grn/2]} M^{-1}_{kj}\frac{4\pi s_j}{\sqrt{\lambda}}
+\order{1/x^{2}}.
\]
Secondly, we have derived the singular behavior at $x=\pm 1$
\footnote{We could also write
$M^{-1}_{k1}$ as
$\sum_j M^{-1}_{kj}V^\indvec_j$, where $V^\indvec_j=(1,0,0,\ldots)$
are the Dynkin labels of the vector representation.
}
\[\label{eq:Hsing}
\Hrsing_k(x)=\frac{1}{x\mp 1}\,\frac{2\pi M^{-1}_{k1}\dimn}{\sqrt{\lambda}}
+\bigorder{1/(x\mp 1)^0}.
\]
Finally, we will use the assumption 
\eqref{eq:qkSym} for the symmetry of $\Hrsing_k(x)$
under the map $x\mapsto 1/x$
\footnote{Again, $\Hrsing_1=\sum_j \Hrsing_j V^\indvec_j$ could be
written in a more `covariant' way.}
\[\label{eq:Hsym}
\Hrsing_k(1/x)=\Hrsing_k(x)-2M^{-1}_{k1}\Hrsing_1(x)+4\pi n_0M^{-1}_{k1}.
\]

\subsection{Integral Representation}
\label{sec:Bethe.Ansatz}

We make properties \eqref{eq:Hsing} and \eqref{eq:Hsym} manifest by defining
\<\label{eq:Hansatz}
\Hrsing_k(x)\eq\Hresolv_k(x)+\Hresolv_k(1/x)
-2M^{-1}_{k1}\Hresolv_1(1/x)
\nl
+\frac{1}{x-1/x}\,\frac{4\pi D}{\sqrt{\lambda}}\,M^{-1}_{k1}
+c_k
-c_1 M^{-1}_{k1}
+2\pi n_0 M^{-1}_{k1}
,
\>
where the $c_k$'s are a set of constants.
The resolvents $\Hresolv_k(x)$ are assumed to be analytic
except at a collection of branch cuts $\contour_k$
and approach zero at $x=\infty$.
Note that this representation of $\Hrsing_k(x)$ is
\emph{ambiguous}. We can add to $\Hresolv_k$
an antisymmetric function
\[\label{eq:Ambiguity}
\Hresolv_k(x)\to
\Hresolv_k(x)+f_k(x)-f_k(1/x)+2M_{k1}^{-1}f_1(1/x),
\]
this has no effect on the physical
function $\Hrsing_k(x)$.

Let us introduce the density $\rho_k(x)$ which
describes the discontinuity across a cut
\[\label{eq:Dens}
\rho_k(x)=
\frac{1-1/x^2}{2\pi i}\bigbrk{\Hresolv_k(x-i\epsilon)-\Hresolv_k(x+i\epsilon)}
\qquad \mbox{for } x\in \contour_k.
\]
The factor of $1-1/x^2$ was introduced for later convenience
and will allow the interpretation of $\rho_k(x)$ as a density.
The apparent pole at $x=0$ is irrelevant as
long as the cuts do not cross this point.
We could also demand positivity of the density, $dx\,\rho_k(x)>0$.
This would fix the position of the cuts $\contour_k$ in
the complex plane, but will not be essential for the
treatment of the classical sigma model.
{}From $\rho_k(x)$ we can reconstruct the function $\Hresolv_k(x)$
\[\label{eq:ResDens}
\Hresolv_k(x)=\int_{\contour_k}\frac{dy\,\rho_k(y)}{1-1/y^2}\,\frac{1}{y-x}\,.
\]
%

\subsection{Asymptotic Behavior}
\label{sec:Bethe.Asymp}

We should now relate the $\grp{SO}(\grn)$ representation of a state
to the cuts and densities.
For that purpose, we note the expansion of the resolvents $\Hresolv_k(x)$
at $x=\infty$
\[\label{eq:ResInfty}
\Hresolv_k(x)=
-\frac{1}{x}\lrbrk{\frac{4\pi\,K_k}{\sqrt{\lambda}}+\Hresolv'_k(0)}
+\order{1/x^2}
\]
and $x=0$
\[\label{eq:ResZero}
\Hresolv_k(x)=
\Hresolv_k(0)
+x\,\Hresolv'_k(0)
+\order{x^2}
\]
where we have defined the normalizations or fillings
of the densities
\[\label{eq:Filling}
K_k=\frac{\sqrt{\lambda}}{4\pi}\int_{\contour_k}dy\,\rho_k(y).
\]
For $\Hrsing_k(x)$ we find the asymptotic behavior at $x\to\infty$
\<\label{eq:AsympRes}
\Hrsing_k(x)\eq
c_k-c_1M^{-1}_{k1}+2\pi n_0M^{-1}_{k1}
+\Hresolv_k(0)
-2M^{-1}_{k1}\Hresolv_1(0)
\nl
+\frac{1}{x}\lrbrk{
\frac{4\pi D}{\sqrt{\lambda}}\,M^{-1}_{k1}
-2H'_1(0)M^{-1}_{k1}
-\frac{4\pi\,K_k}{\sqrt{\lambda}}}
+\order{1/x^2}\,,
\>
We compare this to \eqref{eq:Hatinfty}
and find
the relation between fillings and
Dynkin labels
\[\label{eq:FixFilling}
K_k
=
M^{-1}_{k1}
\lrbrk{
\frac{4\pi D}{\sqrt{\lambda}}\,
-2\Hresolv'_1(0)
}
-\sum_{j=1}^{[\grn/2]}
M^{-1}_{kj}\frac{4\pi s_j}{\sqrt{\lambda}}.
\]
as well as the constants
\[\label{eq:FixConst}
c_k=c_1 M^{-1}_{k1}-2\pi n_0 M^{-1}_{k1}-\Hresolv_k(0)+2M^{-1}_{k1}\Hresolv_1(0).
\]
In fact, this equation for $k=1$ cannot be solved for $c_1$,
it drops out.
This leads to an additional condition
for the resolvents, the momentum constraint
\[\label{eq:MomConstraint}
\Hresolv_1(0)=2\pi n_0.
\]
whereas $c_1$ is not fixed.
When substituting the constants into \eqref{eq:Hansatz}
we obtain
\<\label{eq:HSingFinal}
\Hrsing_k(x)\eq
\Hresolv_k(x)
+\Hresolv_k(1/x)
-\Hresolv_k(0)
\nl
+M^{-1}_{k1}
\lrbrk{
-2\Hresolv_1(1/x)
+2\Hresolv_1(0)
+\frac{1}{x-1/x}\,\frac{4\pi D}{\sqrt{\lambda}}
}
.
\>
Now the expansion of the functions
$\Hrsing_k$ is fixed at the points
$x=\infty$ and $x=0$ and it turns out that the
the ambiguity \eqref{eq:Ambiguity}
must not modify $\Hresolv_k(x)$
at $x=\infty$.%
\footnote{$\Hresolv_k(x)$
was assumed to be zero at $x=\infty$ anyway.}

\subsection{Bethe Equations}
\label{sec:Bethe.Eq}

The singular resolvents $\Hrsing_k$ now satisfy the desired
symmetries and expansions at specific points.
They however have branch cuts along the curves $\contour_k$.
These must not be seen in the transfer matrices, which are
analytic except at the special points.
This leads us to the Bethe equations,
which are manifestations of the 
analyticity conditions for the monodromy matrix, 
see \secref{sec:Sigma.Anal}
\[\label{eq:SigmaBethe}
\sum_{j=1}^{[\grn/2]}
M_{kj}\Hrsingsl_j(x)
= 2\pi n_{k,a},
\qquad \mbox{for }x\in \contour_{k,a}.
\]
As explained in \secref{sec:Spins.Props,sec:Sigma.Anal},
they ensure that across a cut only
the labelling of sheets and the
branch of the logarithm changes.
The slash through a resolvent implies a
principal part prescription,
\[\label{eq:SigmaPrincipal}
\Hresolvsl_k(x)=\half \Hresolv_k(x+i\epsilon)
+\half \Hresolv_k(x-i\epsilon).
\]
Here we have split up the curves $\contour_{k}$ into
their connected components
$\contour_{k,a}$.
For each connected curve we have introduced a mode number
$n_{k,a}$ due to the allowed shift by multiples of $2\pi i$
in the exponent.
When we substitute \eqref{eq:HSingFinal} the Bethe equations read
\<\label{eq:SigmaBetheRes}
2\pi n_{k,a}\eq
\sum_{j=1}^{[\grn/2]}
M_{kj}\bigbrk{\Hresolvsl_j(x)+\Hresolv_j(1/x)-\Hresolv_j(0)}
\nl
\quad
+\delta_{k1}\lrbrk{
-2\Hresolv_1(1/x)
+2\Hresolv_1(0)
+\frac{1}{x-1/x}\,\frac{4\pi D}{\sqrt{\lambda}}
},
\quad \mbox{for }x\in \contour_{k,a}.
\>
Note the explicit appearance of the dimension/energy $D$
which constitutes the physical quantity of main interest.
For a given set of mode numbers $n_{k,a}$ 
and fillings
\[
K_{k,a}=\frac{\sqrt{\lambda}}{4\pi}\int_{\contour_{k,a}}dy\,\rho_k(y),
\]
the equations \eqref{eq:SigmaBetheRes} should
only have a solution if $D$ has the appropriate value.
The Bethe equations of the spin chain in
\secref{sec:Spins.Props} are qualitatively different:
They should always be soluble and the
dimension is subsequently read off from
\eqref{eq:LocalThermo,eq:EngThermo}.

It is useful to go to the
$u$-plane which is related to the $x$-plane by
\cite{Beisert:2004hm}
\[\label{eq:xandu}
x(u)=\half u+\half \sqrt{u^2-4}\,,\qquad
u(x)=x+1/x.
\]
We can introduce a resolvent in the $u$-plane by
\[\label{eq:uRes}
\bar\Hresolv_k(u)
=
\int \frac{dy\,\rho_k(y)}{y+1/y-u}
=
\int \frac{dv\,\rho_k(v)}{v-u}\,.
\]
Note that $\rho_k(x)$ transforms as a density,
i.e.~$dx\,\rho_k(x)=du\,\rho_k(u)$.
It is related to
a symmetric combination of the resolvents
in the $x$-space
\[\label{eq:uResx}
\Hresolv_k(x)
+\Hresolv_k(1/x)
=
\bar\Hresolv_k(x+1/x)
+\Hresolv_k(0).
\]
This allows us to write the Bethe equations in
the $u$-plane
\[\label{eq:SigmaBetheu}
\sum_{j=1}^{[\grn/2]}
M_{kj}\bar\Hresolvsl_j(u)
+\delta_{k1}F\indup{string}(u)
=2\pi n_{k,a}
\qquad \mbox{for }u\in \bar\contour_{k,a}
\]
with
\[\label{eq:SigmaBethePot}
F\indup{string}(u)=
 \frac{1}{\sqrt{u^2-4}}\,\frac{4\pi D}{\sqrt{\lambda}}
+2\Hresolv_1(0)
-2\Hresolv_1\bigbrk{1/x(u)}.
\]
It might be favorable to replace the dimension $D$,
which is intended to be the final result of the computation,
by some known quantities.
We can rewrite the 
the definition of the length 
\eqref{eq:Ldef}
as an energy formula
\[\label{eq:EneFormula2}
\dimn
=L+\frac{\sqrt{\lambda}}{2\pi}\int_{\contour_1}\frac{dy\,\rho_1(y)}{1-1/y^2}\,\frac{1}{y^2}
=L+\frac{\sqrt{\lambda}}{2\pi}\,\Hresolv'_1(0)\,.
\]
When we substitute this in
\eqref{eq:SigmaBethePot} we obtain
\<
F\indup{string}(u)\eq
\frac{1}{\sqrt{u^2-4}}
\,\frac{4\pi L}{\sqrt{\lambda}}
+2\Hresolv_1(0)
+\frac{2\Hresolv_1'(0)}{\sqrt{u^2-4}}
-2\Hresolv_1\bigbrk{1/x(u)}.
\>
%

\subsection{The Sigma-Model on $\Real\times S^5$}
\label{eq:RS5}

Let us now apply our results to the case $\grn=6$, 
i.e.~the sigma model on $\Real\times S^5$. Here we shall adopt
a $\grp{SU}(4)$ notation instead of the one for
$\grp{SO}(6)$. 
The benefit of $\grp{SU}(4)$ is that
it is manifestly a subgroup of $\grp{SU}(2,2|4)$,
the full supergroup of the superstring on 
$AdS_5\times S^5$.
The change merely amounts to swapping the
labels of the first two simple roots,
c.f.~\figref{fig:so6su4}.
We introduce the singular resolvents
$\rsing_k(x)$ corresponding to the simple roots of $\grp{SU}(4)$
by
\<
\rsing_1(x)\eq \Hrsing_2(x),
\nln
\rsing_2(x)\eq \Hrsing_1(x),
\nln
\rsing_3(x)\eq \Hrsing_3(x).
\>
We also interchange labels $1,2$ for 
the densities $\rho_k$ and fillings $K_k$.

The Bethe equations are written as
\<\label{eq:pkBE}
2\rsingsl_1(x)-\rsing_2(x)=
\sheetsl_1(x)-\sheetsl_2(x)\eq 2\pi n_{1,a},
\qquad x\in \contour_{1,a},
\nln
2\rsingsl_2(x)-\rsing_1(x)-\rsing_3(x)
=\sheetsl_2(x)-\sheetsl_3(x)\eq 2\pi n_{2,a},
\qquad x\in \contour_{2,a},
\nln
2\rsingsl_3(x)-\rsing_2(x)
=\sheetsl_3(x)-\sheetsl_4(x)\eq 2\pi n_{3,a},
\qquad x\in \contour_{3,a}.
\>
Now one can draw the Riemann sheets picture
as in Section \ref{sec:Spins.Props}.
The Riemann surface consists of four sheets
each of which corresponds to $p_k$
while resolvents $\rsing_k$ describe
how to connect the sheets with cuts.
The main difference is that
due to the inversion symmetry
all the cuts appear in pairs
as depicted in \figref{fig:Sigma4}.

\subsection{Comparison to Gauge Theory}
\label{sec:Bethe.Weak}

Let us now consider the limit where the 
Dynkin labels $r_k$ and the dimension $D$ are large
with respect to $\sqrt{\lambda}$. For this purpose
we rescale according to 
\[
\set{x,u}\to \frac{4\pi L}{\sqrt{\lambda}}\,\set{x,u},
\qquad
\set{D,r_k,K_k}\to L\set{D,r_k,K_k}
\]
while keeping $\rho_k(x),\resolv_k(x)$ fixed.
Here $L$ is defined to be the limiting value of $D$ (before rescaling) 
at $\lambda=0$ corresponding to the classical dimension in gauge theory.%
\footnote{The `length' $L$ 
was conjectured to be an
action variable in \cite{Mikhailov:2004au}.
If true, it would be interesting to relate it to our definition.
See also \cite{Kruczenski:2004cn} on the definition of `length'
in the coherent approach.}
For convenience, we define the effective coupling constant $g$ as
\[
g^2=\frac{\lambda}{8\pi^2L^2}=\frac{\gym^2 N}{8\pi^2L^2}\,.
\]

The Bethe equations \eqref{eq:SigmaBetheu} are left invariant,
but the function $F\indup{string}(u)$ changes to%
\footnote{This equation along with 
the generic form of the Bethe equations
\eqref{eq:SigmaBetheu}
was proposed independently by M.~Staudacher
\cite{Staudacher:2004qq}. 
He also showed that the solutions
discussed in \secref{sec:Spins.Exam} for this 
deformation of the equations yield precisely 
the energies computed from the string equations of motion 
\cite{Minahan:2002rc,Frolov:2003qc,Minahan:2004ds,Kruczenski:2004cn}.
We would like to thank him for insightful discussions.}
\[
F\indup{string}(u)=
 \frac{1}{\sqrt{u^2-2g^2}}
+\lrbrk{2\resolv_2(0)
+\frac{g^2\resolv_2'(0)}{\sqrt{u^2-2g^2}}
-2\resolv_2\bigbrk{g^2/2x(u)}}
\]
where now $x(u)=\frac{1}{2}u+\frac{1}{2}\sqrt{u^2-2g^2}$.
When we expand in $g$ we obtain 
\[
F\indup{string}(u)=
F\indup{gauge}(u)
+\order{g^4},
\]
This means that the functions
$F\indup{string}(u)$ and 
$F\indup{gauge}(u)$ agree up to and including order $g^2$ 
corresponding to two loops for the scaling dimension $D$.
We have thus demonstrated the generic two-loop matching 
of scaling dimensions in gauge theory%
\footnote{
By higher-loop gauge theory we mean 
the higher-loop Bethe ansatz for the $\alg{so}(6)$ sector
in the thermodynamic limit \cite{Minahan:2004ds}. 
It has not yet been shown that this ansatz indeed matches
gauge theory at two or higher loops.}
and energies of spinning strings 
in the $\grp{SO}(6)$ sector.%
\footnote{In both theories we have focussed on the low-lying modes of the 
spectrum. States which have no expansion in the effective coupling $g$ 
are disregarded.}
This complies with the one-loop results of \cite{Kruczenski:2004cn}
in the coherent state approach to spinning strings
\cite{Kruczenski:2003gt,Kruczenski:2004kw,Hernandez:2004uw,Stefanskijr.:2004cw,Hernandez:2004kr}
and also with the matching of integrable charges in special cases
\cite{Engquist:2004bx}.


\section{Discussion}
\label{sec:Concl}

In this work we continued the investigation of 
integrability of the multi-color $\superN=4$ SYM theory 
and its close relation (and hopefully equivalence) to the $AdS_5\times S^5$
string sigma model. The general solutions of the one-loop SYM theory
and of the classical sigma model, constructed here in the
$\Real\times S^5$ or $\alg{so}(6)$ sector, give, as expected,
the same result in the
weakly coupled region of the classical, BMN limit. Elsewhere, the
algebraic curves of the two models appear to have a very similar
structure and differ only in the details. We hope that a  quantized
version of the sigma model will reproduce the known SYM perturbative
data precisely and give in addition the complete non-perturbative
information on the gauge theory, compatible with these perturbative
data.

We see the most natural way to prove  this complete AdS/CFT duality
in construction of the full algebraic curve of the model, with the
following quantization based on this curve. Often the quantization
means an appropriate discretization of  the model and the curve
gives a hint on the right procedure. For example, the matrix model
with a finite size $N$ of a matrix,  is often completely defined by
its large-$N$ algebraic curve and can be considered as its quantum
counterpart. The other example is the discrete Bethe ansatz
equations, as those considered here, which provide the right
quantization of the classical algebraic curve. This procedure of
quantization is carried out in one-loop here. There were recently
some interesting attempts to find the quantum version of the
$AdS_5\times S^5$ sigma model \cite{Arutyunov:2004vx}, though it is
too early to claim that we are close to the whole resolution of this
formidable problem.

As a next important step in this program we would consider the
generalization of the present construction, to the algebraic curve
of the one-loop SYM theory for the full dilatation Hamiltonian of
\cite{Beisert:2003jj}, on the one hand, and of the
full $AdS_5\times S^5$ classical sigma model, on the other hand. The
present paper, together with \cite{Kazakov:2004nh,Kazakov:2004qf},
provides most of the necessary technique for the completion of this
task.

\subsection*{Acknowledgements}

We would like to thank Gleb Arutyunov, Andrei Mikhailov, 
Arkady Tseytlin, Kostya Zarembo 
and especially Matthias Staudacher for helpful discussions and remarks.

N.~B.~would like to thank the Ecole Normale Sup\'erieure
and the Kavli Institute for Theoretical Physics 
for kind hospitality during parts of this project.
The work of N.~B.~is supported in part by the U.S.~National Science
Foundation Grants No.~PHY99-07949 and PHY02-43680. Any opinions,
findings and conclusions or recommendations expressed in this
material are those of the authors and do not necessarily reflect the
views of the National Science Foundation. V.~K.~would  like to thank
the Princeton Institute for Advanced Study for the kind hospitality
during a part of this work. The work of V.~K.~was partially supported
by European Union under the RTN contracts HPRN-CT-2000-00122 and
00131 and by NATO grant PST.CLG.978817. The work of K.~S.~is
supported by the Nishina Memorial Foundation.

\appendix

\section{Vector Spin Chains of $\alg{so}(\grn)$}
\label{sec:SpinSOm}

In this appendix we present the generalizations of several
expression of \secref{sec:Spins} to the case
of $\alg{su}(\grn)$. The R-matrix of two
vectors ($\rep{V}$) \cite{Reshetikhin:1983vw,Reshetikhin:1985vd}
or of a vector and a spinor ($\rep{S}$)
are given by
\<\label{eq:SOmRvec}
\Rmatrix^{\indvec,\indvec}(u)\eq
\opproj^{\rep{T}}
+\frac{u-i}{u+i}\,\opproj^{\rep{A}}
+\frac{(u-i)(u-\frac{i}{2}\grn+i)}{(u+i)(u+\frac{i}{2}\grn-i)}\,\opproj^{\rep{1}}
\nln\eq
\frac{i}{u+i}\,\opperm
+\frac{u}{u+i}\,\opident
-\frac{iu}{(u+i)(u+\frac{i}{2}\grn-i)}\,\optrace^{\indvec,\indvec}
\>
and
\[\label{eq:SOmRspin}
\Rmatrix^{\indspin,\indvec}(u)=
\opproj^{\rep{\indvec\indspin}}
+\frac{u-\frac{i}{4}\grn}{u+\frac{i}{4}\grn}\,\opproj^{\indspin}
=
\opident
+\frac{u-\frac{i}{4}\grn}{\frac{i}{2}}\,\optrace^{\indspin,\indvec}.
\]
The operators
$\opproj^{\rep{T}},\opproj^{\rep{A}},\opproj^{\rep{1}}$
project to the symmetric-traceless ($\rep{T}$),
antisymmetric/adjoint ($\rep{A}$) and singlet ($\rep{1}$)
modules which appear in the tensor product of two vectors,
while $\opproj^{\indvec\indspin},\opproj^{\indspin}$
project to the traceless vector-spinor ($\rep{VS}$) and the spinor
($\rep{S}$) in the product of a vector and a spinor.
These can be written using the operators
$\opident,\opperm,\optrace^{\indvec,\indvec},\optrace^{\indvec,\indspin}$
which are the identity, permutation and trace operators, respectively
\[\label{eq:SOmProjVec}
\opproj^{\rep{T}}=\half\opident+\half\opperm-\sfrac{1}{\grn}\,\optrace^{\indvec,\indvec},\qquad
\opproj^{\rep{A}}=\half\opident-\half\opperm,\qquad
\opproj^{\rep{1}}=\sfrac{1}{\grn}\,\optrace^{\indvec,\indvec}
\]
and
\[\label{eq:SOmProjSpin}
\opproj^{\indvec\indspin}=\opident-\sfrac{1}{\grn}\,\optrace^{\indspin,\indvec},\qquad
\opproj^{\indspin}=\sfrac{1}{\grn}\,\optrace^{\indspin,\indvec}
\]
with
$(\optrace^{\indvec,\indvec})^{ij}{}_{kl}=\delta^{ij}\delta_{kl}$
and
$(\optrace^{\indspin,\indvec})^{\beta j}{}_{\alpha i}=(\gamma^j\gamma_i)^\beta{}_\alpha$.

For the monodromy and transfer matrices it is convenient to define
\<\label{eq:SOmMono}
\opmono^{\indvec}_{a}(u)\eq
\frac{(u-\sfrac{i}{4}\grn+\sfrac{3i}{2})^L(u+\sfrac{i}{4}\grn-\sfrac{i}{2})^L}{u^{2L}}\,
\Rmatrix_{a1}^{\indvec,\indvec}(u-\sfrac{i}{4}\grn+\sfrac{i}{2})
\cdots
\Rmatrix_{aL}^{\indvec,\indvec}(u-\sfrac{i}{4}\grn+\sfrac{i}{2}),
\nln
\opmono^{\indspin}_{a}(u)\eq
\frac{(u+\sfrac{i}{2})^L}{u^L}\,
\Rmatrix^{\indspin,\indvec}_{a1}(u-\sfrac{i}{4}\grn+\sfrac{i}{2})
\Rmatrix^{\indspin,\indvec}_{a2}(u-\sfrac{i}{4}\grn+\sfrac{i}{2})\cdots
\Rmatrix^{\indspin,\indvec}_{aL}(u-\sfrac{i}{4}\grn+\sfrac{i}{2}).
\>
Then the transfer matrices for the Bethe ansatz have a rather symmetric form.
\section{R-Matrices}
\label{sec:Rmatrices}

In this appendix we present the R-matrices
between two spinors of $\alg{so}(6)$.
Together with the R-matrices in
\secref{sec:Spins.Ops} this is a complete
set for the representations $\rep{6},\rep{4},\rep{\bar 4}$.
For instance one can now explicitly prove the
YBE \eqref{eq:YBE} for all combinations of these
representations.
The R-matrices of two fundamentals are
\<\label{eq:R44}
\Rmatrix^{\rep{4},\rep{4}}(u)\eq
\opproj^{\rep{10}}
+\frac{u-\frac{i}{2}}{u+\frac{i}{2}}\,\opproj^{\rep{6}}
=
\frac{\frac{i}{2}}{u+\frac{i}{2}}\,\opperm
+\frac{u}{u+\frac{i}{2}}\,\opident
=
\frac{u+\frac{i}{4}}{u+\frac{i}{2}}\,\opident
+\frac{i}{u+\frac{i}{2}}\,\oprot^{\rep{4},\rep{4}},
\nln
\Rmatrix^{\rep{\bar 4},\rep{\bar 4}}(u)\eq
\opproj^{\rep{\overline{10}}}
+\frac{u-\frac{i}{2}}{u+\frac{i}{2}}\,\opproj^{\rep{6}}
=
\frac{\frac{i}{2}}{u+\frac{i}{2}}\,\opperm
+\frac{u}{u+\frac{i}{2}}\,\opident
=
\frac{u+\frac{i}{4}}{u+\frac{i}{2}}\,\opident
+\frac{i}{u+\frac{i}{2}}\,\oprot^{\rep{\bar 4},\rep{\bar 4}},
\nln
\Rmatrix^{\rep{4},\rep{\bar 4}}(u)\eq
\opproj^{\rep{15}}
+\frac{u-2i}{u+2i}\,\opproj^{\rep{1}}
=
\opident
-\frac{i}{u+2i}\,\optrace^{\rep{4},\rep{\bar 4}}
=
\frac{u+\frac{7i}{4}}{u+2i}\opident
+\frac{i}{u+2i}  \oprot^{\rep{4},\rep{\bar 4}}.
\>
The projectors for $\rep{4}\times\rep{4}=\rep{10}+\rep{6}$ are given by
\[\label{eq:Proj44}
\opproj^{\rep{10}}
=\half \opident+\half \opperm
,\qquad
\opproj^{\rep{6}}
=\half \opident-\half \opperm
\]
while for $\rep{\bar 4}\times\rep{\bar 4}=\rep{\overline {10}}+\rep{6}$
one finds essentially the same expressions
\[\label{eq:Proj4b4b}
\opproj^{\rep{\bar{10}}}
=\half \opident+\half \opperm
,\qquad
\opproj^{\rep{6}}
=\half \opident-\half \opperm.
\]
For the mixed product $\rep{4}\times\rep{\bar 4}=\rep{15}+\rep{1}$
we get
\[\label{eq:Proj44b}
\opproj^{\rep{15}}=
\opident-\sfrac{1}{4}\optrace^{\rep{4},\rep{\bar 4}}
\opproj^{\rep{1}}=
\sfrac{1}{4}\optrace^{\rep{4},\rep{\bar 4}}
\]
with $(\optrace^{\rep{4},\rep{\bar 4}})^{\beta\dot\alpha}{}_{\delta\dot\gamma}=
\delta^{\beta\dot\alpha}\delta_{\delta\dot\gamma}$.
The rotation generators can be written as follows
\[\label{eq:Rot44}
\oprot^{\rep{4},\rep{4}}=
\opperm-\sfrac{1}{4}\opident,
\qquad
\oprot^{\rep{\bar 4},\rep{\bar 4}}=
\opperm-\sfrac{1}{4} \opident,
\qquad
\oprot^{\rep{4},\rep{\bar 4}}=
\sfrac{1}{4}\opident-\optrace^{\rep{4},\rep{\bar 4}}.
\]

\section{Antisymmetric Transfer Matrices of $\alg{su}(4)$}
\label{sec:Antisym}

There is a nice formula to obtain expressions for the
transfer matrices for the Bethe ansatz in all totally antisymmetric products
of the fundamental representation of $\alg{su}(m)$, see \cite{Krichever:1997qd}.
In this appendix we shall present it for our case of interest, $\alg{su}(4)$.
It allows us to obtain the transfer matrices in
$\rep{4},\rep{6},\rep{\bar 4}$.
It is based on a differential operator $\detshift^{\rep{4}}_u$
\<
\label{eq:DetShiftOp}
\detshift^{\rep{4}}_u\eq
\lrbrk{
\exp(i\partial_u)-
\frac{\bits_1(u)}{\bits_1(u+i)}\,
\frac{\bitp(u+2i)}{\bitp(u+\frac{3i}{2})}\,
}
\nl
\cdot
\lrbrk{
\exp(i\partial_u)-
\frac{\bits_2(u-\frac{i}{2})}{\bits_2(u+\frac{i}{2})}\,
\frac{\bits_1(u+i)}{\bits_1(u)}\,
\frac{\bitp(u+i)}{\bitp(u+\frac{i}{2})}\,
}
\nl
\cdot
\lrbrk{
\exp(i\partial_u)-
\frac{\bits_3(u-i)}{\bits_3(u)}\,
\frac{\bits_2(u+\frac{i}{2})}{\bits_2(u-\frac{i}{2})}\,
\frac{\bitp(u-i)}{\bitp(u-\frac{i}{2})}\,
}
\nl
\cdot
\lrbrk{
\exp(i\partial_u)-
\frac{\bits_3(u)}{\bits_3(u-i)}\,
\frac{\bitp(u-2i)}{\bitp(u-\frac{3i}{2})}\,
}.
\>
The operator is slightly
modified from \cite{Krichever:1997qd} to
accommodate for a non-fundamental spin representation
and a different normalization of rapidities.
When this operator is expanded in powers of
$\exp(i\partial_u)$, which shifts $u$ by $i$, it should yield
\<
\label{eq:DetShiftExpand}
\detshift^{\rep{4}}_u\eq
\frac{V(u+2i)V(u+i)V(u-i)V(u-2i)}
     {V(u+\frac{3i}{2})V(u+\frac{i}{2})
      V(u-\frac{i}{2})V(u-\frac{3i}{2})}
\nl
-\exp(\sfrac{i}{2}\partial_u)\,
\frac{\bitp(u+\sfrac{3i}{2})}{\bitp(u+i)}\,
\frac{\bitp(u-\sfrac{3i}{2})}{\bitp(u-i)}\,
\transfer_{\rep{\bar 4}}(u)\,\exp(\sfrac{i}{2}\partial_u)
\nl
+
\exp(i\partial_u)\,
\frac{\bitp(u)}{\bitp(u+\frac{i}{2})}\,
\frac{\bitp(u)}{\bitp(u-\frac{i}{2})}\,
\transfer_{\rep{6}}(u)\,\exp(i\partial_u)
\nl
-\exp(\sfrac{3i}{2}\partial_u)\,
\transfer_{\rep{4}}(u)\,\exp(\sfrac{3i}{2}\partial_u)
\nl
+\exp(4i\partial_u).
\>
Here we can read off the expressions for
$\optrans_{\indrep}(u)$, they agree with
\eqref{eq:BetheTransfer6,eq:BetheTransfer4,eq:BetheTransfer4bar}.
Alternatively one can use the conjugate operator
\<
\label{eq:DetShiftOpBar}
\detshift^{\rep{\bar 4}}_u\eq
\lrbrk{
\exp(i\partial_u)-
\frac{\bits_3(u)}{\bits_3(u+i)}\,
\frac{\bitp(u+2i)}{\bitp(u+\frac{3i}{2})}\,
}
\nl
\cdot
\lrbrk{
\exp(i\partial_u)-
\frac{\bits_2(u-\frac{i}{2})}{\bits_2(u+\frac{i}{2})}\,
\frac{\bits_3(u+i)}{\bits_3(u)}\,
\frac{\bitp(u+i)}{\bitp(u+\frac{i}{2})}\,
}
\nl
\cdot
\lrbrk{
\exp(i\partial_u)-
\frac{\bits_1(u-i)}{\bits_1(u)}\,
\frac{\bits_2(u+\frac{i}{2})}{\bits_2(u-\frac{i}{2})}\,
\frac{\bitp(u-i)}{\bitp(u-\frac{i}{2})}\,
}
\nl
\cdot
\lrbrk{
\exp(i\partial_u)-
\frac{\bits_1(u)}{\bits_1(u-i)}\,
\frac{\bitp(u-2i)}{\bitp(u-\frac{3i}{2})}\,
}
\>
which expands as follows
\<
\label{eq:DetShiftBarExpand}
\detshift^{\rep{\bar 4}}_u\eq
\frac{V(u+2i)V(u+i)V(u-i)V(u-2i)}
     {V(u+\frac{3i}{2})V(u+\frac{i}{2})
      V(u-\frac{i}{2})V(u-\frac{3i}{2})}
\nl
-\exp(\sfrac{i}{2}\partial_u)\,
\frac{\bitp(u+\sfrac{3i}{2})}{\bitp(u+i)}\,
\frac{\bitp(u-\sfrac{3i}{2})}{\bitp(u-i)}\,
\transfer_{\rep{4}}(u)\,\exp(\sfrac{i}{2}\partial_u)
\nl
+
\exp(i\partial_u)\,
\frac{\bitp(u)}{\bitp(u+\frac{i}{2})}\,
\frac{\bitp(u)}{\bitp(u-\frac{i}{2})}\,
\transfer_{\rep{6}}(u)\,\exp(i\partial_u)
\nl
-\exp(\sfrac{3i}{2}\partial_u)\,
\transfer_{\rep{\bar 4}}(u)\,\exp(\sfrac{3i}{2}\partial_u)
+\exp(4i\partial_u)
\>
In the thermodynamic limit, see \secref{sec:Spins.Thermo},
we find the following limits for the operators
\<
\label{eq:DetShiftThermo}
\detshift^{\rep{4}}_u\earel{\to}
\bigbrk{\exp(i\partial_u/L)-\exp(ip_1)}
\nl
\cdot
\bigbrk{\exp(i\partial_u/L)-\exp(ip_2)}
\nl
\cdot
\bigbrk{\exp(i\partial_u/L)-\exp(ip_3)}
\nl
\cdot
\bigbrk{\exp(i\partial_u/L)-\exp(ip_4)}
\>
and the conjugate one
\<
\label{eq:DetShiftBarThermo}
\detshift^{\rep{\bar 4}}_u\earel{\to}
\bigbrk{\exp(i\partial_u/L)-\exp(-ip_4)}
\nl
\cdot
\bigbrk{\exp(i\partial_u/L)-\exp(-ip_3)}
\nl
\cdot
\bigbrk{\exp(i\partial_u/L)-\exp(-ip_2)}
\nl
\cdot
\bigbrk{\exp(i\partial_u/L)-\exp(-ip_1)}.
\>

\section{Higher Charges of the Sigma Model}
\label{sec:Charges}

Here we shall continue the expansion of $q_k$ at the
singular points $x=\pm 1$ to higher orders.
Note, first of all, that the straight
perturbative diagonalization
fails if there are degenerate eigenvalues.
In the case at hand the eigenvalue zero
is indeed degenerate.
However, the zero subspace can be decoupled completely
from the non-zero eigenvectors in perturbation theory.
This is not a problem, because the zero subspace
does not display singular behavior by definition;
there the monodromy matrix behaves as for any other
point $x$ and we cannot expect to be able to find a simple
expression for $q_k(x)$, $k\neq 1$ at $x=\pm 1$.
For $q_1(x)$ the situation is different, it starts with a
pole whose residue is non-degenerate. We would now like to find
a solution of the equation
\[
\oplax(x,\sigma)\vec{V}(x,\sigma)=i f(x,\sigma) \vec{V}(x,\sigma)
\]
such that the eigenvalue $f(x,\sigma)$ is singular at $x=+1$.
As explained around \eqref{eq:LaxDiag}, the leading-order
eigenvector $\vec{V}(x,\sigma)$ is an eigenvector of $j_+(\sigma)$. 
It therefore must be a linear combination 
of $\Xvec$ and $\Xvec_+:=(\sqrt{\lambda}/D)\partial_+\Xvec$,
we find $\Xvec-\sfrac{i}{2}\Xvec_+$
with eigenvalue $f(x,\sigma)=D/\sqrt{\lambda}(x-1)+\cdots$\,.
When we substitute this in the above equation
for $\oplax$ we can solve for the subleading terms.
In the first few orders we find the eigenstate
\<\label{eq:SingularVector}
\vec{V}(x)\eq
\bigbrk{\Xvec-i\Xvec_+}
+\bigbrk{-\sfrac{i}{2} \Xvec_+ - \sfrac{1}{2}\Xvec_{++}}(x-1)
\nl
+\bigbrk{
+ \sfrac{i}{8}\Xvec_+
+\sfrac{i}{4}\Xvec_{3+}
-\sfrac{1}{8}(\Xvec\cdott\Xvec_{4+})\Xvec
+\sfrac{i}{4}(\Xvec\cdott\Xvec_{4+})\Xvec_+
}(x-1)^2
\nl
+
\bigbrk{
-\sfrac{i}{16} \Xvec_+
-\sfrac{i}{8}\Xvec_{3+}
+ \sfrac{1}{8}\Xvec_{4+}
+ \sfrac{3}{16}(\Xvec\cdott\Xvec_{4+})\Xvec_{++}
+ \sfrac{1}{20}(\Xvec\cdott\Xvec_{5+})\Xvec_+
}(x-1)^3
\nl
+\bigorder{(x-1)^4}
\>
where $\Xvec_{n+}$ is defined as
$\Xvec_{n+}:=(\sqrt{\lambda}/D)^n\partial_+^n \Xvec$.
The value $q_1(x)$ is now given as the integral of $f(x,\sigma)$ 
over the closed string
\<\label{eq:SingularValue}
q_1(x)\eq\int_0^{2\pi} d\sigma \,f(x,\sigma)
\nln
\eq\frac{D}{\sqrt{\lambda}}\int_0^{2\pi}d\sigma\bigg[
\frac{1}{x-1}+\half
+(x-1)
\bigbrk{-\sfrac{1}{8}+\sfrac{1}{4} \Xvec\cdott\Xvec_{+-}+ \sfrac{1}{8}\Xvec\cdott\Xvec_{4+}}
\nlnum\nn
\qquad \qquad \qquad \quad
+(x-1)^2
\bigbrk{\sfrac{1}{16}-\sfrac{1}{8} \Xvec\cdott\Xvec_{+-}- \sfrac{1}{16}\Xvec\cdott\Xvec_{4+}}
+\bigorder{(x-1)^3}
\bigg].
\>
We make use of the following two identities
\[\label{eq:SingularIdentity}
\Tr j_{\indvec,+}j_{\indvec,-}=\frac{8D^2}{\lambda}\,\Xvec\cdott\Xvec_{+-},
\qquad
\Tr \partial_\pm j_{\indvec,\pm} \,\partial_\pm j_{\indvec,\pm}
=\frac{8D^4}{\lambda^2}\bigbrk{1-\Xvec\cdott\Xvec_{4\pm}}
\]
to express the expansion of $q_1(x)$ in terms of the currents
$j_\indvec$
\[\label{eq:SingularValueQ}
q_1(x)=
\frac{2\pi D}{\sqrt{\lambda}\,(x\mp 1)}\pm
\frac{\pi D}{\sqrt{\lambda}}
+(x\mp 1)\frac{\lambda^{3/2}\,Q_\pm}{64D^3}
\mp (x\mp 1)^2\frac{\lambda^{3/2}\,Q_\pm}{128D^3}
+\bigorder{(x\mp 1)^3}
.
\]
with
\[\label{eq:SingularCharge}
Q_\pm=
\int_0^{2\pi} d\sigma\,
\lrbrk{
\frac{2D^2}{\lambda}\Tr j_{\indvec,+}j_{\indvec,-}
-\Tr \partial_\pm j_{\indvec,\pm}\,\partial_\pm j_{\indvec,\pm}
}.
\]
Here we have also included the expansion around $x=-1$.
The charges $Q_\pm$ are the first two non-trivial
local commuting charges of the sigma model.

\section{Cartan Matrices}
\label{sec:Cartan}

The Cartan metric for $\alg{so}(\grn)$ is given by
\[\label{eq:CartanSOeven}
M_{jk}=\matr{cccccc}{
+2&-1&  \\
-1&\ddots&\ddots  \\
  &\ddots&\ddots &-1 \\
  &  & -1 & +2 & -1 & -1 \\
  &  &    & -1 & +2 &    \\
  &  &    & -1 &    & +2
},\qquad\mbox{for $\grn$ even}
\]
and by
\[\label{eq:CartanSOodd}
M_{jk}=\matr{ccccc}{
+2&-1&  \\
-1&\ddots&\ddots  \\
  &\ddots&\ddots &-1 \\
  &  & -1 & +2 & -2    \\
  &  &    & -2 & +4    \\
},\qquad\mbox{for $\grn$ odd}.
\]
The inverse metric is given by
\[\label{eq:ICartanSOeven}
M^{-1}_{jk}=\matr{ccccccc}{
1&1&1&\cdots& 1 & \half & \half \\
1&2&2&\cdots& 2 & 1 & 1 \\
1&2&3&\cdots& 3 & \sfrac{3}{2} & \sfrac{3}{2} \\
\vdots & \vdots & \vdots & \ddots & \vdots & \vdots & \vdots  \\
1 & 2 & 3& \cdots & \sfrac{1}{2}(\grn-4) & \sfrac{1}{4}(\grn-4) & \sfrac{1}{4}(\grn-4) \\
\half & 1&\sfrac{3}{2}& \cdots   & \sfrac{1}{4}(\grn-4) & \sfrac{1}{8}(\grn-0) & \sfrac{1}{8}(\grn-4)  \\
\half & 1&\sfrac{3}{2}& \cdots   & \sfrac{1}{4}(\grn-4) &  \sfrac{1}{8}(\grn-4) & \sfrac{1}{8}(\grn-0)
},\qquad\mbox{for $\grn$ even}
\]
and by
\[\label{eq:ICartanSOodd}
M^{-1}_{jk}=\matr{cccccc}{
1&1&1&\cdots& 1 & \half \\
1&2&2&\cdots& 2  & 1 \\
1&2&3&\cdots& 3 & \sfrac{3}{2} \\
\vdots & \vdots& \vdots & \ddots & \vdots & \vdots  \\
1 & 2 & 3& \cdots & \sfrac{1}{2}(\grn-3) & \sfrac{1}{4}(\grn-3) \\
\half & 1&\sfrac{3}{2}& \cdots   & \sfrac{1}{4}(\grn-3) & \sfrac{1}{8}(\grn-1)
},\qquad\mbox{for $\grn$ odd}.
\]
For $\alg{so}(6)$ this reduces to
\[\label{eq:CartanSO6}
M_{jk}=\matr{ccc}{
 +2 & -1 & -1 \\
 -1 & +2 &    \\
 -1 &    & +2
},\qquad
M^{-1}_{jk}=\matr{ccccccc}{
1 & \sfrac{1}{2} & \sfrac{1}{2} \\
\sfrac{1}{2} & \sfrac{3}{4} & \sfrac{1}{4}  \\
\sfrac{1}{2} &  \sfrac{1}{4} & \sfrac{3}{4},
}
\]
while for the $\alg{su}(4)$ notation we need to
permute the first two rows and columns
\[\label{eq:CartanSU4}
M_{jk}=\matr{ccc}{
 +2 & -1 &    \\
 -1 & +2 & -1 \\
    & -1 & +2
},\qquad
M^{-1}_{jk}=\matr{ccccccc}{
\sfrac{3}{4} & \sfrac{1}{2} & \sfrac{1}{4}  \\
\sfrac{1}{2} &            1 & \sfrac{1}{2} \\
\sfrac{1}{4} & \sfrac{1}{2} & \sfrac{3}{4}
}.
\]
%

\section{The Sigma-Model on $\Real\times S^2$}
\label{sec:RS2}

In this section we apply the general results
obtained in section \secref{sec:Sigma}
to the case of the Sigma-Model on $\Real\times S^2$. 

\subsection{Properties of the Monodromy}

The isometry group $\grp{SO}(3)$ of $S^2$ is
locally isomorphic to $\grp{SU}(2)$.
The spinor representation $\rep{2}$ of $\grp{SO}(3)$
can be viewed as the fundamental representation of
$\grp{SU}(2)$.
Correspondingly the monodromy matrix $\mono^\indspin$
can be regarded as the $\grp{SU}(2)$ monodromy matrix
which is diagonalized as
\[
\mono^\indspin\simeq\diag (e^{ip},e^{-ip}).
\]
This quasi-momentum $p$ is identified as
\[
p=\half q_1.
\]
It exhibits the inversion symmetry
\[\label{eq:InvSymS2}
p(1/x)=-p(x)+2\pi n_0,
\]
has the pole structure
\[\label{eq:1pSing}
p(x) =
\frac{\pi\dimn}{\sqrt{\lambda}\,(x\mp 1)}+
\bigorder{(x\mp 1)^0}
\qquad \mbox{for }x\to\pm 1
\]
as well as the asymptotic behavior
\[\label{eq:pasymp}
p(x)=\frac{1}{x}\,\frac{2\pi J}{\sqrt{\lambda}}
+\order{1/x^2}.
\]
where $r=2J$ is the Dynkin label of $\grp{SU}(2)$.
The analyticity condition reads
\[\label{eq:BeqS2}
2\sheetsl(x)=2\pi n_{a},
\qquad x\in \contour_{a}.
\]

\subsection{Algebraic Curve}
\label{sec:Alg.sdf}

Here we repeat the counting of moduli of the algebraic curve
for the case at hand. 
We make the most general ansatz
\[\label{eq:ElSpC2}  
p'(x)= \frac{Q(x+1/x)}{x(x-1/x)^2\sqrt{\prod_{b=1}^{A}(x-x^\ast_b)(1/x-x^\ast_b)}}\,,
\]
where $Q(u)$ is a polynomial of degree $\half A+1$. 
This corresponds to an inversion-symmetric two-sheeted algebraic curve 
with $p'_1+p'_2=0$ and $A$ cuts and the correct singular and
asymptotic behavior described in the previous
section. Note that $A$ must be even to satisfy the general behavior.

There are in total $\sfrac{3}{2}A+2$ free parameters which we shall
now constrain. 
Firstly, the residues of double poles at $x=\pm 1$ need to
be equated \eqref{eq:1pSing} giving one condition. 
The absolute value represents the dimension which 
we shall not fix directly.
Single poles, which would give rise to undesired logarithmic
behavior in $p(x)$ are automatically absent due 
to the symmetry \eqref{eq:InvSymS2}.
Vanishing of $\contour[A]$-cycles yields $\half A$ 
conditions (c.f.~\figref{fig:cycles} for
an illustration of the cycles corresponding to
a branch cut $\contour_a$). 
Note that the sum of $\contour[A]$-cycles around
a symmetric pair of cuts vanishes due to the
symmetry \eqref{eq:InvSymS2}
\[
0=\oint_{\contour[A]_a} dp=
-\oint_{1/\contour[A]_a} dp
\]
Therefore there sum of all $\contour[A]$-cycles is automatically
zero and the number of conditions is not reduced
from the known behavior at $x=\infty$.
Finally, integrality of $\contour[B]$-periods gives 
$\half A+1$ constraints, because each symmetric pair is
related 
\[
2\pi n_a=\int_{\contour[B]_a}dp=4\pi n_0-\int_{1/\contour[B]_a}dp,
\qquad
\int_0^\infty dp=-2\pi n_0.
\]
up to the total momentum $2\pi n_0$.
In total we have constrained $1+\half A+(\half A+1)$ 
coefficients and end up with $\half A$ moduli. 
These correspond to one fillings for each pair
of cuts.

\subsection{Example}
\label{sec:Alg.EqSTWO}

In the simplest case, the curve is described by two cuts $A=2$.
\[\label{eq:ElSpC}  
p'(x)= -\frac{2\pi}{\sqrt{\lambda}}\, \frac{J\sqrt{ab}\,(x-1/x)^2+D(1+ab)(x+1/x)-2D(a+b)}
                                          {x(x-1/x)^2\sqrt{(x-a)(1/x-a)(x-b)(1/x-b)}}\,.
\]
where we fixed the all parameters of $Q(u)$ in \eqref{eq:ElSpC2}
but two branch points $a,b$, from
\eqref{eq:1pSing,eq:pasymp}.
We can fix the branch points and the dimension $D$ from the integrals
\[\label{eq:Acyc}
\oint_{\contour[A]_1} dp=0,\qquad
\oint_{\contour[B]_1} dp=2\pi n_1,\qquad
\oint_{\contour[B]_2} dp=2\pi n_2.
\]
Note that the cycle $\contour[A]_2$ is equivalent to $\contour[A]_1$
under the symmetry and the mode numbers are related to
the total momentum by $n_1+n_2=2n_0$. The filling is directly
related to the charge $J$ and does not require a further condition.
The most general solution to these equations 
corresponds to the solution of the Neumann-Rosochatius
system, see \cite{Arutyunov:2003za}, restricted to a single
spin. Three particularly simple solutions have 
$n_1=\pm n_2$ or $n_2=0$ corresponding to 
two folded and one circular string.
They are obtained by setting $a=-b$. These were investigated 
in \cite{Beisert:2004hm}, Appendices C.1.1 and C.2.1. 
There the sigma model on $\Real\times S^3$ as used
which effectively does not have the $x\to 1/x$ symmetry,
see \secref{sec:RS3}. Instead, the four branch points
were assumed to be at $\pm a,\pm b$. When one spin
is sent to zero to reduce to the $\Real\times S^2$ model 
($\alpha=0,1$)%
\footnote{The folded solution with $\alpha=1$ corresponds
to the one found in \cite{Gubser:2002tv}.}
the solution recovers the inversion symmetry:
One finds that $a$ and $b$ are related by $b=\pm 1/a$.
This is a nice confirmation of the symmetry property.
Unfortunately all these solutions have 
a singular/fractional/trivial
weak-coupling expansion as proposed by Frolov and Tseytlin
and can therefore not be compared directly to gauge theory.

\bibliography{bks}
\bibliographystyle{nb}

\end{document}